\documentclass[fleqn,usenatbib]{mnras}

\usepackage{newtxtext,newtxmath}
\usepackage[T1]{fontenc}

\DeclareRobustCommand{\VAN}[3]{#2}
\let\VANthebibliography\thebibliography
\def\thebibliography{\DeclareRobustCommand{\VAN}[3]{##3}\VANthebibliography}

\usepackage{graphicx}	% Including figure files
\usepackage{amsmath}	% Advanced maths commands
\usepackage{multirow}
\usepackage{siunitx}

\defcitealias{LeConte_2024}{Paper I}
\defcitealias{Guo_2025}{G25}
\defcitealias{Salcedo_2025}{ES25}
\defcitealias{Géron_2025}{GZ CEERS}
\defcitealias{Kim_2021}{K21}
\defcitealias{Gadotti_2011}{G11}
\defcitealias{Liang_2024}{L24}

\title[]{The evolution of the bar fraction and bar lengths in the last 12 billion years}

\author[Z. A. Le Conte et al.]{Zoe A. Le Conte,$^{1}$\thanks{E-mail: zoe.a.le-conte@durham.ac.uk}
Dimitri A. Gadotti,$^{1}$
Leonardo Ferreira,$^{2}$
Christopher J. Conselice,$^{3}$
\newauthor
Camila de Sá-Freitas,$^{4}$
Taehyun Kim,$^{5}$
Justus Neumann,$^{6}$
Francesca Fragkoudi,$^{7}$
E. Athanassoula,$^{8}$
\newauthor
and
Nathan J. Adams$^{3}$
\\
$^{1}$Centre for Extragalactic Astronomy, Department of Physics, Durham University, South Road, Durham DH1 3LE, UK\\
$^{2}$Instituto de Matemática Estatística e Física, Universidade Federal do Rio Grande, Rio Grande, RS, Brazil\\
$^{3}$Jodrell Bank Centre for Astrophysics, University of Manchester, Oxford Road, Manchester M13 9PL, UK\\
$^{4}$European Southern Observatory, Santiago, Vitacura, Chile\\
$^{5}$Department of Astronomy and Atmospheric Sciences, Kyungpook National University, Daegu, 41566, Republic of Korea\\
$^{6}$Max-Planck-Institut f\"{u}r Astronomie, K\"{o}nigstuhl 17, D-69117 Heidelberg, Germany\\
$^{7}$Institute for Computational Cosmology, Department of Physics, Durham University, South Road, Durham DH1 3LE, UK\\
$^{8}$Aix Marseille Univ, CNRS, CNES, LAM, Marseille, France\\
}

\date{Accepted 2025 November 10. Received 2025 November 10; in original form 2025 September 05}

\pubyear{\the\year{}}

\begin{document}
\label{firstpage}
\pagerange{\pageref{firstpage}--\pageref{lastpage}}
\maketitle

\begin{abstract}
We investigate the evolution of the bar fraction and length using an extended JWST NIRCam imaging dataset of galaxies at $1 \leq z \leq 4$. We assess the wavelength dependence of the bar fraction and bar length evolution by selecting a nearly mass-complete CEERS disc sample and performing visual classifications on the short (F200W) and long (F356W+F444W) wavelength channels. A similar bar fraction is observed for both samples, and combined, we find a declining bar fraction from $0.16^{+0.03}_{-0.03}$ to $0.07^{+0.03}_{-0.01}$ over the redshift range. No evolution in the F356W+F444W bar length is measured, with a mean of 3.6\,kpc. A slight increase of $\sim 1$\,kpc towards $z = 1$ is measured in the F200W sample, with a mean of 2.9\,kpc. We find that the correlation between bar length and galaxy mass, for massive galaxies at $z < 1$, is unseen at $z > 1$. By incorporating barred galaxies at $z<1$, we show that there is a modest increase in the bar length ($\approx 2$\,kpc) towards $z=0$, but bars longer than $\approx8$\,kpc are only found at $z<1$. We show that bars and discs grow in tandem, for the bar length normalised by disc size does not evolve. Not only is a significant population of bars forming beyond $z = 1$, but our results also show that some of these bars are as long and strong as the average bar at $z\approx0$.
\end{abstract}


%%%%%%%%%%%%%%%%%%%%%%%%%%%%%%%%%%%%%%%%%%%%%%%%%%

%%%%%%%%%%%%%%%%% BODY OF PAPER %%%%%%%%%%%%%%%%%%

\section{Introduction}
\label{Sec: intro}
The majority of disc galaxies in the local Universe are hosts of striking stellar bars \citep[e.g.,][]{Eskridge_2000,Marinova_2007,Aguerri_2009,Buta_2015}. Occurring in more than a third of disc galaxies \citep[e.g.,][]{Barazza_2009,Nair_2010b,Masters_2011}{}{}, bars are one of the most abundant structures found in observational studies, and their occurrence increases to two-thirds or more when weaker bars are included, particularly in the near-infrared \citep[e.g.,][]{Vaucouleurs_1991,Menendez_Delmestre_2007,Sheth_2008,Erwin_2018}{}{}. Bars are the primary drivers of internal secular evolution as they effectively redistribute the angular momentum, baryonic, and dark-matter content of galaxies through the torque of the bar \citep[e.g.,][]{Lynden_1972,Athanassoula_2003,Athanassoula_2005,Regan_2006,Menendez_Delmestre_2007,Matteo_2013,Fragkoudi_2018}{}{}. Bar-driven evolution impacts star formation across the disc, most notably in the central region with the formation of a nuclear disc \citep[e.g.,][]{Sanders_1980,Knapen_1995,Allard_2006,Coelho_2011,Lorenzo_2012,Bittner_2020,Gadotti_2020}{}{}. Even the fuelling of the Active Galactic Nucleus (AGN) can be linked to bar-driven gas funnelling towards the inner kiloparsec of a galaxy \citep[e.g.,][]{Knapen_1995,Alonso_2013,Cisternas_2015,Alonso_2018,Silva-Lima_2022,Garland_2023,Garland_2024}{}{}.

Through theoretical efforts, it is understood that the presence of a bar indicates that a disc is sufficiently self-gravitating and dynamically cold \citep[e.g.,][]{Hohl_1971,Kalnajs_1972,Ostriker_1973,Sellwood_1993}{}{}. The relatively quick formation of a bar, typically on a timescale of a hundred million years, is induced by disc instabilities in these presumed-to-be-settled massive disc galaxies. Hence, this provides galaxy evolution studies with a marker for the commencement of internal secular evolution. Investigations into the abundance of bars at different epochs are critical to our understanding of the evolutionary process of galaxies in the early Universe. 

Several observational studies have reported the incidence of stellar bars out to $z \approx 1$ and found a decreasing trend from the local out to the higher redshift Universe, with the methodology for bar identification varying; from visual classifications \citep[e.g.,][]{Vaucouleurs_1991,Athanassoula_1990,Eskridge_2000,Nair_2010b,Cheung_2013,Simmons_2014,Buta_2015}{}{} to more automated processes with isophotal analysis and Fourier analysis \citep[e.g.,][]{Wozniak_1995,Buta_1998,Elmegreen_2004,Jogee_2004,Marinova_2007,Guo_2023,Guo_2025}{}{}. 

Between $0.0 < z < 1.5$, \cite{Abraham_1999} found the bar fraction to decline for a sample of disc galaxies in the Hubble Deep Field-North and -South. Within the range $0.20 < z < 0.84$, \cite{Sheth_2008} reported a decrease in the bar fraction for visually and ellipse fit classified galaxies using the 2 deg$^{2}$ Cosmic Evolution Survey (COSMOS). From $z \sim 0.4$ to $z \sim 1.0$, \cite{Melvin_2014} found the bar fraction to decrease by a factor of two in visually classified COSMOS galaxies. Extending to $z \simeq 1.5$, \cite{Simmons_2014} observes the decline in the incidence of barred galaxies and suggests a constant bar fraction of 10 per cent at $z \geq 1$. The capabilities of the Hubble Space Telescope (HST) have furthered bar studies to $z \leq 1$, and these results have a consensus that the bar fraction drops to $\leq 10$ per cent by $z \sim 1$. Hence, this implies that bar-driven evolution commences once the Universe has aged by $\sim 6$ Gyr. 

%Another notable area of study is the disc attributes which are more likely to host a bar, which are thought to be massive, gas-poor and red discs at low-$z$ \citep[e.g.,][]{Masters_2011,Gavazzi_2015,Cervantes_2017}, but \cite{Erwin_2018} contradicts this by suggesting SDSS samples under predicted the bar fraction at lower stellar masses, preferentially missing blue gas-rich galaxies. 

%Bars are discerned by an over-density of aged stellar populations on eccentric orbits at the central region of massive disc galaxies \citep[e.g.,][]{Weinburg_1985,Contopoulos_1989,Athanassoula_1992,Kormendy_2004}{}{}. Their non-axisymmetric nature is due to elongated orbits parallel to the semi-major axis of the bar, namely the $x_{1}$ orbital family and higher multiplicity families \citep{Contopoulos_1980,Wang_2022}. The appearance of bars can range from long with high ellipticity and rectangular to short and more oval. The orientation and inclination of the disc determine the observed shape.

The improved sensitivity and broader wavelength range of the James Webb Space Telescope (JWST) have expanded bar fraction studies out to $z \approx 4$. In an earlier work \citep[][hereafter referred to as \citetalias{LeConte_2024}]{LeConte_2024}, we used JWST to explore for the first time the abundance of barred galaxies out to $z = 3$. \citetalias{LeConte_2024} identified that the bar fraction observed in JWST NIRCam images is twice that viewed in bluer HST filters of similar resolution, but still, a declining trend was observed in the bar fraction, from $\approx 17.8^{+ 5.1}_{- 4.8}$ per cent at $1 \leq z < 2$ to $\approx 13.8^{+ 6.5}_{- 5.8}$ at $2 \leq z \leq 3$. We showed that despite the similarity in spatial resolution, it was due to the improved sensitivity and longer wavelengths available with NIRCam, which enabled us to find twice as many barred galaxies as previously found with HST. The longer wavelengths observed with JWST could facilitate these results, as the NIR is a better tracer of the underlying stellar mass distribution of barred galaxies, which are mainly composed of older stellar populations \citep[][]{Gadotti_2006,Marinova_2007,Sanchez_2011,deSaFreitas_2023}. 

Our study already informed us that bar-driven internal evolution commenced $\sim 11$ Gyrs ago, by virtue of 30 newly classified barred galaxies in the redshift range $1 \leq z \leq 3$. The Point Spread Function (PSF) full-width half maximum (FWHM) limited this study to bars with radii greater than $\sim 2.5$ kpc, leading us to conclude that our results could still be a lower limit to the bar fraction. 

Since our first study, several JWST bar fraction investigations have supported our results of a bar fraction twice that of HST \citep[][]{Salcedo_2025,Géron_2025}. \citet[][hereafter referred to as \citetalias{Guo_2025}]{Guo_2025} found a decline in the JWST CEERS bar fraction within the agreement of \citetalias{LeConte_2024}, across the redshift range $0.5 \leq z \leq 4$ in a mass-limited, $M_{\star} > 10^{10} M_{\odot}$ sample. Using ellipse fit techniques, the results showed a decrease in the bar fraction from $28.2 \pm 3.6$ per cent at $z \sim 0.5$ to $6.4 \pm 2.4$ per cent at $z \sim 3.5$. They evaluated the impact of observed wavelength on bar length evolution; however, they observed minimal correlation and did not report any signs of bar length growth across the redshift range in their F444W sample, but identified a slight growth toward lower redshifts in their F200W sample. 

In-depth studies have found bar-driven gas inflow in individual galaxies at high-$z$ \citep[e.g.,][]{Huang_2023,Huang_2025,Pastras_2025}. Early Universe bars are now recognised as long-lived and robust structures that exhibit the same mechanisms present in local Universe barred galaxies, such as bar-driven gas inflows.

%The statement that bar-driven evolution has commenced at $z \approx 3$ is supported not only by these bar fraction studies, but also by in-depth studies of individual barred galaxies at high-$z$. The prominent stellar bar at $z \sim 2.5$ found in \cite{Huang_2023} and \cite{Huang_2025} indicates that bar-driven star formation is present in a gas-rich massive disc galaxy, for which the funnelling of molecular gas is observed towards the galactic centre. Early Universe bars are now seen to be long-lived and robust structures which exhibit the same mechanisms present in local Universe barred galaxies, such as bar-driven gas inflows. Active star formation from bar-driven rapid gas inflow at Cosmic Noon is found in a massive gas-rich disc galaxy at $z \sim 1.5$ studied by \cite{Pastras_2025}. 

Several barred galaxies at high redshift have been observed recently with ALMA. A massive barred galaxy was confirmed by \cite{Amvrosiadis_2025} at $z = 3.8$, 1.7 Gyr after the Big Bang, in a gas-rich system. Additionally, these young barred galaxies have long bars. In the case of a barred galaxy at $z = 4.4$, \cite{Tsukui_2024} measured the bar to be 3.3 kpc long, which is comparable to the bars ranging in length from $\sim 1$ to 10 kpc in the local Universe \citep[e.g.,][]{Erwin_2005,Gadotti_2011}. These detailed studies push the boundary of bar identification studies and highlight that some bars form significantly earlier than previously thought possible.

Theory suggests that over time, a bar will evolve and grow in length, as the rotational pattern speed of the bar slows, extending resonances to greater radii, thus allowing the trapping of stars on elongated orbits to reach these new limits \citep[][]{Athanassoula_2003}. In the Auriga cosmological simulations, bars that form at $z \leq 1$ form short in length and grow with time, whereas those formed at high-$z$ form very long (with a semi-major axis of about 6\,kpc), possibly due to an interaction-driven bar formation, and do not show an evolution in length \citep[][]{Fragkoudi_2025}, i.e., they form `saturated' in length. Alternatively, some cosmological simulations find weak interaction-induced bars at Cosmic Noon \citep[e.g.,][]{Bi_2022}, which weaken and reform throughout their evolution.

Observations at $z \approx 0$ have suggested that bars grow with time. In a sample of 300 nearby barred galaxies, \citet[][hereafter referred to as \citetalias{Gadotti_2011}]{Gadotti_2011} found longer and stronger bars in galaxies with a higher bulge-to-total luminosity ratio (indicating more evolved galaxies), suggesting that the bars grow longer and stronger with time. A similar argument was put forward by \citet{Erwin_2018}, finding that more massive galaxies host longer bars in a sample of nearby barred galaxies from the Spitzer Survey of Stellar Structure in Galaxies \citep[S$^4$G][]{Sheth_2010}. Recently, with direct measurements of bar ages in nearby galaxies of the TIMER sample \citep{Gadotti_2019}, \citet{Freitas_2025} found that older bars are longer. It should be noted that these observational results are, to date, still consistent with the Auriga results, whereby bar growth depends on bar formation epoch. 

However, in direct observations beyond the nearby Universe, \citet[][hereafter referred to as \citetalias{Kim_2021}]{Kim_2021} found no evolution between $0.2 < z < 0.8$ and a mean bar length at those redshifts similar to the mean bar length from \citetalias{Gadotti_2011} at $z \approx 0$. \citetalias{Guo_2025} is the only study to have investigated the properties of a sample of barred galaxies beyond $z = 1$. The average projected bar length reported in \citetalias{Guo_2025} increased from $\sim 1.7$ – 2.5 kpc at $z \sim 2.5$ to $\sim 2.6$ – 3.8 kpc at $z \sim 0.75$, in the F200W sample, but no evolution was observed in the F444W sample.

In this paper, we investigate the abundance of barred galaxies and their properties at $z \geq 1$ using JWST NIRCam images (\S~\ref{sec:reduction}) selected by redshift and stellar mass (\S~\ref{sec:sampleselection}). In contrast with \citetalias{LeConte_2024}, our parent sample here is significantly enlarged (by nearly a factor of three in sky area), and this study considers galaxies up to $z=4$, also exploring wavelength dependencies. The following methodology is performed independently on both short (F200W) and long (F356W+F444W) wavelength NIRCam images. The technique of ellipse fits is performed to reduce the sample to less inclined, more face-on galaxies (\S~\ref{Sec: opt}). The disc sample is selected by removing point sources through the automated process of fitting the galaxy with a 2D S\'ersic model (\S~\ref{Sec: disc ser}). Subsequently, the visual classifications are determined by five independent participants, of which the final disc and barred galaxy samples are obtained (\S~\ref{Sec: disc vis}). 

Our results show strong agreement with the bar fraction from \citetalias{LeConte_2024}, along with the finding of three barred galaxies at $z \geq 3$ (\S~\ref{Sec: bar frac}). We report no strong correlation between the bar fraction and stellar mass(\S~\ref{Sec: bar prop}). Using ellipse fit techniques, we measure the evolution of bar length and compare the differing trends of short and long wavelengths (\S~\ref{Sec: bar length}). Similarly, the strength of the bar differs between the two filters (\S~\ref{Sec: bar strength}). We discuss the implications of our findings on when bar-driven evolution commences and the impact of observed wavelength on such studies in \S~\ref{Sec: discussion}, and present our conclusions in \S~\ref{Sec: conclusions}. Throughout this study, we assume the latest Planck flat $\Lambda$CDM cosmology with H$_{0}$ = 67.36, $\Omega_{m}$ = 0.3153, and $\Omega_{\Lambda}$ = 0.6847 \citep{Planck_2020}.

\section{The sample}
\label{Sec: parent sample}
\begin{figure}
	\includegraphics[width=\columnwidth]{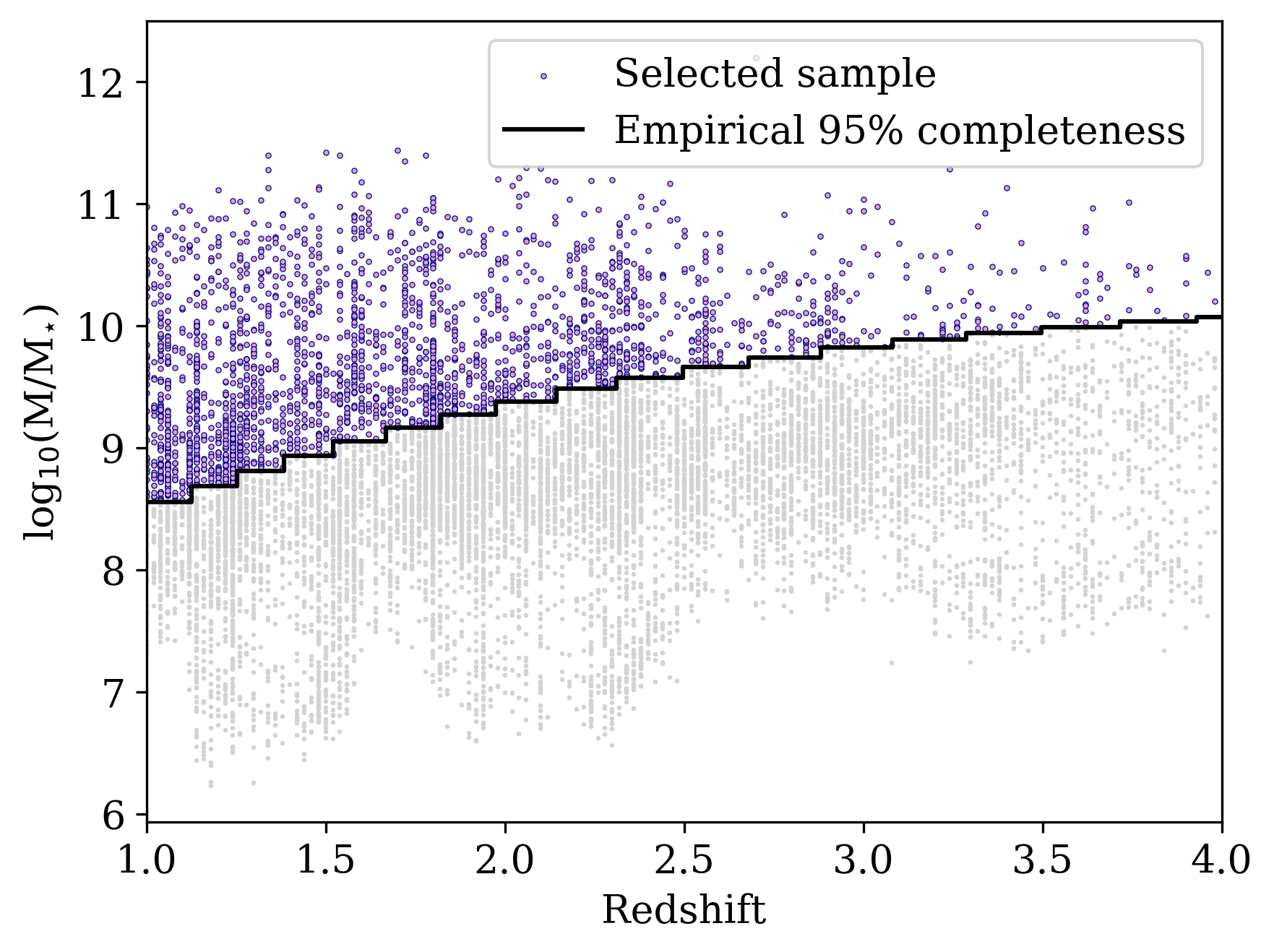}
    \caption{Stellar mass-redshift distribution of CEERS galaxies in the redshift range $1 \leq z \leq 4$. We adopt the 95\% empirical completeness of \citet[][black step function]{Duncan_2019}, originating from CANDELS data. Our sample is selected to be those with stellar masses greater than the redshift-dependent stellar mass limit (purple).}
    \label{fig:mass cut}
\end{figure}

To investigate the epoch at which bar-driven secular evolution commences, we use the ten public NIRCam JWST observations from the Cosmic Evolution Early Release Science Survey (CEERS; PI: Filkelstein, ID=1345, \citealt{Finkelstein_2023}), combining June 2022 observations (CEERS1, CEERS2, CEERS3 and CEERS6) and January 2023 observations (CEERS4, CEERS5, CEERS7, CEERS8, CEERS9 and CEERS10), which overlap with 50 per cent of the Cosmic Assembly Near-IR Deep Extragalactic Legacy Survey (CANDELS; \citealt{Grogin_2011, Koekemoer_2011}) on the Extended Groth Strip field (EGS). Together, the data cover $\sim \ 88$ arcmin$^2$ of an area with CANDELS HST overlap. 

\subsection{Data reduction pipeline}
\label{sec:reduction}
We reprocessed all raw NIRCam exposures retrieved from the \texttt{Mikulski Archive for Space Telescopes (MAST)} with the official JWST pipeline (v1.8.2, CRDS v1084), following the procedures of \cite{Ferreira_2022} and \cite{Adams_2023} but incorporating several refinements. After Stage 1 calibration and application of the latest post-flight reference files, we remove large-scale "wisp" artefacts with second-generation Space Telescope Science Institute (STScI) templates. Stage 2 processing includes the 1/f-noise correction \citep[][]{Adams_2025}, while Stage 3 is split so that sky subtraction—flat, then 2-D with photutils—is tuned on a frame-by-frame basis before the full mosaicking step.

Background-corrected frames are then combined using the STScI \texttt{DrizzlePac} software, anchoring the astrometry to GAIA DR3 via \texttt{tweakreg}; each filter is subsequently re-projected onto the F444W grid at $0.03^{\prime\prime}$pixel$^{-1}$ to ensure sub-pixel alignment across the dataset. A comprehensive description of the workflow and its validation is given in \cite{Adams_2024, Conselice_2024a} and \cite{Harvey_2024}.

\subsection{Sample selection}
\label{sec:sampleselection}
For this work, we use the robust photometric redshifts and stellar masses originating from CANDELS-based catalogues \citep[][]{Duncan_2019}. Measurements are taken from HST, Spitzer and ground-based observations and calibrated by spectroscopic redshifts, with an average outlier fraction of $\frac{|\Delta z|}{1+z_{spec}} \sim 5\%$ (see \citealt{Duncan_2019} for details). We do not use JWST-based photometric redshifts, for \citet[][]{Conselice_2024} report essentially no difference in results at $z < 3$.

We select a sample of sources that lie within the CEERS footprint and have photometric redshifts and stellar masses. We limit our sample selection to the high redshift range $1 \leq z \leq 4$. No magnitude or signal-to-noise cut is applied, but rather an empirical 95 per cent mass completeness limit dependent on redshift is adopted from \citet[][see their Sect. 3.4.1 for further details]{Duncan_2019}. For full details on how the mass-completeness limit is determined by the flux limit of the HST, Spitzer and ground-based surveys, we refer the reader to the overall method in \cite{Duncan_2019}. Our redshift and stellar mass cuts result in a sample of 2438 galaxies selected for this paper, for which the stellar mass-redshift distribution is shown in Fig. \ref{fig:mass cut}. For our sample, we produce $30 \ \rm mas$ 128x128 pixel cutouts for the JWST filters, namely F200W, F356W and F444W (the NIRCam filter selection is discussed in \S~\ref{Sec: filters}).

\section{Methodology}
\label{Sec: method}
This study aims to observe the commencement of bar-driven secular evolution at high z, through measurements of the bar fraction and bar properties. The bar fraction $f_{bar}$ is defined as the fraction of disc galaxies which are hosts to a stellar bar:

\begin{equation}
    f_{bar} = \frac{N_{bar}}{N_{disc}},
\end{equation}

where $N_{disc}$ is the observed sample size of disc galaxies and $N_{bar}$ is the number of barred galaxies identified in the disc galaxy sample. Hence, in this work, we seek to identify a disc galaxy sample and subsequently a barred galaxy sample. 

The parent sample is refined through a series of galaxy removal procedures. Considering stellar bars are distinguishable in moderately inclined disc galaxies, and become challenging to define in highly inclined galaxies, the sample must be reduced to remove near edge-on galaxies. With a moderately inclined sample, point sources can be removed from the sample through constraints on the distribution of Sérsic indices. We explore the possibility of using the global Sérsic index to define the disc sample, but determine that at high redshifts,  the Sérsic index is not adequate to differentiate between the morphological types of discs and spheroids. Hence, we conduct visual classifications for morphological type and bar identification. We discuss the NIRCam imaging selected for our sample in \S~\ref{Sec: filters}, the removal of highly inclined objects through an optimisation routine in \S~\ref{Sec: opt}, using the global Sérsic index to remove point sources in \S~\ref{Sec: disc ser}, followed by disc and bar visual classifications in \S~\ref{Sec: disc vis}.

\subsection{NIRCam imaging}
\label{Sec: filters}
To observe prominent bars in the early Universe, we select the wide and long-wavelength JWST NIRCam filters F356W and F444W to capture the underlying stellar mass distribution of the disc whilst reducing effects from dust obscuration and clumpy star formation, which can hamper the identification of bars. To observe NIRCam images with comparable rest frame wavelengths, the filter F356W is used for galaxies between $1 \leq z < 2$; obtaining a rest-frame wavelength of $1.4\mu m$ for the median redshift of the bin $z=1.5$. Furthermore, F444W is used for galaxies between $2 \leq z \leq 4$; with a rest-frame wavelength of $1.1\mu m$ for the median redshift $z=3$. The NIRCam empirical PSF FWHM\footnote{PSF FWHM taken from the JWST user documentation: https://jwst-docs.stsci.edu/jwst-near-infrared-camera/nircam-performance/nircam-point-spread-functions} is amplified at longer wavelengths and is $0.116^{\prime\prime}$ and $0.145^{\prime\prime}$ for the filters F356W and F444W, respectively. At the median of the two redshift bins, the corresponding linear resolutions are 1.01 kpc and 1.14 kpc. With regards to bar detection, the minimum identifiable bar length is 2 x FWHM \citep[e.g.,][]{Erwin_2005}, which for the NIRCam long-wavelength channels are 2.02 kpc and 2.28 kpc, respectively. Hence, we are likely missing bars shorter than $\approx 2$ kpc in this study, when using the long wavelength filters. 

To evaluate the effects of rest-frame wavelength, we also employ images of all of the galaxies across $1 \leq z \leq 4$ in the NIRCam short wavelength filter F200W. This filter corresponds to a rest-frame wavelength of $0.67\mu m$ at $z=2$ and has a PSF FWHM of $0.066^{\prime\prime}$, resulting in a linear resolution of 0.6 kpc. Hence, the shortest bars identified with these NIRCam images are 1.2 kpc in length. However, at this rest-frame wavelength, dust and star formation effects may hamper the identification of bars. Hence, our methodology is independently repeated for the NIRCam imaging in the short-wavelength filter F200W and long-wavelength filters F356W+F444W. We note that in a few cases, galaxies fall between the gaps of NIRCam detectors, and therefore, we could not retrieve images in both F200W and F356W/F444W, but we still include these sources in our sample when available, as from this point onward, the samples are independent.

\begin{figure}
	\includegraphics[width=\columnwidth]{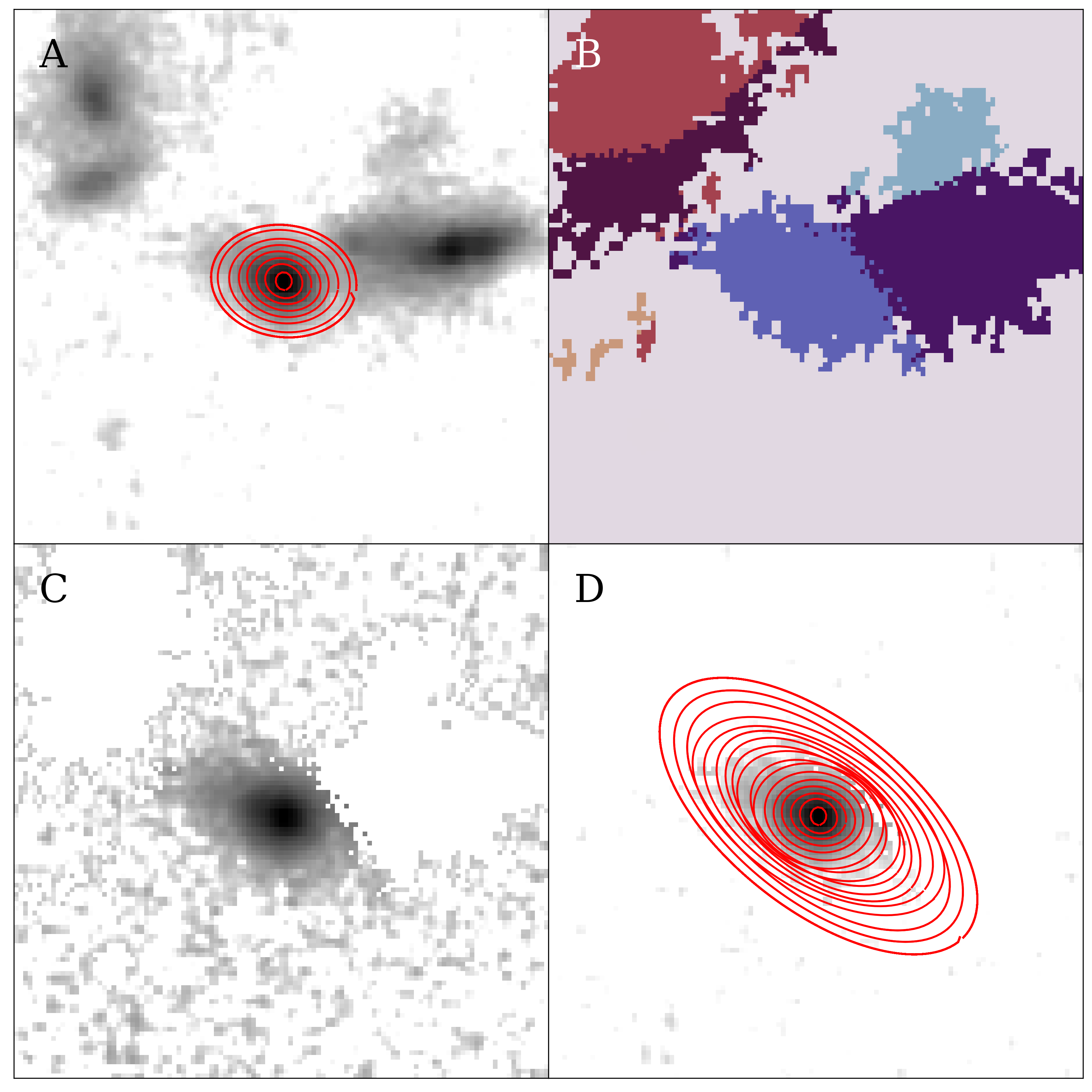}
    \caption{Masking of neighbouring sources with SExtractor performed on the example galaxy EGS$\_1246$. Top left - JWST NIRCam F444W image with fitted elliptical isophotes. Top right - SExtractor segmentation map. Bottom left - Resulting image with identified neighbouring sources masked out. Bottom right - Masked image with newly fitted elliptical isophotes. This figure shows the influence of nearby sources on the orientation and elongation of isophotal ellipse fits.}
    \label{fig:sextractor}
\end{figure}

Before progressing with our sample optimisation, at this point in our methodology, Sextractor is employed to mask neighbouring objects \citep[][]{Bertin_1996}. Figure \ref{fig:sextractor} demonstrates the masking process on the example galaxy EGS$\_1246$ in the F444W NIRCam image. This process is particularly important, as the sample optimisation procedure depends on the ellipticity of the isophotes that trace the outer disc, which in this case has changed significantly due to the masking of sources.

\subsection{Sample optimisation}
\label{Sec: opt}
Barred galaxies can be clearly identified in face-on galaxies; however, at higher inclinations, this becomes increasingly challenging. Thus, to avoid this bias, we remove highly inclined $i > 65$ galaxies from the sample by an optimisation process. A detailed description can be found in \citetalias{LeConte_2024}, but the three stages are summarised below. The inclination of a galaxy is measured by fitting elliptical isophotes to the image, using \texttt{photutils.isophote} from Python's astropy package \citep{Bradley_2022}. An ellipticity limit, corresponding to $i > 65$, is then applied to the outer isophotes.

\begin{description}
    \item[\textbf{Phase 1}] The first run of elliptical isophotes fits is performed, with the pixel coordinates of the centre of the galaxy as free parameters. For a given galaxy, fittings are performed with a selection of initial parameters until the galaxy is successfully fit, even if the fits do not perfectly align with the galaxy. Approximately 90\% and 75\% of the parent sample had successful ellipse fits in the F356W+F444W and F200W filters, respectively. The remaining $\sim$ 10 - 25\% of galaxies that failed ellipse fittings are overly disturbed, poorly resolved, and/or low surface brightness systems and are removed from the sample.
    \item[\textbf{Phase 2}] \texttt{photutils.isophote} selects the central position of a galaxy by assessing a 10x10 window about a specified centre. We specify the centre of each galaxy as the average central pixel for the inner 10 - 40\% of the isophotes fitted in phase 1. The lower isophote limit is chosen to be greater than 2 x FWHM, simultaneously avoiding bright star-forming regions near the galactic centre, and the upper isophote limit avoids tracing structures toward the outer regions of the galaxy, such as spiral arms, which may lead to a mistaken identification of the galaxy centre.
    
    Isophote fitting is repeated on the sample of galaxies that pass phase 1, with the adaptation of a fixed centre. Phase 2 achieved a 75\% and 57\% success rate in the F356W+F444W and F200W filters, respectively. Those galaxies that failed, typically overly irregular or ambiguous in morphology, were removed.
    \item[\textbf{Phase 3}] We define the inclination of a galaxy by measuring the ellipticity of the outermost fitted ellipse,
    \begin{equation}
    e = 1 - \frac{b}{a} ,
    \end{equation}
    where $b$ is the minor axis length and $a$ is the major axis length. The inclination is defined as
    \begin{equation}
    \cos{i} = \frac{b}{a} .
    \end{equation}
    The success rate of phase 3 was 64\% and 63\% in the F356W+F444W and F200W filters, respectively. Galaxies that were determined to be too highly inclined were removed from the sample, resulting in overlapping optimised samples of 1073 and 649 galaxies, respectively, all of which had successful ellipse fits.
\end{description}

In summary, the optimisation phase aimed to fit elliptical isophotes to the NIRCam images and measure eccentric outer isophotes as having an approximate ellipticity $e > 0.5$, implying that the galaxy is highly inclined, leading to its removal from our sample. Overall, of the 2438 galaxies in our mass complete sample, the following number of galaxies were removed at each phase for F356W+F444W: 223 at phase 1, 540 at phase 2 and 597 at phase 3; resulting in an optimised sample of 1073 galaxies. As for F200W: 616 at phase 1, 782 at phase 2 and 386 at phase 3; resulting in an optimised sample of 649 galaxies. The optimisation routine reduced the F200W sample significantly more than the F356W+F444W sample, which we deem to be due to the short wavelength filter tracing star formation, appearing clumpy and becoming challenging for ellipse fitting.

\subsection{Morphological classification}
\label{Sec: discs}
Classification studies at high redshift have been able to differentiate galaxies into three distinct types: disc, spheroids and irregular. \cite{Ormerod_2024} studied the evolution of structural parameters with CEERS galaxies at $0.5 \leq z \leq 8$, to which they found shallower light profiles and diffuse brightness distributions for all morphological types at high-$z$, when compared to their low-$z$ counterparts. Therefore, uncertainties arise when trying to distinguish between galaxy types when using surface brightness profiles. Hence, we followed a two-step method to select the disc sample of this study. Firstly, we fit single S\'ersic models to obtain structural information and use this to select and remove point sources from our sample. We then had five classifiers visually classify and identify the disc and barred galaxies.

\subsubsection{S\'ersic fits}
\label{Sec: disc ser}

\begin{figure*}
	\includegraphics[width=\textwidth]{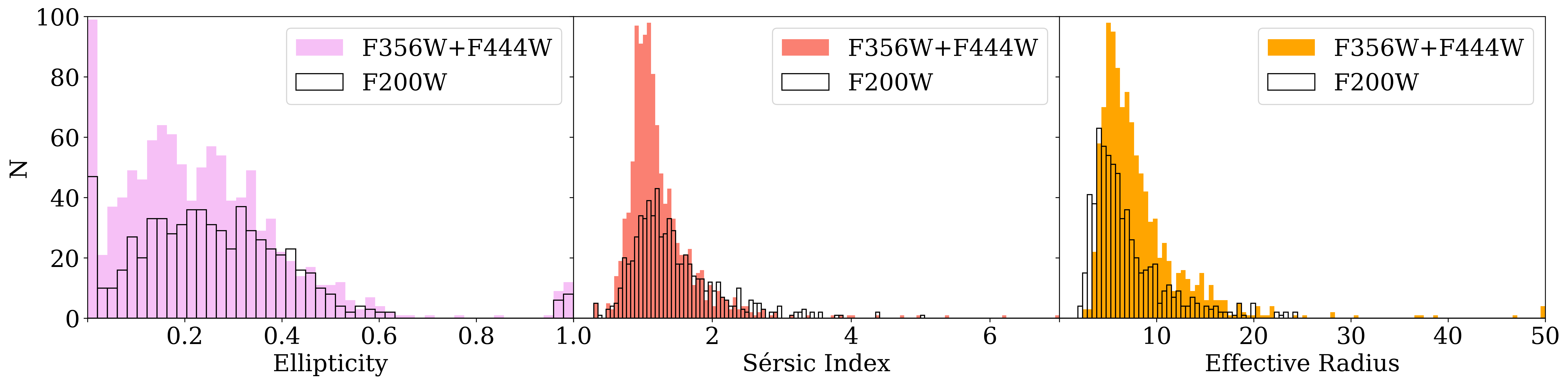}
    \caption{Distribution of the best-fit structural parameters from a single 2D S\'ersic function fitting using \texttt{Imfit}, performed on the optimised samples for galaxies between $1 \leq z \leq 4$. Models fit to F356W and F444W NIRCam images are filled distributions, and those as a result of fits to F200W NIRCam images are unfilled distributions with black edges. Left: single S\'ersic best-fit ellipticity, $e$. Middle: single S\'ersic best-fit S\'ersic index, $n$. Right: single S\'ersic best-fit effective radius, $r_{e}$.}
    \label{fig:imfit}
\end{figure*}

A 2D elliptical S\'ersic profile can model the light distribution in disc galaxies, and the program \texttt{Imfit} \citep[][]{Erwin_2015} is designed to fit such functions to astronomical images. The program produces a best-fitting model by minimising the Cash statistic using a differential evolution (DE) algorithm \citep[][]{storn_1997}, after convolving the model image with oversampled PSF images created by the Space Telescope PSF (STPSF) package\footnote{https://www.stsci.edu/jwst/science-planning/proposal-planning-toolbox/psf-simulation-tool}. The stochastic nature of the DE approach is beneficial because it can search large areas of the parameter space whilst being less likely to be trapped in local minima. In addition, this approach does not require initial parameter values, but only lower and upper limits, and has been shown to produce the best results when compared to iterative, human-supervised fits (Gadotti, subm.). 

We fitted a single S\'ersic component to all galaxy images, and the best-fitting model parameters that we focus on are ellipticity ($e$), S\'ersic index ($n$) and effective radius ($r_{e}$). The distributions of these parameters for all galaxies in the optimised samples of F356W+F444W and F200W are shown in Fig. \ref{fig:imfit}. For all parameters, the distributions show similar trends between the shorter and longer wavelength NIRCam filters. This is to be expected, as for each parameter, the model is convolved with the corresponding PSF, meaning that the single-component model is to a certain degree independent of the observed wavelength. The slight difference in $r_{e}$ between the two samples is less than $1\sigma$. Hence, the result is statistically insignificant and not explored further. The occurrence of only a few galaxies with $e > 0.6$ demonstrates the efficiency of the optimisation process described in \S~\ref{Sec: opt} to remove highly inclined galaxies. 

There is a noticeable peak in sources with $e \approx 0$, which we attribute mostly to point sources and unresolved sources. To reduce our sample for visual classification, we use the best-fit parameters $n$ and $r_{e}$ to define a point source sample. Point sources have discrete near-circular brightness distributions and are selected by the limits $e \leq 0.2$ and $n \leq 1$, selecting 152 and 37 galaxies in F356W+F444W and F200W, respectively. From this selected sample of point sources, the co-authors of this paper, DG and ZLC, visually inspected the NIRCam images and light profiles to check the accuracy of this process and reintroduced galaxies which we disagreed with being point sources. After this step, the optimised samples consist of 927 and 628 galaxies in F356W+F444W and F200W, respectively, from which the disc sample is determined.

In the local Universe, the morphological types of galaxies can be separated by their global $n$, with massive ellipticals having a high $n$ and a high degree of central concentration, whilst massive discs have a more diffuse central concentration and lower values of $n$. Figure \ref{fig:imfit} shows over 90 per cent of the sample to have values of $n$ between 0 and 2. Hence, we decided not to use limits on the S\'ersic index as a selection process for the disc sample, as the S\'ersic index distribution of these high redshift galaxies is significantly lower than ellipticals in the local Universe \citep[see also][]{Ormerod_2024}. Instead, we decided to make disc selection a part of the visual classification process.

\subsubsection{Visual classifications}
\label{Sec: disc vis}

\begin{table*}
    \centering
    \renewcommand{\arraystretch}{1.5} % Default value: 1
    \begin{tabular}{c|ccc|ccc|ccc}
    \hline
    Redshift & \multicolumn{9}{c}{Sample Size} \\
    \hline
     & \multicolumn{3}{c|}{F200W} & \multicolumn{3}{c|}{F356W+F444W} & \multicolumn{3}{c}{F200W+F356W+F444W} \\
     & SB & WB & UD & SB & WB & UD & SB & WB & UD \\
    \hline
    $1 \leq z < 2$ & 15 & 17 & 196 & 11 & 32 & 274 & 23 & 43 & 358 \\
    $2 \leq z < 3$ & 2 & 2 & 61 & 5 & 5 & 103 & 5 & 7 & 137 \\
    $3 \leq z \leq 4$ & 0 & 1 & 14 & 0 & 1 & 14 & 0 & 2 & 25 \\
    \hline
    \end{tabular}
    \caption{The sample sizes for the visually classified samples in F200W, F356W+F444W and F200W+F356W+F444W in three redshift bins. The sample size is presented for strongly barred galaxies (SB), weakly barred (WB) and unbarred disc galaxies (UD).}
    \label{tab: sample}
\end{table*}

This study incorporates disc identification into our visual classification method to find barred and unbarred disc galaxies. Five of the co-authors of this paper, ZLC, DG, CdSF, JN and TK, inspected the NIRCam F356W+F444W and F200W images and intensity radial profiles of each galaxy in the sample of 927 and 628 galaxies, respectively. The visual classification categories were \textit{Barred, Maybe Barred, Unbarred Disc, Spheroid, Other} and \textit{Unresolved}. Each classifier voted each galaxy into one of these six categories and independently repeated the process for short and long-wavelength NIRCam filters. An outline of the categories is as follows:
\begin{description} 
    \item[\textbf{Barred}] These galaxies present an ordered outer disc with a prominent elongated central stellar bar. The outer disc may have spiral arm features or rings.
    \item[\textbf{Maybe Barred}] Similar in structure to the barred galaxy category, but for deviations from a prominent barred structure. These include rounder, thicker bars, as well as systems with dominating spiral arms, which obscure a barred structure.
    \item[\textbf{Unbarred Disc}] These galaxies show a well-ordered disc, ranging from near circular to somewhat flattened. They have exponential surface brightness profiles and can contain features such as spiral arms, rings and centrally concentrated structures, but will be identified in one of the preceding categories if a bar is present.
    \item[\textbf{Spheroid}] Appearing smooth, round-like and featureless, these systems share the likeness of elliptical galaxies in the local Universe, sometimes with high degrees of central concentration.
    \item[\textbf{Other}] Some galaxies at high-$z$ appear peculiar and clumpy in nature, and these can often be disguised interaction or merger systems. If disciness cannot be observed, then the galaxy is assigned to this category.
    \item[\textbf{Unresolved}] Despite our efforts to remove point sources, a few can be found in our samples and would fall into this category at this stage. Occasionally, a bright central feature in a galaxy caused PSF diffraction spikes to dominate the image, hampering a proper morphological classification.
\end{description}

The ultimate classification of a galaxy is attributed to a consensus vote, in which 3 out of 5 classifiers agree on a category. When a consensus cannot be reached, the source is allocated to the \textit{Unclassified} category. The disc sample is compiled of galaxies attributed to the categories \textit{Unbarred Disc}, \textit{Barred} and \textit{Maybe Barred}, meaning 3 out of 5 votes from one to any combination of all three of these categories would proceed to a disc classification. Examples of unbarred disc classifications include, but are not limited to, galaxies which obtained 3 unbarred disc votes; 2 maybe-barred and 1 unbarred disc vote; 1 barred, 1 maybe-barred and 1 unbarred disc vote. We establish two barred galaxy classifications:
\begin{description}
    \item[\textbf{Strongly Barred}] These are disc galaxies that have prominent elongated stellar bars, and accordingly have achieved above the consensus threshold of 3 out of 5 votes for \textit{Barred}.
    \item[\textbf{Weakly Barred}] These disc galaxies have rounder and less apparent bars, subsequently achieve the threshold of 3 out of 5 for \textit{Maybe Barred} or by combining \textit{Barred} and \textit{Maybe Barred} votes.
\end{description}

This classification scheme is similar to that used in \citetalias{LeConte_2024}, for which the process worked well, and our results from \citetalias{LeConte_2024} were consistent with subsequent bar fraction studies. We deemed that the bar fraction should be determined using a similar methodology to identify the disc sample as well as the barred sample. Hence, as the disc sample cannot be established through limits on structural parameters, we have decided again to conduct visual classifications to identify barred galaxies at high-$z$. Morphological analysis of sources at high-$z$ is challenged by the faintness of the targets and their very small angular size. Thus, we include two barred galaxy classifications to provide upper and lower limits to the bar fraction.

The bar fraction is determined as:
\begin{equation}
    f_{bar} = \frac{N_{bar}}{N_{disc}} = \frac{N_{SB}+N_{WB}}{N_{UD}+N_{SB}+N_{WB}}
\end{equation}

where $N_{SB}$ is the observed sample size of strongly barred galaxies, $N_{WB}$ is the observed sample size of weakly barred galaxies, and $N_{UD}$ is the observed sample size of unbarred discs. The sample sizes for each redshift are presented in Table \ref{tab: sample}.

Strong agreement between classifiers is considered when the vote threshold is met. Visual classifications from the NIRCam filters F356W+F444W resulted in a high agreement between classifiers in 74 per cent of cases, and are similar for F200W, with 70 per cent of the cases in high agreement. 

We evaluated the reliability of agreement between the classifiers of this study in categorical assignment using the descriptive statistic Randolph's kappa, $\kappa$ \citep[][]{Randolph_2005}. The free-marginal multirater $\kappa$ measures the nominal agreement between multiple classifiers, employing 1 / number of categories as the proportion of agreement expected by chance. The statistic $\kappa$ takes values from 1 to -1; values between 1 and 0 indicate agreement better than chance, 0 indicates a level of agreement expected by chance, and values between 0 and -1 indicate agreement worse than chance. Taking into account all categories in our visual classifications, we achieve a Randolph $\kappa$ of 0.30 in F356W+F444W and 0.28 in F200W, implying agreement beyond chance.

\section{The bar fraction}
\label{Sec: bar frac}
\begin{figure*}
	\includegraphics[width=\textwidth]{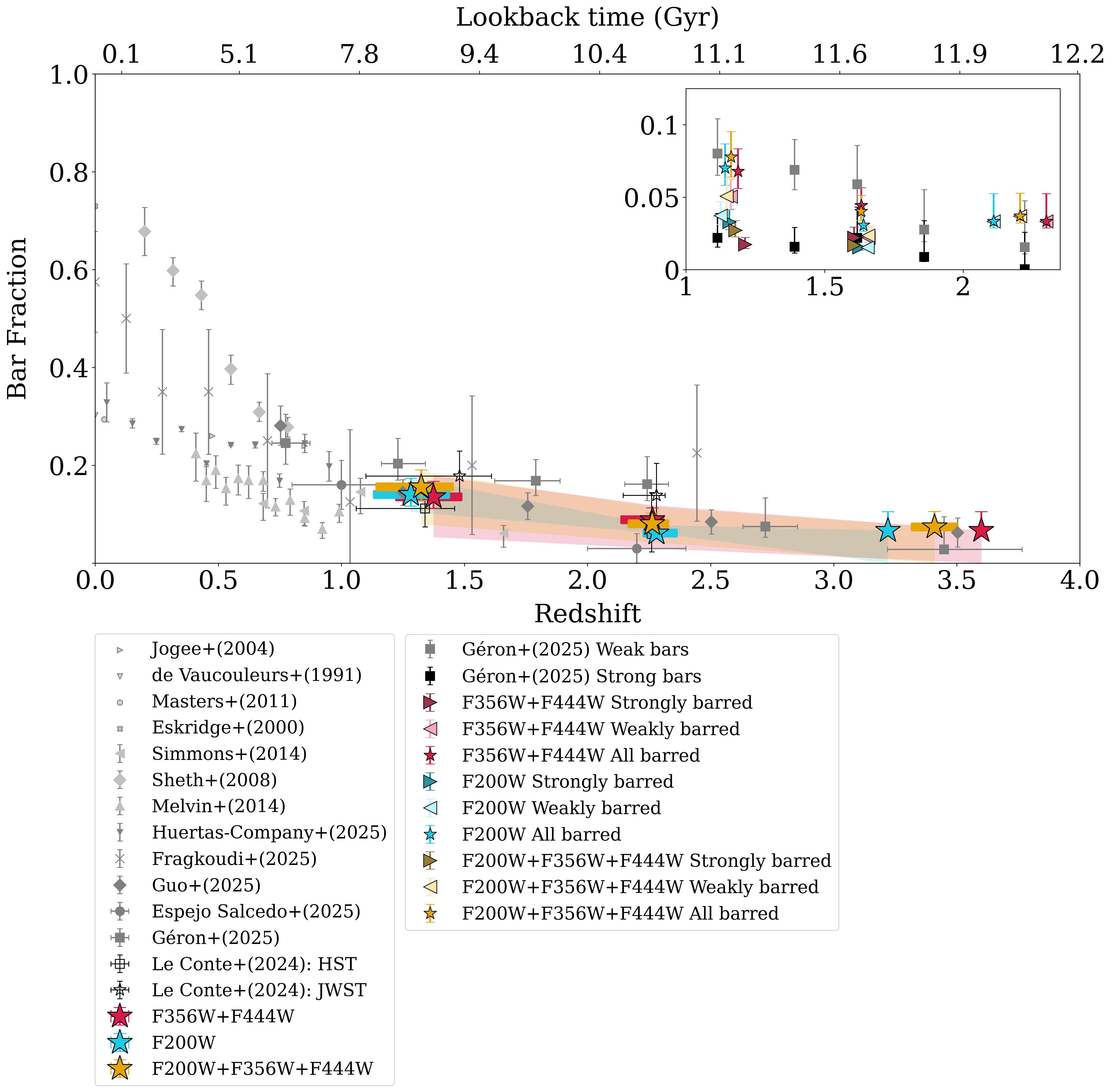}
    \caption{Evolution of the bar fraction in disc galaxies. At high-$z$ the bar fraction is found in three redshift bins, $1 \leq z < 2$, $2 \leq z < 3$ and $3 \leq z \leq 4$, using the visual classification of the images from the JWST NIRCam filters F356W+F444W (red stars) and F200W (blue stars) and combining these results to find a total bar fraction (yellow stars). Error bars in $f_{bar}$ are the $1\sigma$ bimodal interval, and the shaded area is the upper and lower bounds of the bar fraction (see text for details). Dashed horizontal error bars show the full range in $z$ of the identified bars, while thick horizontal solid lines show the corresponding 25\%-75\% inter-quartile range. The insert shows the bar fraction for the breakdown of strongly (right-pointing triangle) and weakly (left-pointing triangle) barred galaxies. The results of Paper 1 are black unfilled squares (HST) and stars (JWST), and in grey are the JWST bar fraction of \citetalias{Guo_2025} (diamond), \citetalias{Salcedo_2025} (circle) and \citetalias{Géron_2025} (square).}
    \label{fig:bar fraction}
\end{figure*}

\begin{table*}
    \centering
    \renewcommand{\arraystretch}{1.5} % Default value: 1
    \begin{tabular}{c|ccc|ccc|ccc}
    \hline
    Redshift & \multicolumn{9}{c}{Bar Fraction} \\
    \hline
     & \multicolumn{3}{c|}{F200W} & \multicolumn{3}{c|}{F356W+F444W} & \multicolumn{3}{c}{F200W+F356W+F444W} \\
     & SB & WB & B & SB & WB & B & SB & WB & B \\
    \hline
    $1 \leq z < 2$ & $0.07^{+0.02}_{-0.01}$ & $0.07^{+0.02}_{-0.01}$ & $0.14^{+0.03}_{-0.02}$ & $0.03^{+0.01}_{-0.01}$ & $0.10^{+0.02}_{-0.02}$ & $0.14^{+0.03}_{-0.02}$ & $0.05^{+0.01}_{-0.01}$ & $0.10^{+0.02}_{-0.02}$ & $0.16^{+0.03}_{-0.03}$ \\
    $2 \leq z < 3$ & $0.03^{+0.01}_{-0.01}$ & $0.03^{+0.01}_{-0.01}$ & $0.06^{+0.02}_{-0.01}$ & $0.04^{+0.01}_{-0.01}$ & $0.04^{+0.01}_{-0.01}$ & $0.09^{+0.02}_{-0.01}$ & $0.03^{+0.01}_{-0.01}$ & $0.05^{+0.01}_{-0.01}$ & $0.08^{+0.02}_{-0.01}$ \\
    $3 \leq z \leq 4$ & - & $0.07^{+0.04}_{-0.01}$ & $0.07^{+0.04}_{-0.01}$ & - & $0.07^{+0.04}_{-0.01}$ & $0.07^{+0.04}_{-0.01}$ & - & $0.07^{+0.03}_{-0.01}$ & $0.07^{+0.03}_{-0.01}$ \\
    \hline
    \end{tabular}
    \caption{The bar fraction for the visually classified samples in F200W, F356W+F444W and F200W+F356W+F444W in three redshift bins. The $f_{bar}$ is presented for strongly barred galaxies (SB), weakly barred (WB) and all barred galaxies (B). No strongly barred galaxies were found beyond $z = 3$. The errors presented are the statistical $1\sigma$ bimodal interval (see text for details).}
    \label{tab: Bar frac}
\end{table*}

This study uses the improved sensitivity and longer wavelengths of JWST to observe the evolution of the bar fraction - the number of barred galaxies in the number of disc galaxies across the redshift range $1 \leq z \leq 4$. We split the redshift range into three bins, and those redshift bins are specified as follows $1 \leq z < 2$, $2 \leq z < 3$ and $3 \leq z \leq 4$. The disc and barred galaxy samples have been visually classified in the NIRCam filters F356W+F444W and F200W (two independent samples), across the redshift range (F356W for $1 \leq z < 2$, F444W for $2 \leq z \leq 4$ and F200W for $1 \leq z \leq 4$; for further details see \S~\ref{Sec: filters}), to assess the impact of rest-frame wavelength and resolution on the bar fraction. 

Firstly, a brief comment on the general outcome of the classification scheme. The $f_{disc}$, based on a 3/5 vote threshold, is found to be 0.48 and 0.43, considering the whole sample in F356W+F444W and F200W, respectively. Our results of $f_{disc}$ agree with the visual classifications of \cite{Ferreira_2022}, with which our parent sample overlapped, which obtained $f_{disc} \approx 0.48$ for the same redshift range. We found a significantly smaller spheroidal population, $f_{spheroid} \approx 0.07$ and $f_{spheroid} \approx 0.06$ for the samples in F356W+F444W and F200W, respectively. This is less than the fraction classified by \cite{Ferreira_2022}, who found $f_{spheroid} \approx 0.15$. The lower spheroidal fraction found in this study could be due to the additional categories included in our visual classifications, such as \textit{Other} and \textit{Unresolved}.

We consider a combined sample, where we combine the F356W+F444W and F200W disc sample, such that if a disc were to be identified in the short-wavelength filter but not in the long-wavelength filters, it was added to the sample, creating a sample of disc galaxies classified in either short or long wavelength NIRCam filters (hereafter referred to as the combined sample). We consider this combined sample to capture the shorter bars missed by the lower resolution of the longer wavelength filters, whilst observing those barred galaxies, which might have highly significant star formation and dust obscuring the bar. The combined sample has a total of 520 disc galaxies.

The bar fraction, $f_{bar}$, is measured in each of the three redshift bins, for the short-, long-wavelength and combined samples. Figure \ref{fig:bar fraction} shows the evolution of the bar fraction across the 12 Gyr range, comparing the bar fractions observed with JWST in this study to those found in other theoretical and observational work \citep[][]{deVaucouleurs_1991,Eskridge_2000,Jogee_2004,Sheth_2008,Masters_2011,Melvin_2014,Simmons_2014,Euclid_2025,Guo_2025,Fragkoudi_2025,Géron_2025,Salcedo_2025}. 

The bar fraction is found to decrease towards higher redshifts in both short and long wavelength filters. The F356W+F444W sample has bar fractions $f_{bar, long} \approx 0.14, 0.09$ and $0.07$ for redshift bins $1 \leq z < 2$, $2 \leq z < 3$ and $3 \leq z \leq 4$. Similarly, the F200W sample has bar fractions $f_{bar, short} \approx 0.14, 0.06$ and $0.07$ for the redshift bins $1 \leq z < 2$, $2 \leq z < 3$ and $3 \leq z \leq 4$. The combined sample has a slightly higher bar fraction $f_{bar, comb} \approx 0.16, 0.08$ and $0.07$ for the redshift bins $1 \leq z < 2$, $2 \leq z < 3$ and $3 \leq z \leq 4$. 

The bar fraction has been investigated at $z < 1$ in \textit{Euclid} Q1 \citep[][]{Euclid_2025}, finding the largest barred galaxy sample beyond the local Universe. Over 7,700 barred galaxies were identified using \texttt{Zoobot} classifications \citep[][]{euclid2_2025}, obtaining a mean bar fraction of 0.2 to 0.4, for $log_{10}(M_{\star}/M_{\odot}) > 10$ galaxies, at $z=1$ to $z=0$, respectively. The decreasing trend found in \textit{Euclid}, shown in Figure \ref{fig:bar fraction} agrees with HST observations and simulations; however, by $z=1$, the bar fraction is twice that previously measured in HST, and corroborates the conclusions of \citetalias{LeConte_2024}. The \textit{Euclid} mission has achieved unprecedented spatial resolution for very large surveys, which will be vital for population statistics.

At the two Cosmic Noon bins, $0.8 < z < 1.3$ and $2.0 < z < 2.5$, \citet[][hereafter referred to as \citetalias{Salcedo_2025}]{Salcedo_2025} obtained the visually classified bar fractions of $\approx 16$ per cent and 3 per cent with JWST NIRCam imaging. The Galaxy Zoo CEERS project \citep[][hereafter referred to as \citetalias{Géron_2025}]{Géron_2025} observed a decreasing bar fraction in a volume-limited sample that ranged from 25 per cent at $0.5 < z < 1.0$ to 3 per cent at $3.0 < z < 4.0$. The decreasing trends of \citetalias{Salcedo_2025}, \citetalias{Guo_2025} and \citetalias{Géron_2025} agree with our results within uncertainty. These JWST studies agree with the bar fraction being twice that found in HST and with the fact that the onset of bar formation occurs earlier than we previously thought.

The bar fraction is broken down into its weakly and strongly barred counterparts across the redshift range in the insert of Figure \ref{fig:bar fraction}, compared to the strong and weak bar fractions of \citetalias{Géron_2025}. A lower bar fraction is seen for our strongly barred galaxies, which remains $\sim 0.05$ across the redshift range $1 \leq z < 3$. At the lowest redshift bin, a higher bar fraction for weakly barred galaxies is found, which decreases towards higher redshifts. This difference between strongly and weakly barred galaxies is also found in \citetalias{Géron_2025}.

As detailed in \citetalias{LeConte_2024}, we use the Jeffreys interval \citep[][]{Brown_2001,Gelman_2003}{}{} to determine the statistical uncertainty on the bar fraction. We refer the reader to \citetalias{LeConte_2024} for our reasoning for using the beta distribution quantile technique, as argued for in \cite{Cameron_2011}. The sample used in this study is approximately mass complete, meaning we do not account for incomplete sampling in the uncertainty estimates. The visually classified derived bar fractions with statistical uncertainties for the single-band NIRCam images in either F200W, F356W+F444W or combined can be found in Table \ref{tab: Bar frac}. 

Systematic uncertainties arise from difficulties in morphological classification and can be described by the agreement between classifiers. We introduced Randolph $\kappa$ in \S~\ref{Sec: disc vis} and state the strong agreement between classifiers in over 70 per cent of cases. However, since the $\kappa$ thresholds for agreement better than chance are not universally agreed upon, we present a different method for the upper and lower bounds of the bar fraction. The shaded regions in Figure \ref{fig:bar fraction} represent the upper and lower bar fractions, obtained from a more lenient and a more restrictive threshold, respectively. The upper $f_{bar}$ is determined by a barred and disc sample with a 2 out of 5 voter threshold, whereas the lower $f_{bar}$ requires a 4 out of 5 consensus threshold. We include these variations on the voting threshold to emphasise the impact of classification definitions and consensus on the results. These upper and lower $f_{bar}$ uncertainties encompass our results in this work and \citetalias{LeConte_2024}, and also the work of \citetalias{Guo_2025} and \citetalias{Salcedo_2025}.

\section{Bar properties}
\label{Sec: bar prop}
Thus far, we have responded to the question of when the onset of bar formation occurs by exploring the bar fraction at $1 \leq z \leq 4$. Now we wish to analyse these high-$z$ barred galaxies in more detail and understand how they differ from barred galaxies at $z < 1$ and $z \approx 0$. We explore the mass dependence of the bar fraction in \S~\ref{Sec: mass} and measure the evolution of the bar length in \S~\ref{Sec: bar length} and strength in \S~\ref{Sec: bar strength}.

\subsection{Mass dependence}
\label{Sec: mass}
\begin{figure}
	\includegraphics[width=\columnwidth]{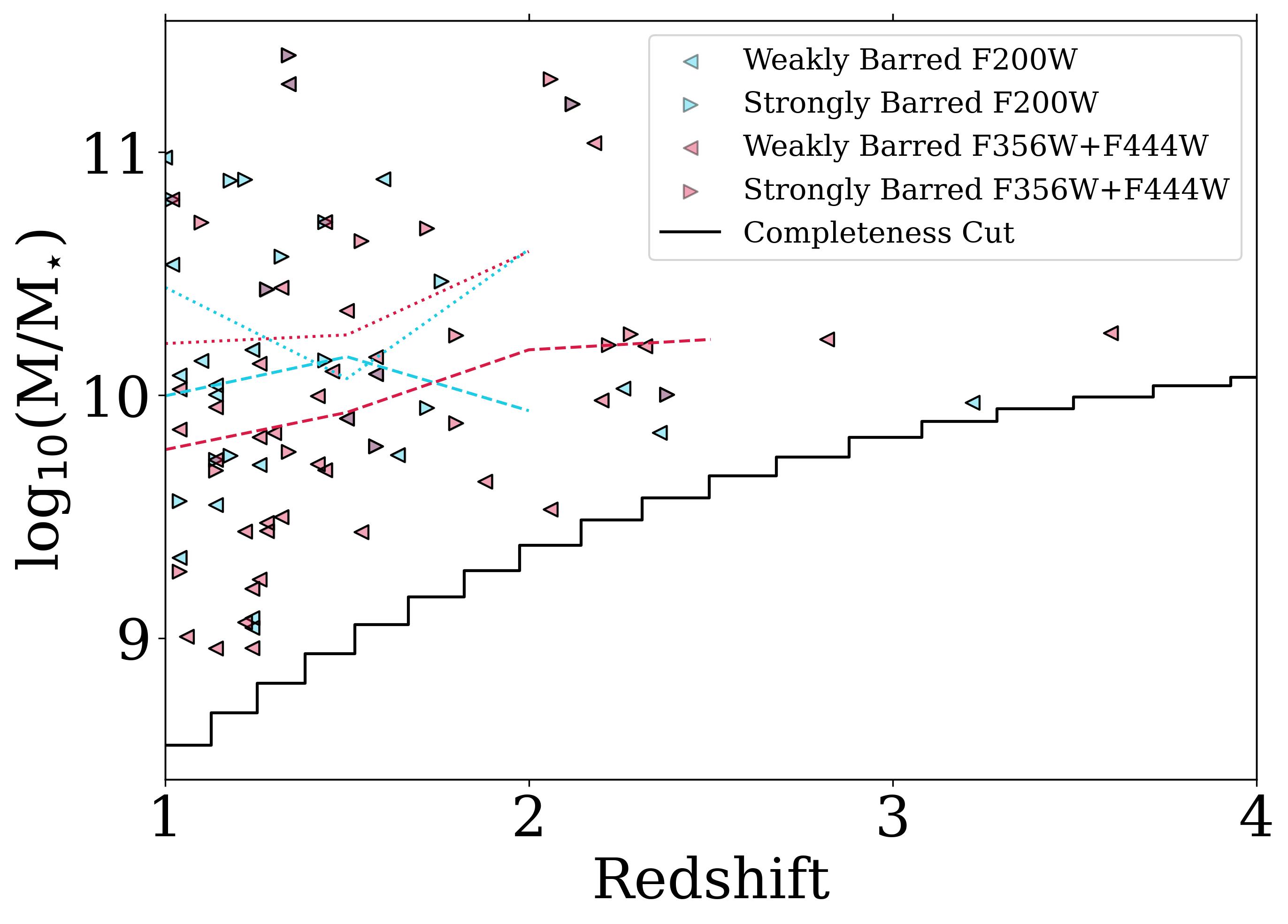}
    \caption{Stellar mass - redshift distribution of barred galaxies found by visual classification. For the independent studies using NIRCam filters F200W (blue) and F356W+F444W (red), galaxies are defined as strongly barred (right-pointing triangle, dotted) and weakly barred (left-pointing triangle, dashed). For the strongly and weakly barred galaxies, we show the mean stellar mass across the redshift range as dotted and dashed lines, respectively. The 95\% empirical completeness adopted from \citet{Duncan_2019} (black step function) sets the lower limit to the stellar mass distribution.}
    \label{fig: bar mass}
\end{figure}
\begin{figure}
	\includegraphics[width=\columnwidth]{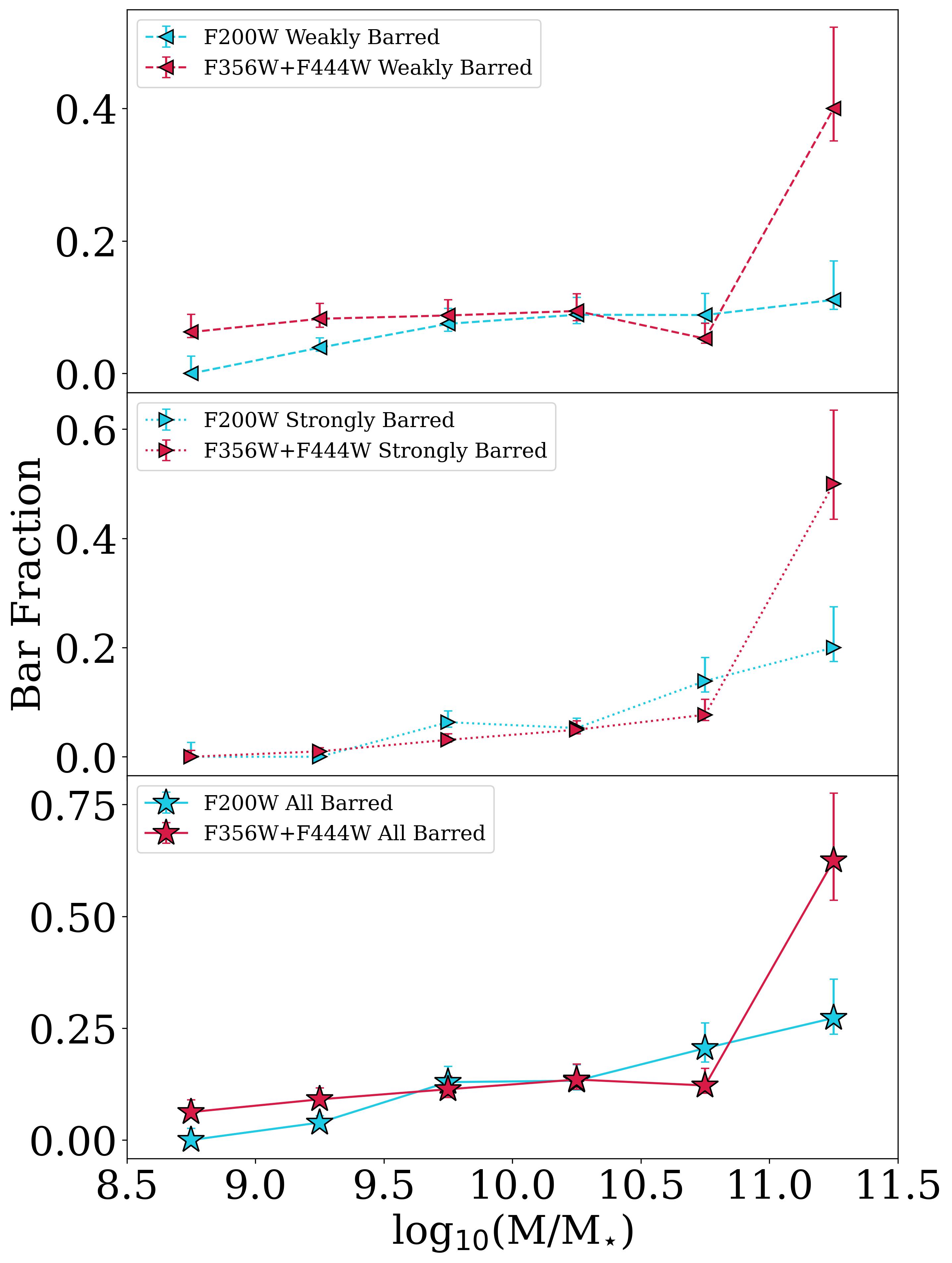}
    \caption{The fraction of bars in disc galaxies for a given stellar mass in the range $8.5 \leq$ log$(M_{\star}/M_{\odot}) \leq 11.5$. We show results for the NIRCam filters F200W (blue) and F356W+F444W (red). Three different samples are shown: weakly barred (top panel), strongly barred (middle panel), and the sum of both weakly and strongly barred galaxies (bottom panel). Error bars in $f_{bar}$ show the $1\sigma$ bimodal interval.}
    \label{fig: bar mass frac}
\end{figure}

Observational studies have reported on a bar fraction dependence on stellar mass from the local to high redshift Universe \citep[e.g.,][]{Sheth_2008,Erwin_2018}. We extend these investigations to $z \geq 1$ for our 95\% empirically complete sample (discussed in Section \ref{Sec: parent sample}) between the redshift range $1 \leq z \leq 4$. In Figure \ref{fig: bar mass} we show the mean stellar mass across the redshift range for our bars identified in F200W (blue) and F356W+F444W (red), separated into the weakly and strongly barred samples. The adopted completeness limit results in less massive galaxies at lower redshifts hosting bars, whereas towards higher redshifts, bars are present in the most massive galaxies. This is, of course, driven by the redshift-dependent completeness limits, as the plot shows that all masses probed within these limits (from $10^9$ to $10^{10} M_{\odot}$) include barred galaxies. Nevertheless, we find that strongly barred galaxies tend to have a higher stellar mass at all redshifts. The mean is similar for both wavelength samples, with the spread in stellar masses being slightly greater in the longer wavelength sample.

We show the bar fraction dependence on stellar mass in Figure \ref{fig: bar mass frac} for the two JWST samples. A low $f_{bar}$ -- about 0.1 or less -- is seen in log$(M_{\star}/M_{\odot}) \leq 10$. In F200W, a steady increase in the bar fraction with mass is observed for both weak and strong bars, rising to $f_{bar} \approx 0.2$ at log$(M_{\star}/M_{\odot}) \approx 10.75$. At the highest stellar masses, log$(M_{\star}/M_{\odot}) \approx 11.25$, $f_{bar}$ peaks, with $f_{bar} > 0.2$ for the short wavelength, and with $f_{bar} > 0.5$ for the long wavelengths, but it is worth noting that this stellar mass bin is less statistically significant compared to the others, as it contains a small sample of only eight disc galaxies. Nevertheless, this trend of increasing bar fraction for higher galaxy masses agrees with the results using HST from \citet{Sheth_2008}, for galaxies at $0.37 < z < 0.84$ (see their Fig. 3) and \citet{Melvin_2014}, for galaxies at $0.4 < z < 1.0$ (their Fig. 5).
%Compared to bar fractions reported at $z = 0$, the results shown here appear to be very low, but it is important to remember the lower $f_[bar] \approx 0.2$ found at $z \geq 1$.

\subsection{The bar length}
\label{Sec: bar length}
Another open question we aim to address is at what length bars form, and does the bar length evolve over cosmic time? For the remainder of this study, we measure and analyse the lengths of our high-$z$ barred galaxy sample in short and long NIRCam wavelengths to address these questions and test the impact of the rest-frame wavelength on the observed evolution of the bar length.

Visual and automated bar fitting techniques are accepted to measure the bar length, but in the local Universe, barred galaxy studies have found automated bar identification techniques, such as isophotal and Fourier analysis, to be favourable when used on large samples. At higher redshifts, we find that automated processes are challenging despite the improved sensitivity of JWST. Ellipse fitting to star-forming massive galaxies at cosmic noon, and thus, clumpy and irregular systems, is used in this study, but it is challenging with the number of initial free parameters and requires tailoring for each galaxy to avoid spurious bar length measurements. Consequently, we identify the barred galaxy sample through visual classification, but use ellipse fit techniques to quantify bar properties. We emphasise that adjustments are made to the initial parameters on a case-by-case basis and involve visual inspection of the fitting for every galaxy. In some cases, this also involved adjusting the fitting parameters until a satisfactory fit was obtained, namely, one in which the fitted ellipses outlining the bar agree with our visual assessment.

To perform this structural analysis, the strongly and weakly barred galaxies from the visual classification are fitted by elliptical isophotes using \texttt{photutils.isophote} (for further details, see Section \ref{Sec: opt}) with fixed central coordinates obtained from the initial ellipse fits. The stellar bar can be traced by ellipse fits, which remain at a constant position angle, PA (with a change in $PA$, $\Delta PA \leq 10^\circ$) and a gradual increase in ellipticity, $e$, towards the end of the bar. The end of the bar is the point at which the position angle changes by more than 10 degrees ($\Delta PA > 10^{\circ}$) and the ellipticity peaks (with $e_{peak} > 0.2$). We refer the reader to Figure 1 of \citetalias{LeConte_2024} for an example of the ellipse fit radial profiles discussed here. Some studies have found that the end of the bar can also be parameterised as a minimum in ellipticity, which follows on from the peak. Still, we found that this feature could not be observed in all of the barred galaxy ellipticity radial profiles. Hence, in this study, the bar length, $L_{bar}$, is defined as the average of the bar lengths at $\Delta PA$ and $e_{peak}$, i.e., $L_{bar}$ is effectively the semi-major axis of the bar.

The measured projected $L_{bar}$ from ellipse fits is checked and adjusted through visual inspections by ZLC on the radial profiles, and when necessary, constraints on parameter thresholds are applied. Theoretical works show bars form in rotationally supported and cold stellar discs; hence, we assume that these high-$z$ disc galaxies are dynamically settled and vertically thin. This assumption is vital as it enables us to deproject the measured $L_{bar}$, and is a reasonable assumption at least towards the ends of the bar, where it matters most. We thus derive the deprojected semi-major axis, semi-minor axis, $e$ and $PA$ of an observed projected ellipse corresponding to the end of the bar \citep[for full derivation details, we refer the reader to Appendix A of][]{Gadotti_2007}. To carry out this deprojection procedure, we require the inclination of the galaxy, the position angle of the line of nodes and the position angle of the bar, all of which are derived through the ellipse fits.

\begin{figure*}
	\includegraphics[width=\textwidth]{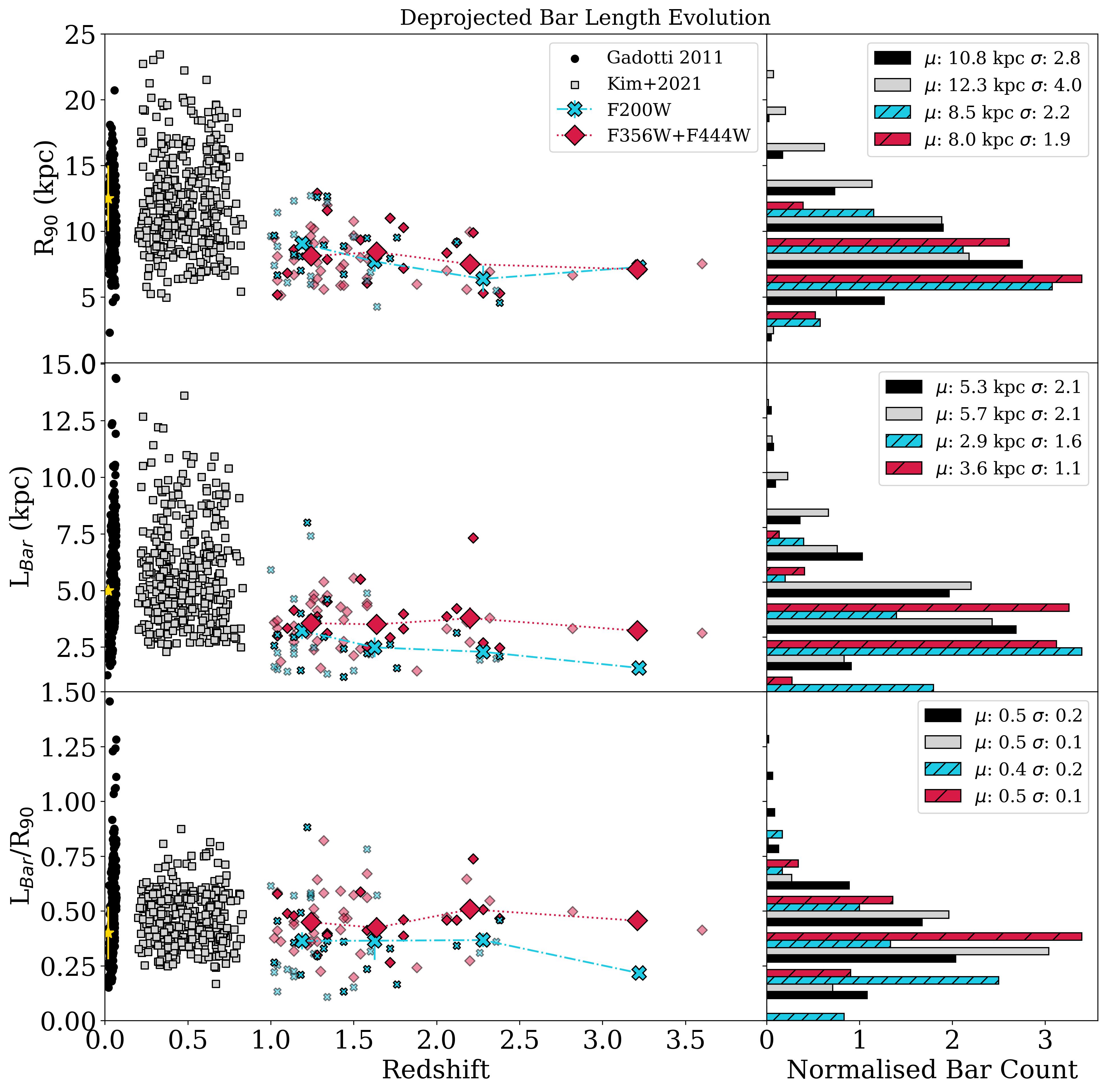}
    \caption{The evolution of the bar length over $0 \leq z \leq 4$. The left panels show lengths measured in this work from JWST NIRCam filters F200W (blue crosses) and F356W+F444W (red diamonds) for bars found in galaxies between the redshift range $1 \leq z \leq 4$, with their normalised distribution (see text for details) shown on the right panels. Weakly barred galaxies are shown as fainter markers and strongly barred galaxies are brighter.} The first row shows the distribution of $R_{90}$, whilst the second row is the deprojected $L_{bar}$ and the third row is the normalised $L_{bar}$, $L_{bar}/R_{90}$. The high redshift sample is compared against a sample of SDSS $i-$band barred galaxies at $z \approx 0$ \citepalias[][black circles]{Gadotti_2011}, the Milky Way \citep[with measurements from][yellow star]{Bland_Hawthorn_2016}, and a sample of barred galaxies at $0.2 < z \leq 0.835$ using F814W images from the COSMOS survey \citepalias[][grey squares]{Kim_2021}. The mean, $\mu$, value for each parameter in each sample is given in the right panel, with the standard deviation, $\sigma$.
    \label{fig:length dist}
\end{figure*}

The deprojected $L_{bar}$ distribution for barred galaxies identified in short and long NIRCam wavelength filters is shown in the second row of Figure \ref{fig:length dist}. We share a similar figure in Appendix \ref{App: A}, showing the distribution of projected $L_{bar}$. On average, shorter bars are measured in F200W, with a mean $L_{bar} \approx 2.9$ kpc, than in F356W+F444W, with mean $L_{bar} \approx 3.6$ kpc, but in both samples, the average $L_{bar}$ has minimal variation across the redshift range, with the F200W values of $L_{bar}$ increasing slightly towards lower redshifts. Additionally, the distributions of the two JWST samples overlap greatly, as can be seen in the right panel of Figure \ref{fig:length dist}. In Figure \ref{fig:length dist}, the bar count is normalised by dividing the count by the number of observations times the bin width to form a probability density. Bars as short as $\sim 1$ kpc are measured in F200W, but bars as long as $\sim 8$ kpc are found in both barred samples, showing that long bars are already present in $z \geq 1$.

The bar length comparison across wavelengths is discussed for the whole barred galaxy sample. However, we separate the samples into strongly and weakly barred galaxies by making their markers in Figure \ref{fig:length dist} brighter and fainter, respectively. For the longer wavelength sample, there is no difference in the mean of the weakly and strongly bar lengths, which are 3.5 kpc and 3.8 kpc, respectively. Furthermore, the shorter wavelength sample shows the same results, with the mean bar length of weakly barred galaxies being 2.8 kpc and 3.0 kpc for strongly barred galaxies.

To understand the origin of this minimal $L_{bar}$ evolution in F200W and no evolution in F356W+F444W, we normalise $L_{bar}$ by galaxy size and assess the evolution of both $R_{90}$ and the ratio $L_{bar}/R_{90}$. We use the value of $R_{90}$ as a proxy of galaxy size, wherefore it is the distance from the centre of the galaxy which contains 90\% of the total light in a given NIRCam filter. The first row of Figure \ref{fig:length dist} shows the evolution of $R_{90}$ for the barred galaxy samples, along with their distribution on the right panel. The measured values of $R_{90}$ cannot be disentangled for the two JWST samples across the redshift range. A slight increase towards lower redshifts can be seen in F200W, whilst F356W+F444W remains approximately constant. To assess if the bar grows with disc size, or if the bar grows at rates greater than the growth of the disc, we observe the evolution of the bar length normalised by disc size, $L_{bar}/R_{90}$, in the bottom row of Figure \ref{fig:length dist}. Within the redshift range $1 \leq z < 2.5$, there is no evolution observed in either barred sample, and the shorter wavelength filter is marginally shorter than the longer wavelength filters. This result indicates that the growing $L_{bar}$ in F200W is accompanied by the growth of the disc size.

Our high redshift sample is compared to a sample of barred galaxies from SDSS at $z \approx 0$ \citep[\citetalias{Gadotti_2011}. See also][for more bar length measurements at $z \sim 0$]{Neumann_2024}. The volume-limited, near-face-on, massive galaxy sample ($0.02 \leq z \leq 0.07$, $b/a \geq 0.9$, $> 10^{10}M_{\odot}$) is fit for a bulge, disc and bar in $i-$band images, to determine a sample of 291 barred galaxies. The SDSS PSF FWHM at $z \approx 0.05$ results in a spatial resolution of $\approx 1.5$ kpc, meaning the resolution of the SDSS sample is comparable to the longer NIRCam wavelength filter, at $z \sim 2$. The mean $L_{bar} \approx 5.3$ kpc is slightly longer than those measured for the JWST sample; however, the distribution strongly overlaps with the JWST samples. Similarly, the values of $R_{90}$ at $z \approx 0$ are $\sim 2$ kpc greater than the high-$z$ sample, but the entire distributions overlap. The normalised bar length is found to be the same at $z \approx 0$ and $z \geq 1$, with the distributions peaking at $L_{bar}/R_{90} = 0.5$. That is, longer bars are found in larger discs in SDSS, as in the JWST samples, resulting in a constant ratio.

The Milky Way stellar mass, disc size and bar length are indicated in Figure \ref{fig:length dist}, and taken from \citet[][]{Bland_Hawthorn_2016}. The Milky Way has features comparable to those of the local barred galaxy population.

We bridge the gap between $z\approx0$ and $z > 1$, using a sample of 379 barred galaxies identified in the F814W images from the COSMOS survey at $0.2 < z \leq 0.835$ with $10.0 \leq log(M_{\star}/M_{\odot}) \leq 11.4$ \citepalias{Kim_2021}. Ellipse fits are used to measure $L_{bar}$, and the physical resolution of COSMOS is $\approx0.6$ kpc at $z \approx 0.5$, making it comparable to the PSF FWHM of F200W in JWST NIRCam at $z\approx2$. The COSMOS sample has the longest mean $L_{bar} = 5.7$ kpc and a greater spread in the distribution. Bars with $L_{bar} < 2$ kpc have not been identified in this sample, possibly due to the selection of more massive galaxies than our high-$z$ sample, namely a mass cut such that $ \log(M_{\star}/M_{\odot})\geq10$. This observation is supported by the association that the most massive galaxies are hosts to the longest structures in the local Universe \citep[e.g.,][]{DiazGarcia_2016,Erwin_2019}. However, it should be noted that the $z\approx0$ SDSS sample of \citetalias{Gadotti_2011} has the same mass cut but does not show a similar dearth of bars with $L_{bar} < 2$ kpc. Many long bars, greater than 6 kpc, are found across the redshift range of \citetalias{Gadotti_2011,Kim_2021}, and such bars are present but not abundant beyond $z = 1$. The disc size of this intermediate sample is shifted towards the larger end, but has a stronger overlap with the high-$z$ sample than the bar length. However, the normalised bar length achieves the same ratio as $z = 0$ and $z \geq 1$, which means that the bar length scales with the size of the disc.

%Incredibly, the high-$z$ sample mirrors the trends previously found with HST and SDSS, and minimal to no evolution is observed in $L_{bar}$ across $0 \leq z \leq 4$. We note that shorter bars are found in F200W compared to F356W+F444W, subsequently, evolution is observed. Furthermore, the extent of $L_{bar}$ at $z \leq 1$ is significantly greater, $L_{bar} < 12$ kpc, than that seen at high-$z$. We state that we find long bars at $z \geq 1$, but a population of $L_{bar} > 6$ kpc has not yet formed, which is seen at $z \leq 1$.

Surprisingly, while we see an increase of $\approx 2$ kpc in the mean values of $L_{bar}$ from $z > 1$ to $z \approx 0$, there is significant overlap in all distributions. Additionally, when the bar length is normalised by the disc size, we see that its value, $L_{bar}/R_{90}$, remains on average approximately constant from $z=4$ to $z=0$. In this context, it is also interesting to note that bars as long as $4-6$ kpc are already abundant at $z > 1$, and this corresponds to the mean bar length at $z\sim0$ \citepalias{Gadotti_2011}. However, we observe a population of bars as long as $8-12$ kpc at $z<1$, which are not seen at higher redshifts. The lack of long bars is due to the smaller size of high-$z$ disc galaxies.

The population of long bars with $L_{bar } > 8$ kpc at low and intermediate redshifts, found in \citetalias{Gadotti_2011} and \citetalias{Kim_2021}, respectively, suggests that in some galaxies a period of pronounced bar growth commences after the formation of high-$z$ bars, which are measured in this study to be $< 8$ kpc (and more typically less than $< 6$ kpc). To assess the validity that bars with $L_{bar} > 8$ kpc do not exist at Cosmic Noon, and that a period of growth occurs, we consider volume effects. The area of SDSS (3324 deg$^2$) and COSMOS (2 deg$^2$) are significantly larger than in CEERS; hence, long bars in our sample could be missed due to their rarity and the smaller volume of CEERS (88 arcmin$^2$). The co-moving volumes of the samples from \citetalias{Gadotti_2011} and \citetalias{Kim_2021} are larger than the co-moving volume of CEERS between $1 \leq z \leq 4$ by a factor of $\sim 11$ and $\sim 12$, respectively. However, in nearby disc galaxies, 37 very long bars were observed, whilst 58 were found at intermediate redshifts. This means that one bar with $L_{bar} > 8$ kpc is observed per $251,824$ Mpc$^3$ and $178,274$ Mpc$^3$ for the two samples, respectively. Since, the co-moving volume of this study is $884,085$ Mpc$^3$, approximately 4 and 5 long bars should be identified if the occurrence follows \citetalias{Gadotti_2011} and \citetalias{Kim_2021}, respectively. Therefore, the lack of long bars at high-$z$ is not determined to be due to volume effects, but rather could indicate a period of bar growth.

\subsection{The bar ellipticity}
\label{Sec: bar strength}
Various techniques and definitions are used to define the strength of a bar. These include, but are not limited to, the axis ratio; the ratio of the fluxes inside and outside of the bar (i.e., the bar contrast); the ratio of the $m = 2$ and $m = 0$ Fourier components. Many of these methods require the deprojection and decomposition of the galaxy image; however, bar and central component modelling is beyond the scope of this paper. Therefore, in this study, the strength of the bar is parameterised as the projected ellipticity of the ellipse located at $L_{bar}$. Nevertheless, we note that the measurement of the bar ellipticity, $e_{bar}$, in high-redshift galaxies can be severely affected by unresolved bright central sources, the prominence of the central component, and PSF and resolution effects, as we indeed show below. Here, we thus simply have an initial assessment of the bar ellipticity at high redshifts and make a brief comparison between short and long-wavelength filters. Additionally, the deprojection technique described previously is not used, for it can spuriously broaden the bar if the position angle of the bar is oriented along the line of nodes.

\begin{figure}
	\includegraphics[width=\columnwidth]{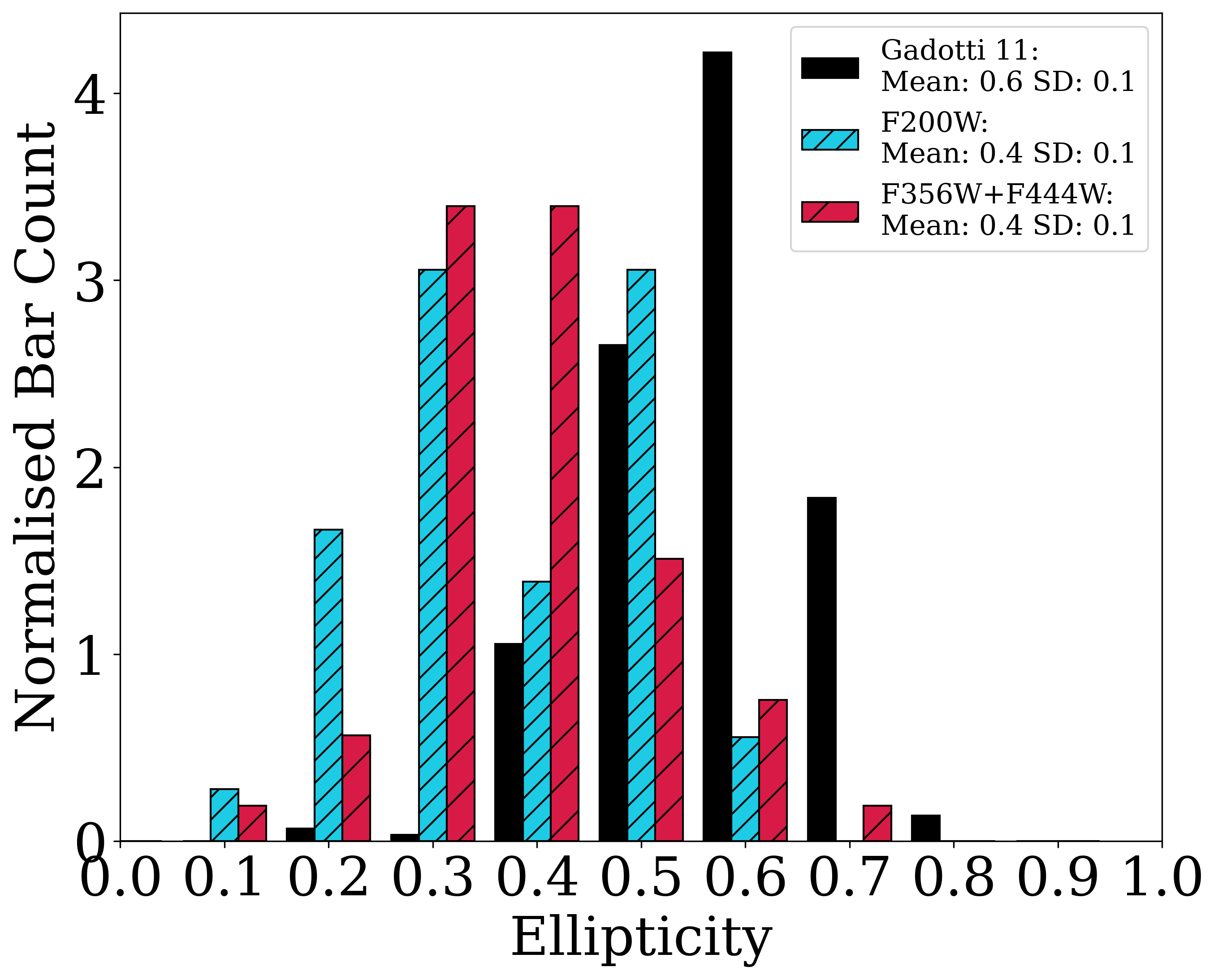}
    \caption{Distribution of the projected bar ellipticity for JWST NIRCam filters F200W (double hatched) and F356W+F444W (hatched) for bars found in galaxies in the redshift range $1 \leq z \leq 4$. The high redshift sample is compared against a sample of SDSS barred galaxies at $z \approx 0$ \citepalias[][solid black]{Gadotti_2011}. The mean ellipticity and standard deviation are given for each sample.}
    \label{fig:ellip dist}
\end{figure}

The distribution of $e_{bar}$ is shown in Figure \ref{fig:ellip dist} for short and long JWST NIRCam wavelengths and is compared to the SDSS barred galaxy sample at $z \approx 0$. The bar count is normalised as in Figure \ref{fig:length dist}. The ellipse fit technique is shown to underestimate the ellipticity by 20 per cent in \citet[][]{Gadotti_2008}; hence, we derive the corrected $e_{bar}$ as $e_{bar} \times 1.2$. The mean $e_{bar} \approx 0.4$ for the sample of barred galaxies in F356W+F444W, but for F200W, there is a greater proportion of elongated bars; however, the entire F200W sample also results in a mean $e_{bar} \approx 0.4$. These high redshift bars thus appear significantly rounder than the near face-on (disc axial ratio $b/a \geq 0.9$) bars observed at $z \approx 0$, for which the mean $e_{bar} \approx 0.6$. We speculate that the barred galaxies in F356W+F444W could be rounder due to lower resolution effects when compared to F200W. Overall, the bars in the high-$z$ sample are rounder than those in the $z \approx 0$ sample; however we cannot rule out the possibility that unresolved bright galactic centres and the broadening of central structures by the PSF may artificially make bars appear rounder.

\section{Discussion}
\label{Sec: discussion}
In this study, we have reported on the evolution of the bar fraction between $1 \leq z \leq 4$. We identified a significant population of high-redshift barred galaxies with the bar fraction decreasing from $f_{bar} \approx 0.16^{+0.03}_{-0.03}$ at $1 \leq z < 2$ to $0.08^{+0.02}_{-0.01}$ at $2 \leq z < 3$ and $0.07^{+0.03}_{-0.01}$ at $3 \leq z \leq 4$. We also measured the evolution of the absolute value of the deprojected bar length from $z = 4$ to $z = 0$, finding shorter bars in the F200W filter and little to no evolution in both samples. Subsequently, we measured the normalised bar length by disc size and found no evolution from the local to the high redshift Universe. 

Identifying barred galaxies at high redshift presents several challenges, particularly due to spatial resolution limitations and cosmological surface brightness dimming. It should be noted that the observational studies ranging from $z = 0 - 4$ used for mapping the evolutionary path of barred galaxies employ different rest-frame wavelengths, which can affect the derived structural parameters. The sample analysed in \citetalias{Gadotti_2011} is based on rest-frame optical data, in the SDSS $i$-band. However, \citetalias{Kim_2021} measure bar lengths using F814W, which corresponds to the rest-frame $g-$ and $r$-bands in the corresponding redshift range. In this study, we use a combination of filters that probe a broader wavelength range: rest-frame $g-$ to $i$-band using the F200W filter, and rest-frame $i$-band to near-infrared ($\sim 1.6 \mu m$, $H$-band equivalent) using the F356W+F444W filters. Below, we discuss our main findings, particularly in the context of the co-evolution of discs and bars.

%We suggest below that the JWST shorter wavelength imaging is often necessary at these redshifts to reveal the shorter bar population, due to the narrower PSF (\S~\ref{Sec: dis-short}). This ties into the broader context of the downsizing scenario, where more massive galaxies are expected to form bars first and thus presumably host stronger bars, which we discuss in \S~\ref{Sec: dis-strong}. Observations also suggest a correlation between bar length and disc size, yet we report that measurements in shorter wavelengths systematically yield shorter bar lengths (\S~\ref{Sec: dis-discs}). This wavelength dependence, coupled with the intrinsic evolution of galaxies, complicates the interpretation of bar properties across cosmic time.

\subsection{Capturing short bars with the F200W filter}
\label{Sec: dis-short}
How confident can we be that we have identified the entire barred galaxy population? The simulation-based work of \citet[][hereafter referred to as \citetalias{Liang_2024}]{Liang_2024} suggests that the bar fraction would be significantly greater, $f_{bar} > 40$ per cent for $z < 3$, when using the shorter wavelength channels of NIRCam for improved spatial resolution. This motivated the repetition of this study in both the F200W and F356W+F444W filters. \citetalias{LeConte_2024} discussed the limitations of using the long wavelength filter, F444W, as bars shorter than $\sim 2-3$ kpc were missed from the sample, since the PSF FWHM is $0.145^{\prime\prime}$; hence, in this study, we aimed to test if we could capture the shorter bars by performing visual classification on F200W as well, with its improved PSF FWHM of $0.07^{\prime\prime}$. 

Surprisingly, we found no deviation in the bar fraction between the two samples (i.e., with the F200W and with the F356W+F444W filters), despite obtaining different barred galaxy populations. We suggest that the similar result in bar fraction is due to the shorter wavelength filter being subject to dust and star formation obscuration, whilst the longer wavelength images trace better the underlying older stellar population of the bar. Thus, we combined the two samples to capture both shorter bars seen in F200W and those only observed in F356W+F444W, which are obscured at shorter wavelengths. This combined sample agrees within the uncertainties with the independent bar fractions of this study and with the few other studies on the high-$z$ bar fraction \citepalias{Guo_2025,Salcedo_2025,Géron_2025}. 

Simulations predict that bars are strengthened and grow with time \citep[e.g.,][]{Athanassoula_2003,Algorry_2017}. In the context of missing shorter bars, this would suggest that toward the highest redshift end, it becomes more difficult to identify young bars, as their size could be below the detection threshold \citep[][]{Erwin_2005}. Through our attempts to mitigate biases, we conclude that we still observe a decreasing bar fraction trend; however, this is not as steep as previously reported by HST investigations. Nevertheless, we acknowledge that this bar fraction could still be a lower limit caused by the impact of redshift, due to lower physical spatial resolution, surface brightness dimming, and possibly enhanced gas/dust content and star formation at high $z$.

\begin{figure}
	\includegraphics[width=\columnwidth]{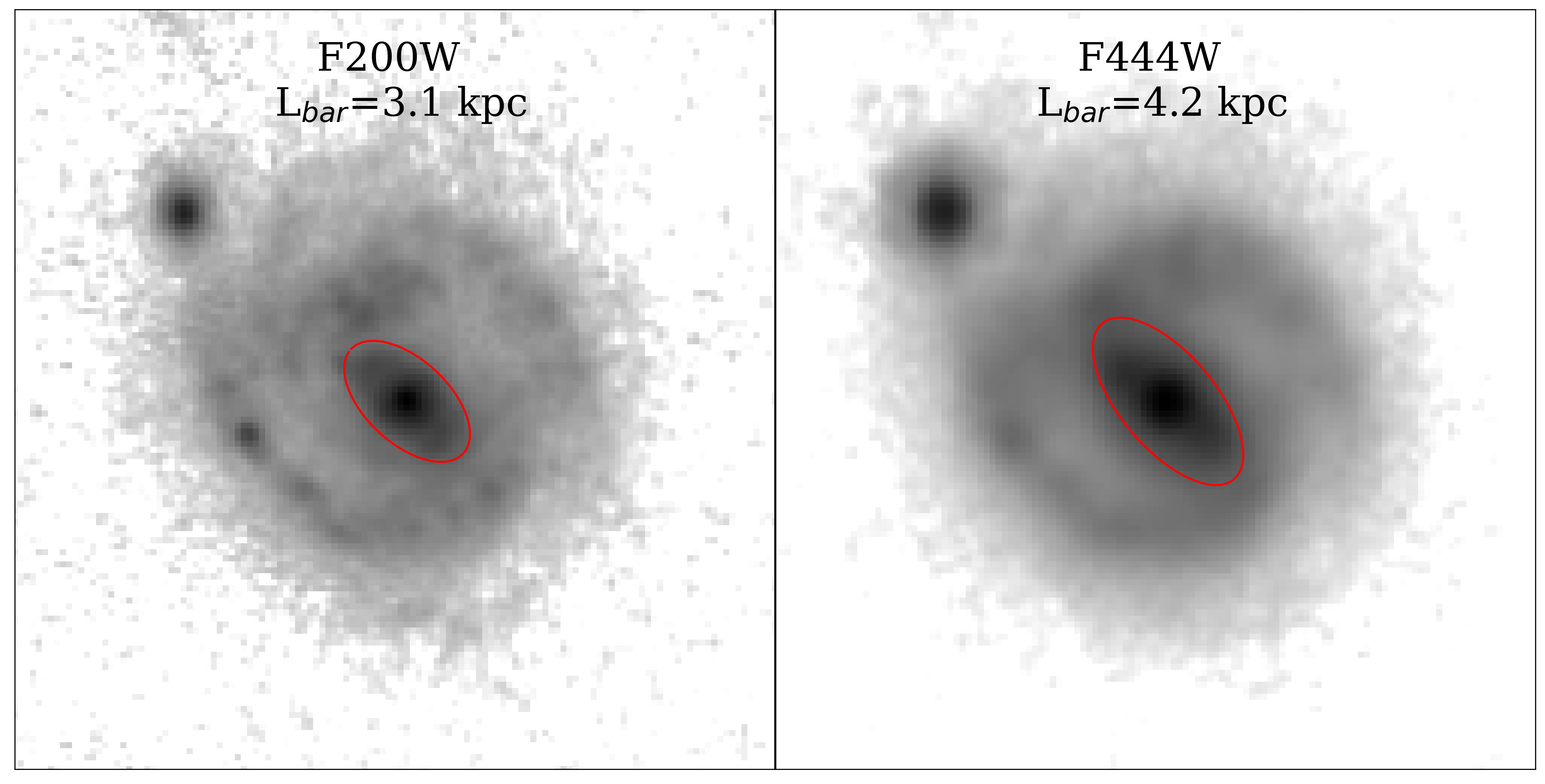}
    \caption{Single-band images of the example galaxy EGS 23205 in the NIRCam filters F200W (left) and F444W (right) at $z \approx 2.1$. The deprojected bar length is specified, and the bar is represented by a red ellipse. This galaxy is chosen to illustrate how bars are measured to be shorter in the short wavelength channels of NIRCam than in the long wavelength channels. The rest-frame wavelength is $\sim 0.645 \mu m$ and $1.4 \mu m$ for the two wavelengths, respectively.}
    \label{fig: filter comp}
\end{figure}

Attempting to disentangle the true length of the bars, we compared the bars in F200W with those of F356W+F444W. Systematically, the bars are measured to be shorter in F200W than in F356W+F444W images. The simulation work from \citetalias{Liang_2024} comments on the robustness of barred studies at high-$z$, and shows that the $PA$ is unchanged by resolution effects and that the bar size is minimally affected under the condition that $L_{bar} > 2 \times FWHM$.

Considering the mean $L_{bar}$ measured in F200W images is 2.9 kpc, the bias-corrected bar length $L_{corr} = 2.93$ kpc \citepalias[For further details see \S~\ref{Sec: ell-corr}, or Equation 4 of ][]{Liang_2024}. Hence, considering resolution effects, the short wavelength bars remain systematically shorter than those found with longer wavelength filters.

For galaxies identified as barred in both the short and long wavelength filters of NIRCam, we measure the ratio of bar length, $\frac{L_{bar,F356W+F444W}}{L_{bar,F200W}}$. We find the bars in F356W+F444W to be 1.4 times longer than those in F200W.

To understand the phenomenon behind this result, we show the example galaxy EGS 23205 in Figure \ref{fig: filter comp} in both filters. In this case, it can be clearly seen that the bar is shorter in the shorter wavelength image. We speculate on the origin of this observation; this galaxy is at $z \approx 2.1$, meaning that the corresponding rest-frame wavelength for F200W is $\sim 6450 \si{\angstrom}$. 

It is known that bars are more apparent in the NIR than in visible wavelengths \citep[e.g.,][]{Marinova_2007}. Hence, the ends of the bar in EGS 23205 are fainter/obscured in F200W due to the bar dust lanes crossing the bar-spiral arm boundary. Therefore, the fact that we find that on average bars tend to be shorter in F200W as compared to F356W+F444W seems to be due to two effects: first, because of the effect just mentioned at the bar-spiral arm boundary, but also, second, because the better spatial resolution in F200W allows us to see some short bars that are not detected in F444W.

In contrast, bars were found to be 9 per cent longer in B-band observations than at $3.6 \mu m$ in 16 nearby S$^4$G galaxies in \citet[][]{Delmestre_2024}. On the other hand, an increase in bar length and ellipticity at bluer wavelengths was not seen in the study by \citet[][]{Goncalves_2025} with 50 barred galaxies in the TNG50 simulations. However when considering only star-forming galaxies, the wavelength dependence becomes prominent. The authors argue that the older stars populating the bar form shorter and rounder bars, whilst younger stars extend to greater radii, hence increasing bar length and ellipticity. This contrasts with our results, where bars are {\em shorter} at shorter wavelengths. However, it is not clear if the effects from dust absorption and scattering are comparable between these studies and ours. In addition, the aforementioned studies have employed much broader wavelength ranges than we do here. Furthermore, it should be noted that star-forming spiral arms become more apparent at bluer wavelengths, which can bias bar length measurements based on ellipse fits to trace bars as longer structures, artificially going over the boundary between the spiral arms and the stellar bar. In this study, we have carefully inspected our ellipse fits individually and taken appropriate action to counter this effect (see Sect.\,\ref{Sec: bar length}).

\subsection{Bar length-stellar mass relation}
\label{Sec: mass relation}
\begin{figure}
	\includegraphics[width=\columnwidth]{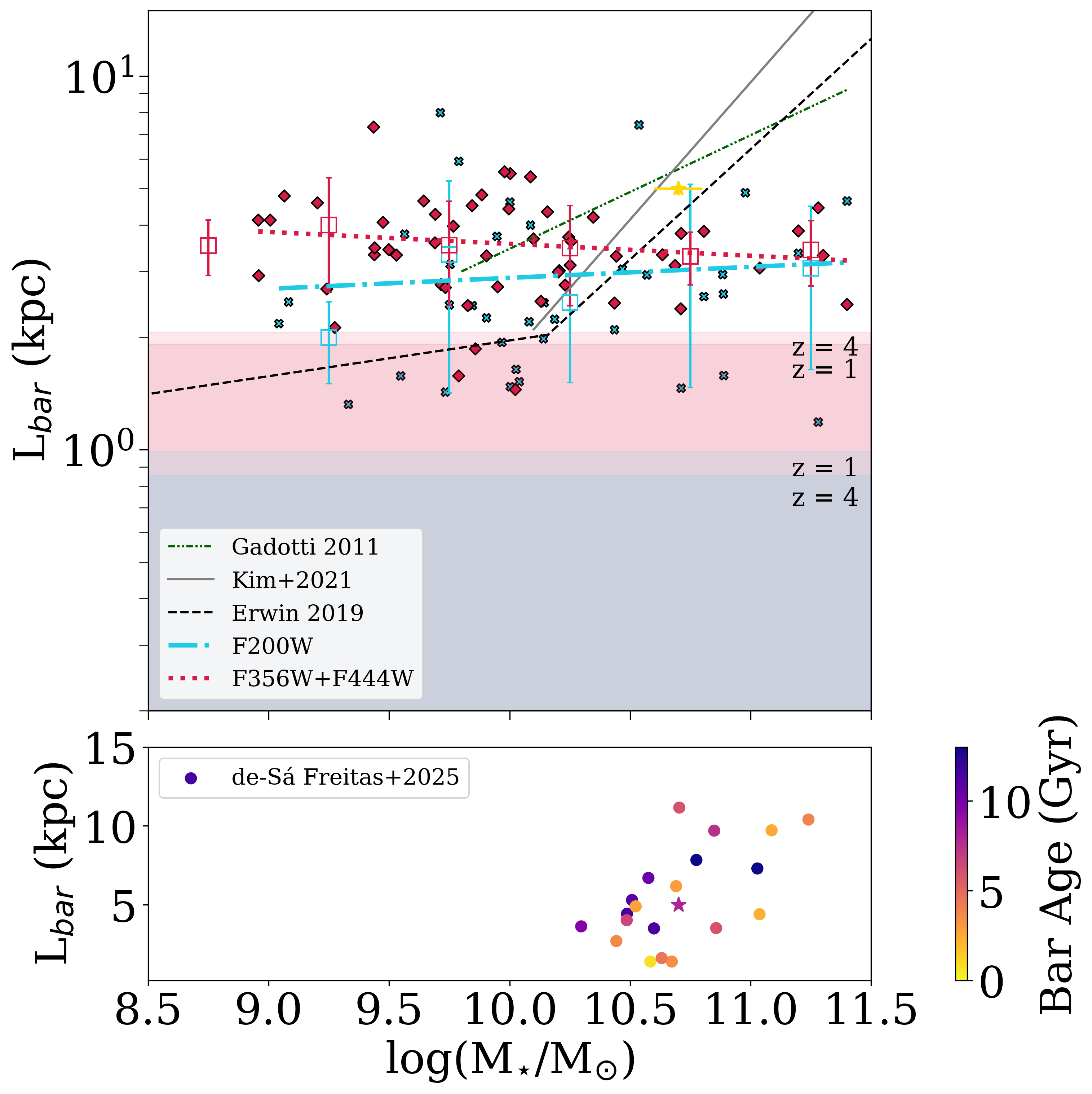}
    \caption{Bar length versus stellar mass for JWST NIRCam barred galaxies at $1 \leq z \leq 4$. The two JWST samples in F200W (crosses) and F356W+F444W (diamonds) with ordinary least squares fits (dash-dotted and dotted, respectively); binned mean L$_{bar}$ as boxes and standard deviation as error bars. The star indicates the Milky Way \citep[see results in][]{Bland_Hawthorn_2016,Wylie_2022}}. The green dash-dot-dotted is the ordinary least squares fit to nearby SDSS barred galaxies from \citetalias{Gadotti_2011}. The grey solid line is the ordinary least squares bisector fit to barred galaxies between $0.2 \leq z \leq 0.8$ from \citetalias{Kim_2021}. The black dashed line shows the locally weighted regression fit for nearby S$^4$G  spiral galaxies from \citet{Erwin_2019}. The red and blue shaded regions show $2 \times FWHM$ for the lower and upper redshift boundaries of the F356W+F444W and F200W samples, respectively. The bottom panel shows the bar length stellar mass relation for local galaxies with bar ages estimated.
    \label{fig: length-mass}
\end{figure}

We showed that strongly barred galaxies are most commonly found in the most massive galaxies in \S~\ref{Sec: mass} in our nearly mass-complete sample of high-$z$ galaxies. To explore further the mass dependence of these high-$z$ barred galaxies, we show the bar length versus the galaxy mass in Figure \ref{fig: length-mass}.

In the local Universe, the understandable association that the bar size scales with stellar mass was shown in S$^4$G galaxies from \citet{DiazGarcia_2016}. \citet[][]{Erwin_2019} reports on a correlation between bar length and galaxy mass, which is nearly flat for log($M_{\star}/M_{\odot} \leq 10$) and becomes steep for greater masses; hence, the most massive disc galaxies are host to the largest structures (see his Figure 1). The Milky Way resides on this relation, with the Galactic stellar mass and bar length taken from \citet[][]{Bland_Hawthorn_2016} and the bar age from \citet[][]{Wylie_2022}.

Our sample does not follow a steep relation for log($M_{\star}/M_{\odot} \geq 10$), for we measure shorter bars in the most massive galaxies at high $z$, as compared to local counterparts. Using an ordinary least-squares fit, we find that the galaxies in the F200W sample follow log($R_{bar}) = 0.20 \times log(M_\star/M_\odot) + 0.91$, and that the galaxies in the F356W+F444W sample follow log($R_{bar}) = -0.26 \times log(M_\star/M_\odot) + 6.15$. The improved PSF FWHM of the F200W sample means we should observe the shorter bars in less massive galaxies; however, our observations reside above the linear relation.

We also report on the fitting to nearby SDSS galaxies in \citetalias{Gadotti_2011}, which follow the trend of S$^4$G galaxies, albeit with a shallower slope. Additionally, the intermediate redshift, log$(M_\star/M_\odot)\geq10$ barred galaxy sample from \citetalias{Kim_2021} scales to $L\_{bar}$ with a slightly steeper linear fit than to the S$^4$G sample. 
%The shaded regions in Figure \ref{fig: length-mass} show the minimum $L_{bar} \leq 2 \times FWHM$. 

We show the steeper relation of \citetalias[][]{Kim_2021} in Figure \ref{fig: length-mass}, which is derived from their entire sample with a redshift range of $0.2 < z < 0.8$. However, they report that the relation depends on the redshift. Hence, considering only galaxies at $z \geq 0.5$, the steepness of the relation decreases; see their Figure 2d. We report that the relation is flat for $z > 1$; hence, between $0.5 < z < 1.5$, we suggest that the relation evolves. Based on the deviation between the high and low-$z$ relations at $> 10^{10} M_{\odot}$, the greatest growth would occur in the most massive galaxies. 

In fact, in the inset of Figure \ref{fig: length-mass}, we show the bar length as a function of galaxy mass for the sample of local galaxies for which bar ages were estimated in \citet[][]{Freitas_2025}. These initial results indicate that older bars are located within the upper envelope of the relation observed locally, generally consistent with this conjecture.

However, further studies with larger samples are required to make definitive statements. Toward the less massive end, the JWST samples sit above the flatter relation. We speculate that these galaxies are progenitors of the more massive galaxies in the local Universe; hence, they would grow in mass and shift towards the steeper linear relation at $z=0$. These results suggest that the bars at z > 1 are forming long, through mechanisms independent of their stellar mass, such as mergers. The lack of a trend is supported by the perspective of simulations on bar length evolution, namely the finding of merger-induced long bars in \citet[][see Section \ref{Sec: dis-discs} for further discussions]{Fragkoudi_2025}.

\subsection{The "downsizing" scenario}
\label{Sec: dis-strong}
In Figure \ref{fig: bar mass}, the mean stellar mass for strongly barred galaxies is greater than that of weakly barred galaxies, identified in either short or long wavelength images across the redshift range. Furthermore, the strongly barred fraction in disc galaxies surpasses the weakly barred fraction toward the most massive end of Figure \ref{fig: bar mass frac}, altogether suggesting that the first bars formed in the most massive galaxies. These results are consistent with the proposed model for "downsizing" in bar formation, for which the lowest mass galaxies could be the most dynamically hot due to the greater impacts of star formation, accretion and interactions. Hence, the most massive disc population has the greatest ability to withstand these interactions and remain dynamically cold \citep{Sheth_2012}, thus being subject to bar instabilities. 

On the other hand, \cite{Freitas_2025} find no correlation between the age of the bar and total present-day stellar mass of the galaxy in the TIMER sample of nearby galaxies, contrasting the downsizing scenario. The possibility of reconciling these seemingly opposing results concerns the effects of the environment. If interactions form bars that grow stronger faster than bars that form with no help from interactions \citep[e.g.,][]{Lokas_2016}, then, by considering that the most massive galaxies tend to reside in denser environments, the most massive galaxies would form stronger bars even without downsizing effects. Testing this possibility with a quantitative measurement of the environmental density for the galaxies in our sample of barred galaxies would thus be of major importance, but is beyond the scope of this paper.

\subsection{Co-evolution of the bar and disc}
\label{Sec: dis-discs}
Prior to JWST, the measurement of the bar length beyond $z = 1$ was restricted, and our understanding of the evolution of the bar length was instructed by simulations. \cite{Athanassoula_2003} reported that the bar length should evolve, in coordination with the slowing down of the pattern speed of the bar, causing the co-rotation radius to expand to greater radii, hence increasing the growth potential for the bar. 

In this study, one sees that there is a slight increase in both bar length and disc size from $z=4$ to $z=1$, which corresponds to a period of approximately 4\,Gyr, but then at the transition between $z\approx1.5$ and $z\approx0.5$ (which also corresponds to about 4\,Gyr) there is a sharp increase in both parameters, which remain approximately constant (on average) all the way to $z\approx0$.

Our results show bars at $z > 1$ formed with lengths between 1 kpc and 8 kpc, with a mean $L_{bar} = 2.9 - 3.6$ kpc depending on the observed wavelength. Across the redshift range $0 \leq z \leq 4$, we observe no evolution in the normalised bar length. Although we do observe some evolution in the mean value of the absolute bar length (2 kpc on average), and the appearance of very long bars (i.e., with $L_{bar} > 8$ kpc; see Figure \ref{fig:length dist}). In addition, while bar length correlates with galaxy mass for bars at $z < 1$, we find that this is not the case beyond $z = 1$ (see Figure \ref{fig: length-mass}). Our results do not necessarily suggest that individual galaxies do not evolve, but that the general population shows, on average, little evolution. Our current sample of bars at $z > 1$ is not large enough to test these dependencies as predicted in simulations, but these would be important follow-up studies. 

The bars identified in this study are at different evolutionary stages, even when within the same redshift bin; thus, the mean values of the properties are a broad representation of the ages and stages of the bar evolution. In simulations, the formation time of the bar, as well as the mechanisms under which the bar forms, are known, and it has been seen that the evolutionary rate of the bar depends on the environment and host galaxy properties \citep[see, e.g.,][]{Athanassoula_2002,Lokas_2016,Fragkoudi_2025}.

Remarkably, the bar length normalised by disc size from the local to the high redshift Universe is constant at $L_{bar}/R_{90} \approx 0.5$ (see Fig. \ref{fig:length dist}). This result indicates that the growth of a bar scales with the disc size. In other words, both bars and discs grow from high redshifts to $z\sim0$, but do so at a similar rate.

In the Auriga suite of magneto-hydrodynamical cosmological zoom-in simulations \citep[][]{Grand_2017}, \citet[][]{Fragkoudi_2025} report on the evolution of 39 Milky-Way-like barred galaxies. Through significant merging events at $z > 1.5$, bars form with lengths that remain at $4 < L_{bar} < 7$ kpc. However, bars which form at later epochs form shorter through disc instabilities and evolve in length. Our results are consistent with these findings, as we see both some evolution of the mean bar length towards $z < 1$, as well as bars at $z > 1$ that are already as long as the mean bar length at $z \sim 0$, namely $\sim5$ kpc, and even some bars at $z > 1$ with lengths up to 8 kpc.

\subsection{Bar ellipticity correction}
\label{Sec: ell-corr}
The parameter most affected by low-resolution images is the bar ellipticity \citepalias{Liang_2024}. In \S~\ref{Sec: bar strength}, we share the results that the average $e_{bar}$ in JWST bars is significantly less than that reported at $z=0$. These values correspond to projected measurements as the deprojection of the bar can systematically lead to lower values of $e_{bar}$ (i.e., rounder bars) if the bar is aligned with the galaxy line of nodes.

\citetalias{Liang_2024} show the impact of resolution levels, defined as $n = L_{bar}/FWHM$, on the mean fractional difference ($\overline{\Delta e/e}$) between true and observed bar ellipticities for simulated data (see their Figure 10). We can use $\overline{\Delta e/e}$ to obtain the bias-corrected bar ellipticity, following Equation 4 in \citetalias[][]{Liang_2024}. 

\begin{figure}
	\includegraphics[width=\columnwidth]{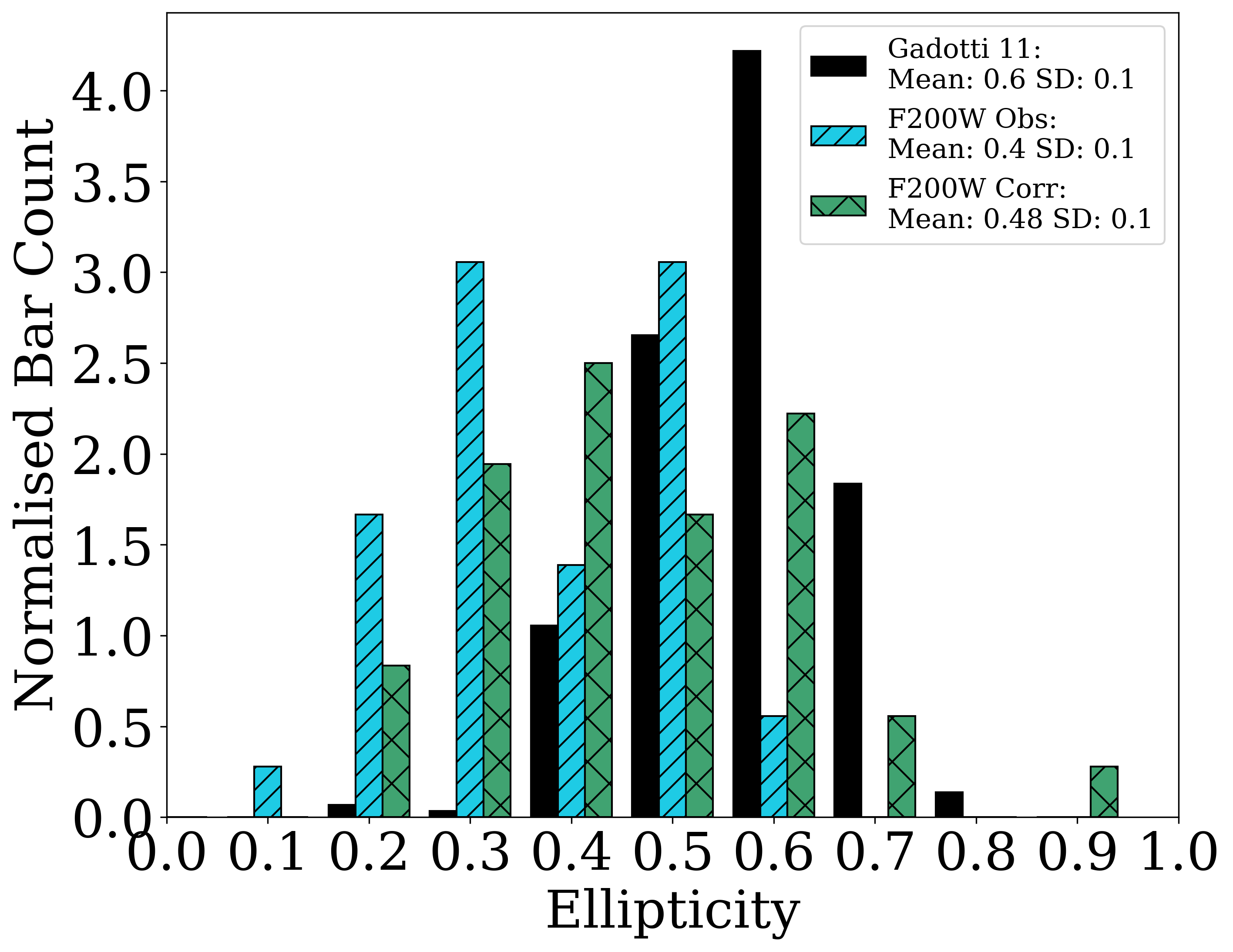}
    \caption{Distribution of the projected bar ellipticity for the JWST NIRCam filter F200W (double hatched) and corrected for resolution effects (cross hatched) for bars found in galaxies in the redshift range $1 \leq z \leq 4$. Bias-corrected bar ellipticities are derived from \citetalias{Liang_2024} (their Equation 4). The high redshift sample is compared against a sample of SDSS barred galaxies at $z \approx 0$ \citetalias{Gadotti_2011} (solid black). The mean ellipticity and standard deviation are given for each sample.}
    \label{fig: ell corr}
\end{figure}

The mean resolution level for the F200W sample is $n = 5.1$, and the standard deviation is 2.8. Hence, the mean $n$ corresponds to the value $\overline{\Delta e/e} = -0.10$. Thus, for the average F200W bar ellipticity, $e_{bar} = 0.4$, the bias-corrected average bar ellipticity becomes $e_{corr} = 0.48$. Figure \ref{fig: ell corr} shows the shift from rounder observed bar ellipticities in the F200W sample, which are significantly rounder than the SDSS sample at $z=0$, to more elongated structures when corrected for resolution effect biases. 

We do not repeat this bias correction in the JWST F356W+F444W sample, as the resolution corrections derived from simulated data are produced from images in the F200W filter in \citetalias{Liang_2024}, but we caution that the correction would most likely be greater, as the PSF FWHM worsens towards longer wavelengths in NIRCam. This implies that the barred structures observed at high-$z$ are commonly seen rounder and broader due to resolution effects, but in reality could resemble the defined and elongated bars in nearby disc galaxies.

\section{Summary and main conclusions}
\label{Sec: conclusions}
We observed bar-driven secular internal evolution to commence earlier than previously thought, only 2 billion years after the Big Bang. This was made possible by the improved sensitivity and longer wavelengths of the JWST NIRCam imaging. Several studies of barred galaxy abundance have observed that the bar fraction follows a declining trend towards high$-z$, however, it is twice that previously measured in HST studies \citepalias{LeConte_2024,Guo_2025,Salcedo_2025,Géron_2025}. To date, only \citetalias{Guo_2025} have explored the evolution of the bar length for a barred galaxy sample at $z > 1$. Hence, we measure the bar fraction and bar length evolution of a sample three times the size of the initial sample of \citetalias{LeConte_2024}.

This study uses NIRCam imaging of all 10 CEERS pointings and selects a 95\% empirically complete sample of 2438 galaxies in the redshift range $1 \leq z \leq 4$. To assess the wavelength dependence of the bar fraction and properties, we applied the methodology to two independent samples in different NIRCam wavelength filters: with the filters F356W+F444W we observe the older stellar population in the bar; the filter F200W has an improved PSF FWHM, to capture bars shorter than $\sim 2-3$ kpc, which may go undetected in the F356W or F444W images. 

Neighbouring sources are removed using SExtractor, then the samples are optimised using ellipse fits to remove highly inclined systems, after which the samples become 1073 and 649 galaxies in the F356W+F444W and F200W filters, respectively. Imfit is used to fit 2D S\'ersic models to the images, and through parameter limits, 152 and 37 point sources are removed from the F356W+F444W and F200W samples, respectively. The disc and barred galaxy samples are obtained by visual classifications from five of the coauthors, and a bar fraction is derived for the F200W and F356W+F444W samples. Using the $\Delta PA$ and $e_{max}$ values from the ellipse fits, we derive the deprojected bar lengths for the barred galaxy samples at Cosmic Noon.

A similar bar fraction is observed for both samples, and by combining the two samples we find a declining trend in the bar fraction: $0.16^{+0.03}_{-0.03}$ at $1 \leq z < 2$; $0.08^{+0.02}_{-0.01}$ at $2 \leq z < 3$; $0.07^{+0.03}_{-0.01}$ at $3 \leq z \leq 4$. These results are in agreement with our previous work and the other recent bar fraction investigations, suggesting that dynamically cold and rotationally supported massive discs (and therefore bar-unstable) are present at Cosmic Noon. Nonetheless, at least a fraction of these observed bars could have formed via interactions triggering the bar instability in otherwise stable discs.

The lowest bar fraction ($< 10$\%) is seen for galaxies with log$(M_{\star}/M_{\odot}) \leq 10$, which then raises to about 20\% for galaxies with log$(M_{\star}/M_{\odot}) \approx 10.75$ and above. These results suggest that the first bars formed in the most massive galaxies, which is consistent with the model for "downsizing" in bar formation. However, we discuss how interactions can trigger stronger bars in more massive galaxies without downsizing effects.

Systematically shorter bars in the short wavelength channel are observed because of the better physical spatial resolution, but we suggest that dust obscuration can also contribute to this result. Interestingly, we find that the correlation between bar length and galaxy mass for galaxies more massive than $10^{10} M_{\odot}$ observed at $z < 1$ is not seen at $z > 1$. 

No evolution in the bar length is measured in the long-wavelength NIRCam images in the range $1 \leq z \leq 4$, with a mean of 3.6\,kpc, but a slight increase of about 1\,kpc from $z = 4$ to $z = 1$ is measured in the short-wavelength images, which have a mean of 2.9\,kpc. By adding samples of barred galaxies studied with images from HST (at $0.2 < z < 0.8$) and SDSS (at $z \approx 0$), from \citetalias{Kim_2021} and \citetalias{Gadotti_2011}, respectively, we find that the absolute value of bar length increases by about 2 kpc on average from $z = 4$ to $z = 0$. However, there is significant overlap in the distributions of bar lengths at all redshifts. In addition, we show that bars and discs grow in tandem, for the bar length normalised by disc size, $L_{bar}/R_{90}$, does not evolve from $z = 4$ to $z = 0$.

Finally, we show that some of the bars at $z > 1$ are already as long and strong as the average bar at $z \approx 0$. On the other hand, we find that the long bars (with $L_{bar} > 8$ kpc) found at $z < 1$ are not seen beyond $z = 1$. Altogether, our results show that, as a population, on average, bars grow only moderately longer from $z > 1$ to $z < 1$, with long bars already present at $z > 1$, but that some bars have gone through significant growth in the same period.

\section*{Acknowledgements}
\label{Sec: ack}
This work was supported by STFC grants ST/X508354/1, ST/T000244/1 and ST/X001075/1. This work used the DiRAC@Durham facility managed by the Institute for Computational Cosmology on behalf of the STFC DiRAC HPC Facility (www.dirac.ac.uk). The equipment was funded by BEIS capital funding via STFC capital grants ST/K00042X/1, ST/P002293/1, ST/R002371/1 and ST/S002502/1, Durham University, and STFC operations grant ST/R000832/1. DiRAC is part of the National e-Infrastructure.

%%%%%%%%%%%%%%%%%%%%%%%%%%%%%%%%%%%%%%%%%%%%%%%%%%
\section*{Data Availability}
\label{Sec: data}
This work used Astropy \citep[][]{Astropy_2013}{}{}, SExtractor \citep[][]{Bertin_1996}, PHOTUTILS \citep[][]{Bradley_2022}{}{} and Imfit \citep[][]{Erwin_2015}. The specific observations analysed can be accessed via \url{https://doi.org/10.17909/xm8m-tt59}.


\bibliographystyle{mnras}
\bibliography{Le_Conte}

@ARTICLE{Bi_2022,
       author = {{Bi}, Da and {Shlosman}, Isaac and {Romano-D{\'\i}az}, Emilio},
        title = "{Modeling Evolution of Galactic Bars at Cosmic Dawn}",
      journal = {\apj},
         year = 2022,
        month = jul,
       volume = {934},
       number = {1},
          eid = {52},
        pages = {52},
          doi = {10.3847/1538-4357/ac779b},
archivePrefix = {arXiv},
       eprint = {2112.09718},
 primaryClass = {astro-ph.GA},
       adsurl = {https://ui.adsabs.harvard.edu/abs/2022ApJ...934...52B}
}

@ARTICLE{Wylie_2022,
       author = {{Wylie}, Shola M. and {Clarke}, Jonathan P. and {Gerhard}, Ortwin E.},
        title = "{The Milky Way's middle-aged inner ring}",
      journal = {\aap},
         year = 2022,
        month = mar,
       volume = {659},
          eid = {A80},
        pages = {A80},
          doi = {10.1051/0004-6361/202142343},
archivePrefix = {arXiv},
       eprint = {2110.03658},
 primaryClass = {astro-ph.GA},
       adsurl = {https://ui.adsabs.harvard.edu/abs/2022A&A...659A..80W}
}

@article{Bland_Hawthorn_2016,
   title={The Galaxy in Context: Structural, Kinematic, and Integrated Properties},
   volume={54},
   ISSN={1545-4282},
   url={http://dx.doi.org/10.1146/annurev-astro-081915-023441},
   DOI={10.1146/annurev-astro-081915-023441},
   number={1},
   journal={Annual Review of Astronomy and Astrophysics},
   publisher={Annual Reviews},
   author={Bland-Hawthorn, Joss and Gerhard, Ortwin},
   year={2016},
   month=sep, pages={529–596} }

@misc{euclid2_2025,
      title={Euclid Quick Data Release (Q1): First visual morphology catalogue}, 
      author={Euclid Collaboration and M. Walmsley and M. Huertas-Company and L. Quilley and K. L. Masters and S. Kruk and K. A. Remmelgas and J. J. Popp and E. Romelli and D. O'Ryan and H. J. Dickinson and C. J. Lintott and S. Serjeant and R. J. Smethurst and B. Simmons and J. Shingirai Makechemu and I. L. Garland and H. Roberts and K. Mantha and L. F. Fortson and T. Géron and W. Keel and E. M. Baeten and C. Macmillan and J. Bovy and S. Casas and C. De Leo and H. Domínguez Sánchez and J. Katona and A. Kovács and N. Aghanim and B. Altieri and A. Amara and S. Andreon and N. Auricchio and H. Aussel and C. Baccigalupi and M. Baldi and A. Balestra and S. Bardelli and A. Basset and P. Battaglia and R. Bender and A. Biviano and A. Bonchi and E. Branchini and M. Brescia and J. Brinchmann and S. Camera and G. Cañas-Herrera and V. Capobianco and C. Carbone and J. Carretero and F. J. Castander and M. Castellano and G. Castignani and S. Cavuoti and K. C. Chambers and A. Cimatti and C. Colodro-Conde and G. Congedo and C. J. Conselice and L. Conversi and Y. Copin and F. Courbin and H. M. Courtois and M. Cropper and A. Da Silva and H. Degaudenzi and G. De Lucia and A. M. Di Giorgio and C. Dolding and H. Dole and F. Dubath and C. A. J. Duncan and X. Dupac and S. Dusini and A. Ealet and S. Escoffier and M. Fabricius and M. Farina and R. Farinelli and F. Faustini and F. Finelli and P. Fosalba and S. Fotopoulou and M. Frailis and E. Franceschi and S. Galeotta and K. George and B. Gillis and C. Giocoli and P. Gómez-Alvarez and J. Gracia-Carpio and B. R. Granett and A. Grazian and F. Grupp and S. Gwyn and S. V. H. Haugan and H. Hoekstra and W. Holmes and I. M. Hook and F. Hormuth and A. Hornstrup and P. Hudelot and K. Jahnke and M. Jhabvala and B. Joachimi and E. Keihänen and S. Kermiche and A. Kiessling and R. Kohley and B. Kubik and K. Kuijken and M. Kümmel and M. Kunz and H. Kurki-Suonio and O. Lahav and Q. Le Boulc'h and A. M. C. Le Brun and D. Le Mignant and P. Liebing and S. Ligori and P. B. Lilje and V. Lindholm and I. Lloro and G. Mainetti and D. Maino and E. Maiorano and O. Mansutti and S. Marcin and O. Marggraf and M. Martinelli and N. Martinet and F. Marulli and R. Massey and S. Maurogordato and H. J. McCracken and E. Medinaceli and S. Mei and M. Melchior and Y. Mellier and M. Meneghetti and E. Merlin and G. Meylan and A. Mora and M. Moresco and L. Moscardini and R. Nakajima and C. Neissner and R. C. Nichol and S. -M. Niemi and J. W. Nightingale and C. Padilla and S. Paltani and F. Pasian and K. Pedersen and W. J. Percival and V. Pettorino and S. Pires and G. Polenta and M. Poncet and L. A. Popa and L. Pozzetti and F. Raison and R. Rebolo and A. Renzi and J. Rhodes and G. Riccio and M. Roncarelli and B. Rusholme and R. Saglia and Z. Sakr and A. G. Sánchez and D. Sapone and B. Sartoris and J. A. Schewtschenko and P. Schneider and T. Schrabback and M. Scodeggio and A. Secroun and G. Seidel and M. Seiffert and S. Serrano and P. Simon and C. Sirignano and G. Sirri and L. Stanco and J. Steinwagner and P. Tallada-Crespí and D. Tavagnacco and A. N. Taylor and H. I. Teplitz and I. Tereno and N. Tessore and S. Toft and R. Toledo-Moreo and F. Torradeflot and I. Tutusaus and E. A. Valentijn and L. Valenziano and J. Valiviita and T. Vassallo and G. Verdoes Kleijn and A. Veropalumbo and Y. Wang and J. Weller and A. Zacchei and G. Zamorani and F. M. Zerbi and I. A. Zinchenko and E. Zucca and V. Allevato and M. Ballardini and M. Bolzonella and E. Bozzo and C. Burigana and R. Cabanac and A. Cappi and D. Di Ferdinando and J. A. Escartin Vigo and L. Gabarra and J. Martín-Fleitas and S. Matthew and N. Mauri and R. B. Metcalf and A. Pezzotta and M. Pöntinen and C. Porciani and I. Risso and V. Scottez and M. Sereno and M. Tenti and M. Viel and M. Wiesmann and Y. Akrami and I. T. Andika and S. Anselmi and M. Archidiacono and F. Atrio-Barandela and C. Benoist and K. Benson and D. Bertacca and M. Bethermin and L. Bisigello and A. Blanchard and L. Blot and H. Böhringer and M. L. Brown and S. Bruton and F. Buitrago and A. Calabro and B. Camacho Quevedo and F. Caro and C. S. Carvalho and T. Castro and F. Cogato and A. R. Cooray and O. Cucciati and S. Davini and F. De Paolis and G. Desprez and A. Díaz-Sánchez and J. J. Diaz and S. Di Domizio and J. M. Diego and P. -A. Duc and A. Enia and Y. Fang and A. G. Ferrari and A. Finoguenov and A. Fontana and A. Franco and K. Ganga and J. García-Bellido and T. Gasparetto and V. Gautard and E. Gaztanaga and F. Giacomini and G. Gozaliasl and M. Guidi and C. M. Gutierrez and A. Hall and W. G. Hartley and S. Hemmati and C. Hernández-Monteagudo and H. Hildebrandt and J. Hjorth and J. J. E. Kajava and Y. Kang and V. Kansal and D. Karagiannis and K. Kiiveri and C. C. Kirkpatrick and J. Le Graet and L. Legrand and M. Lembo and F. Lepori and G. Leroy and G. F. Lesci and J. Lesgourgues and L. Leuzzi and T. I. Liaudat and A. Loureiro and J. Macias-Perez and G. Maggio and M. Magliocchetti and F. Mannucci and R. Maoli and C. J. A. P. Martins and L. Maurin and M. Miluzio and P. Monaco and C. Moretti and G. Morgante and C. Murray and S. Nadathur and K. Naidoo and A. Navarro-Alsina and S. Nesseris and F. Passalacqua and K. Paterson and L. Patrizii and A. Pisani and D. Potter and S. Quai and M. Radovich and P. -F. Rocci and G. Rodighiero and S. Sacquegna and M. Sahlén and D. B. Sanders and E. Sarpa and C. Scarlata and J. Schaye and A. Schneider and M. Schultheis and D. Sciotti and E. Sellentin and F. Shankar and L. C. Smith and K. Tanidis and G. Testera and R. Teyssier and S. Tosi and A. Troja and M. Tucci and C. Valieri and A. Venhola and D. Vergani and G. Verza and P. Vielzeuf and N. A. Walton and E. Soubrie and D. Scott},
      year={2025},
      eprint={2503.15310},
      archivePrefix={arXiv},
      primaryClass={astro-ph.GA},
      url={https://arxiv.org/abs/2503.15310}, 
}

@ARTICLE{Euclid_2025,
       author = {{Euclid Collaboration} and {Huertas-Company}, M. and {Walmsley}, M. and {Siudek}, M. and {Iglesias-Navarro}, P. and {Knapen}, J.~H. and {Serjeant}, S. and {Dickinson}, H.~J. and {Fortson}, L. and {Garland}, I. and {G{\'e}ron}, T. and {Keel}, W. and {Kruk}, S. and {Lintott}, C.~J. and {Mantha}, K. and {Masters}, K. and {O'Ryan}, D. and {Popp}, J.~J. and {Roberts}, H. and {Scarlata}, C. and {Makechemu}, J.~S. and {Simmons}, B. and {Smethurst}, R.~J. and {Spindler}, A. and {Baes}, M. and {Corsini}, E.~M. and {Dom{\'\i}nguez S{\'a}nchez}, H. and {Duran-Camacho}, E. and {Fu}, H. and {Junais}, J. and {Mendez-Abreu}, J. and {Nersesian}, A. and {Shankar}, F. and {Le}, M.~N. and {Vega-Ferrero}, J. and {Wang}, L. and {Aghanim}, N. and {Altieri}, B. and {Amara}, A. and {Andreon}, S. and {Auricchio}, N. and {Baccigalupi}, C. and {Baldi}, M. and {Balestra}, A. and {Bardelli}, S. and {Basset}, A. and {Battaglia}, P. and {Bernardeau}, F. and {Biviano}, A. and {Bonchi}, A. and {Branchini}, E. and {Brescia}, M. and {Brinchmann}, J. and {Camera}, S. and {Capobianco}, V. and {Carbone}, C. and {Carretero}, J. and {Casas}, S. and {Castellano}, M. and {Castignani}, G. and {Cavuoti}, S. and {Chambers}, K.~C. and {Cimatti}, A. and {Colodro-Conde}, C. and {Congedo}, G. and {Conselice}, C.~J. and {Conversi}, L. and {Copin}, Y. and {Courbin}, F. and {Courtois}, H.~M. and {Cropper}, M. and {Da Silva}, A. and {Degaudenzi}, H. and {De Lucia}, G. and {Di Giorgio}, A.~M. and {Dolding}, C. and {Dole}, H. and {Dubath}, F. and {Duncan}, C.~A.~J. and {Dupac}, X. and {Dusini}, S. and {Ealet}, A. and {Escoffier}, S. and {Fabricius}, M. and {Farina}, M. and {Farinelli}, R. and {Faustini}, F. and {Ferriol}, S. and {Finelli}, F. and {Fotopoulou}, S. and {Frailis}, M. and {Galeotta}, S. and {George}, K. and {Gillard}, W. and {Gillis}, B. and {Giocoli}, C. and {Gracia-Carpio}, J. and {Grazian}, A. and {Grupp}, F. and {Gwyn}, S. and {Haugan}, S.~V.~H. and {Hoekstra}, H. and {Holmes}, W. and {Hook}, I.~M. and {Hormuth}, F. and {Hornstrup}, A. and {Hudelot}, P. and {Jahnke}, K. and {Jhabvala}, M. and {Keih{\"a}nen}, E. and {Kermiche}, S. and {Kubik}, B. and {Kuijken}, K. and {K{\"u}mmel}, M. and {Kunz}, M. and {Kurki-Suonio}, H. and {Le Boulc'h}, Q. and {Le Brun}, A.~M.~C. and {Le Mignant}, D. and {Ligori}, S. and {Lilje}, P.~B. and {Lindholm}, V. and {Lloro}, I. and {Maino}, D. and {Maiorano}, E. and {Mansutti}, O. and {Marcin}, S. and {Marggraf}, O. and {Martinelli}, M. and {Martinet}, N. and {Marulli}, F. and {Massey}, R. and {McCracken}, H.~J. and {Medinaceli}, E. and {Melchior}, M. and {Mellier}, Y. and {Meneghetti}, M. and {Merlin}, E. and {Meylan}, G. and {Mora}, A. and {Moresco}, M. and {Moscardini}, L. and {Neissner}, C. and {Nichol}, R.~C. and {Niemi}, S. -M. and {Nightingale}, J.~W. and {Padilla}, C. and {Paltani}, S. and {Pasian}, F. and {Pedersen}, K. and {Percival}, W.~J. and {Pettorino}, V. and {Pires}, S. and {Polenta}, G. and {Poncet}, M. and {Popa}, L.~A. and {Pozzetti}, L. and {Raison}, F. and {Renzi}, A. and {Rhodes}, J. and {Riccio}, G. and {Romelli}, E. and {Roncarelli}, M. and {Saglia}, R. and {Sakr}, Z. and {Sapone}, D. and {Sartoris}, B. and {Schirmer}, M. and {Schneider}, P. and {Scodeggio}, M. and {Secroun}, A. and {Seidel}, G. and {Seiffert}, M. and {Serrano}, S. and {Simon}, P. and {Sirignano}, C. and {Sirri}, G. and {Stanco}, L. and {Steinwagner}, J. and {Tallada-Cresp{\'\i}}, P. and {Taylor}, A.~N. and {Tereno}, I. and {Toft}, S. and {Toledo-Moreo}, R. and {Torradeflot}, F. and {Tutusaus}, I. and {Valenziano}, L. and {Valiviita}, J. and {Vassallo}, T. and {Verdoes Kleijn}, G. and {Wang}, Y. and {Weller}, J. and {Zacchei}, A. and {Zamorani}, G. and {Zerbi}, F.~M. and {Zinchenko}, I.~A. and {Zucca}, E. and {Allevato}, V. and {Ballardini}, M. and {Bolzonella}, M.},
        title = "{Euclid Quick Data Release (Q1), A first look at the fraction of bars in massive galaxies at $z<1$}",
      journal = {arXiv e-prints},
         year = 2025,
        month = mar,
          eid = {arXiv:2503.15311},
        pages = {arXiv:2503.15311},
          doi = {10.48550/arXiv.2503.15311},
archivePrefix = {arXiv},
       eprint = {2503.15311},
 primaryClass = {astro-ph.GA},
       adsurl = {https://ui.adsabs.harvard.edu/abs/2025arXiv250315311E}
}

@ARTICLE{Neumann_2024,
       author = {{Neumann}, Justus and {Thomas}, Daniel and {Maraston}, Claudia and {Gleis}, Damian R. and {Mao}, Chuanming and {Schinnerer}, Eva and {Stuber}, Sophia K.},
        title = "{Azimuthal variations of stellar populations in barred galaxies}",
      journal = {\mnras},
         year = 2024,
        month = nov,
       volume = {534},
       number = {3},
        pages = {2438-2457},
          doi = {10.1093/mnras/stae2252},
archivePrefix = {arXiv},
       eprint = {2409.18180},
 primaryClass = {astro-ph.GA},
       adsurl = {https://ui.adsabs.harvard.edu/abs/2024MNRAS.534.2438N}
}

@ARTICLE{Lokas_2016,
       author = {{{\L}okas}, Ewa L. and {Ebrov{\'a}}, Ivana and {del Pino}, Andr{\'e}s and {Sybilska}, Agnieszka and {Athanassoula}, E. and {Semczuk}, Marcin and {Gajda}, Grzegorz and {Fouquet}, Sylvain},
        title = "{Tidally Induced Bars of Galaxies in Clusters}",
      journal = {\apj},
         year = 2016,
        month = aug,
       volume = {826},
       number = {2},
          eid = {227},
        pages = {227},
          doi = {10.3847/0004-637X/826/2/227},
archivePrefix = {arXiv},
       eprint = {1601.07433},
 primaryClass = {astro-ph.GA},
       adsurl = {https://ui.adsabs.harvard.edu/abs/2016ApJ...826..227L}
}

@INPROCEEDINGS{Garland_2024,
       author = {{Garland}, Izzy},
        title = "{Large-scale bars as a mechanism for triggering AGN}",
    booktitle = {Galaxies at Crossroads: Outflows and IMF in the VLT/ELT/ALMA/JWST Era},
         year = 2024,
        month = sep,
          eid = {19},
        pages = {19},
          doi = {10.5281/zenodo.14901974},
       adsurl = {https://ui.adsabs.harvard.edu/abs/2024gcoi.confE..19G}
}

@ARTICLE{Grand_2017,
       author = {{Grand}, Robert J.~J. and {G{\'o}mez}, Facundo A. and {Marinacci}, Federico and {Pakmor}, R{\"u}diger and {Springel}, Volker and {Campbell}, David J.~R. and {Frenk}, Carlos S. and {Jenkins}, Adrian and {White}, Simon D.~M.},
        title = "{The Auriga Project: the properties and formation mechanisms of disc galaxies across cosmic time}",
      journal = {\mnras},
         year = 2017,
        month = may,
       volume = {467},
       number = {1},
        pages = {179-207},
          doi = {10.1093/mnras/stx071},
archivePrefix = {arXiv},
       eprint = {1610.01159},
 primaryClass = {astro-ph.GA},
       adsurl = {https://ui.adsabs.harvard.edu/abs/2017MNRAS.467..179G}
}

@ARTICLE{Gadotti_2008,
       author = {{Gadotti}, Dimitri Alexei},
        title = "{Image decomposition of barred galaxies and AGN hosts}",
      journal = {\mnras},
         year = 2008,
        month = feb,
       volume = {384},
       number = {1},
        pages = {420-439},
          doi = {10.1111/j.1365-2966.2007.12723.x},
archivePrefix = {arXiv},
       eprint = {0708.3870},
 primaryClass = {astro-ph},
       adsurl = {https://ui.adsabs.harvard.edu/abs/2008MNRAS.384..420G}
}

@ARTICLE{DiazGarcia_2016,
       author = {{D{\'\i}az-Garc{\'\i}a}, S. and {Salo}, H. and {Laurikainen}, E. and {Herrera-Endoqui}, M.},
        title = "{Characterization of galactic bars from 3.6 {\ensuremath{\mu}}m S$^{4}$G imaging}",
      journal = {\aap},
         year = 2016,
        month = mar,
       volume = {587},
          eid = {A160},
        pages = {A160},
          doi = {10.1051/0004-6361/201526161},
archivePrefix = {arXiv},
       eprint = {1509.06743},
 primaryClass = {astro-ph.GA},
       adsurl = {https://ui.adsabs.harvard.edu/abs/2016A&A...587A.160D}
}

@ARTICLE{Conselice_2024,
       author = {{Conselice}, Christopher J. and {Basham}, Justin T.~F. and {Bettaney}, Daniel O. and {Ferreira}, Leonardo and {Adams}, Nathan and {Harvey}, Thomas and {Ormerod}, Katherine and {Caruana}, Joseph and {Bluck}, Asa F.~L. and {Li}, Qiong and {Roper}, William J. and {Trussler}, James and {Irodotou}, Dimitrios and {Austin}, Duncan},
        title = "{EPOCHS Paper V. The dependence of galaxy formation on galaxy structure at z < 7 from JWST observations}",
      journal = {\mnras},
         year = 2024,
        month = jul,
       volume = {531},
       number = {4},
        pages = {4857-4875},
          doi = {10.1093/mnras/stae1180},
archivePrefix = {arXiv},
       eprint = {2405.00376},
 primaryClass = {astro-ph.GA},
       adsurl = {https://ui.adsabs.harvard.edu/abs/2024MNRAS.531.4857C}
}

@ARTICLE{Adams_2025,
       author = {{Adams}, Nathan J. and {Ferrami}, Giovanni and {Westcott}, Lewi and {Harvey}, Thomas and {Estrada-Carpenter}, Vicente and {Conselice}, Christopher J. and {Austin}, Duncan and {Wyithe}, J. Stuart B. and {Goolsby}, Caio M. and {Li}, Qiong and {Rusakov}, Vadim and {Windhorst}, Rogier A. and {Cohen}, Seth H. and {Jansen}, Rolf A. and {Summers}, Jake and {O'Brein}, Roselia and {Koekemoer}, Anton M. and {Driver}, Simon P. and {Frye}, Brenda and {Hathi}, Nimish P. and {Coe}, Dan and {Grogin}, Norman A. and {Marshall}, Madeline A. and {Pirzkal}, Nor and {Ryan}, Jr., Russell E. and {Willmer}, Christopher N.~A. and {Yan}, Haojing and {Holwerda}, Benne W. and {Kamieneski}, Patrick S. and {Broadhurst}, Tom and {Maksym}, W. Peter and {Saikia}, Payaswini and {Gelfand}, Joseph D.},
        title = "{JWSTs PEARLS: NIRCam imaging and NIRISS spectroscopy of a $z=3.6$ star-forming galaxy lensed into a near-Einstein Ring by a $z=1.258$ massive elliptical galaxy}",
      journal = {arXiv e-prints},
         year = 2025,
        month = apr,
          eid = {arXiv:2504.03571},
        pages = {arXiv:2504.03571},
          doi = {10.48550/arXiv.2504.03571},
archivePrefix = {arXiv},
       eprint = {2504.03571},
 primaryClass = {astro-ph.GA},
       adsurl = {https://ui.adsabs.harvard.edu/abs/2025arXiv250403571A}
}

@misc{Conselice_2024a,
      title={EPOCHS I. The Discovery and Star Forming Properties of Galaxies in the Epoch of Reionization at $6.5 < z < 18$ with PEARLS and Public JWST data}, 
      author={Christopher J. Conselice and Nathan Adams and Thomas Harvey and Duncan Austin and Leonardo Ferreira and Katherine Ormerod and Qiao Duan and James Trussler and Qiong Li and Ignas Juodzbalis and Lewi Westcott and Honor Harris and Louise T. C. Seeyave and Asa F. L. Bluck and Rogier A. Windhorst and Rachana Bhatawdekar and Dan Coe and Seth H. Cohen and Cheng Cheng and Simon P. Driver and Brenda Frye and Lukas J. Furtak and Norman A. Grogin and Nimish P. Hathi and Benne W. Holwerda and Rolf A. Jansen and Anton M. Koekemoer and Madeline A. Marshall and Mario Nonino and Aaron Robotham and Jake Summers and Stephen M. Wilkins and Christopher N. A. Willmer and Haojing Yan and Adi Zitrin},
      year={2024},
      eprint={2407.14973},
      archivePrefix={arXiv},
      primaryClass={astro-ph.GA},
      url={https://arxiv.org/abs/2407.14973}, 
}

@ARTICLE{Sanchez_2011,
       author = {{S{\'a}nchez-Bl{\'a}zquez}, P. and {Ocvirk}, P. and {Gibson}, B.~K. and {P{\'e}rez}, I. and {Peletier}, R.~F.},
        title = "{Star formation history of barred disc galaxies}",
      journal = {\mnras},
         year = 2011,
        month = jul,
       volume = {415},
       number = {1},
        pages = {709-731},
          doi = {10.1111/j.1365-2966.2011.18749.x},
archivePrefix = {arXiv},
       eprint = {1103.3796},
 primaryClass = {astro-ph.CO},
       adsurl = {https://ui.adsabs.harvard.edu/abs/2011MNRAS.415..709S}
}

@ARTICLE{Gadotti_2006,
       author = {{Gadotti}, D.~A. and {de Souza}, R.~E.},
        title = "{On the Lengths, Colors, and Ages of 18 Face-on Bars}",
      journal = {\apjs},
         year = 2006,
        month = apr,
       volume = {163},
       number = {2},
        pages = {270-281},
          doi = {10.1086/500175},
archivePrefix = {arXiv},
       eprint = {astro-ph/0511799},
 primaryClass = {astro-ph},
       adsurl = {https://ui.adsabs.harvard.edu/abs/2006ApJS..163..270G}
}

@ARTICLE{Goncalves_2025,
       author = {{Gon{\c{c}}alves}, Gustavo F. and {Machado}, Rubens E.~G. and {Men{\'e}ndez-Delmestre}, Kar{\'\i}n and {Bueno-Dalpiaz}, Thiago},
        title = "{Bar properties as a function of wavelength in Illustris TNG50: analysis of mock images}",
      journal = {arXiv e-prints},
         year = 2025,
        month = jun,
          eid = {arXiv:2506.19373},
        pages = {arXiv:2506.19373},
          doi = {10.48550/arXiv.2506.19373},
archivePrefix = {arXiv},
       eprint = {2506.19373},
 primaryClass = {astro-ph.GA},
       adsurl = {https://ui.adsabs.harvard.edu/abs/2025arXiv250619373G}
}

@ARTICLE{Delmestre_2024,
       author = {{Men{\'e}ndez-Delmestre}, Kar{\'\i}n and {Gon{\c{c}}alves}, Thiago S. and {Sheth}, Kartik and {D{\"u}ringer Jacques de Lima}, Tom{\'a}s and {Kim}, Taehyun and {Gadotti}, Dimitri A. and {Schinnerer}, Eva and {Athanassoula}, E. and {Bosma}, Albert and {Elmegreen}, Debra Meloy and {Knapen}, Johan H. and {Machado}, Rubens E.~G. and {Salo}, Heikki},
        title = "{Bar properties as a function of wavelength: a local baseline with S$^{4}$G for high-redshift studies}",
      journal = {\mnras},
         year = 2024,
        month = feb,
       volume = {527},
       number = {4},
        pages = {11777-11800},
          doi = {10.1093/mnras/stad3662},
archivePrefix = {arXiv},
       eprint = {2312.04545},
 primaryClass = {astro-ph.GA},
       adsurl = {https://ui.adsabs.harvard.edu/abs/2024MNRAS.52711777M}
}

@ARTICLE{Huang_2025,
       author = {{Huang}, Shuo and {Kawabe}, Ryohei and {Umehata}, Hideki and {Kohno}, Kotaro and {Tamura}, Yoichi and {Saito}, Toshiki},
        title = "{Large gas inflow driven by a matured galactic bar in the early Universe}",
      journal = {Nature},
         year = 2025,
        month = may,
       volume = {641},
        pages = {861-865},
          doi = {10.1038/s41586-025-08914-2},
       adsurl = {https://doi.org/10.1038/s41586-025-08914-2}
}

@ARTICLE{Freitas_2025,
       author = {{de S{\'a}-Freitas}, Camila and {Gadotti}, Dimitri A. and {Fragkoudi}, Francesca and {Coelho}, Paula and {de Lorenzo-C{\'a}ceres}, Adriana and {Falc{\'o}n-Barroso}, Jes{\'u}s and {S{\'a}nchez-Bl{\'a}zquez}, Patricia and {Kim}, Taehyun and {Mendez-Abreu}, Jairo and {Neumann}, Justus and {Querejeta}, Miguel and {van de Ven}, Glenn},
        title = "{Bar ages derived for the first time in nearby galaxies: Insights on secular evolution from the TIMER sample}",
      journal = {arXiv e-prints},
         year = 2025,
        month = mar,
          eid = {arXiv:2503.20864},
        pages = {arXiv:2503.20864},
          doi = {10.48550/arXiv.2503.20864},
archivePrefix = {arXiv},
       eprint = {2503.20864},
 primaryClass = {astro-ph.GA},
       adsurl = {https://ui.adsabs.harvard.edu/abs/2025arXiv250320864D}
}

@ARTICLE{Sheth_2012,
       author = {{Sheth}, Kartik and {Melbourne}, Jason and {Elmegreen}, Debra Meloy and {Elmegreen}, Bruce G. and {Athanassoula}, E. and {Abraham}, Roberto G. and {Weiner}, Benjamin J.},
        title = "{Hot Disks and Delayed Bar Formation}",
      journal = {\apj},
         year = 2012,
        month = oct,
       volume = {758},
       number = {2},
          eid = {136},
        pages = {136},
          doi = {10.1088/0004-637X/758/2/136},
archivePrefix = {arXiv},
       eprint = {1208.6304},
 primaryClass = {astro-ph.CO},
       adsurl = {https://ui.adsabs.harvard.edu/abs/2012ApJ...758..136S}
}

@ARTICLE{Pastras_2025,
       author = {{Pastras}, Stavros and {Genzel}, Reinhard and {Tacconi}, Linda J. and {Schuster}, Karl and {Neri}, Roberto and {F{\"o}rster Schreiber}, Natascha M. and {Naab}, Thorsten and {Barfety}, Capucine and {Burkert}, Andreas and {Cao}, Yixian and {Chen}, Jianhang and {Combes}, Fran{\c{c}}oise and {Davies}, Ric and {Eisenhauer}, Frank and {Espejo Salcedo}, Juan M. and {Garc{\'\i}a-Burillo}, Santiago and {Herrera-Camus}, Rodrigo and {Jolly}, Jean-Baptiste and {Lee}, Lilian L. and {Lee}, Minju M. and {Liu}, Daizhong and {Lutz}, Dieter and {Nestor Shachar}, Amit and {Parlanti}, Eleonora and {Price}, Sedona H. and {Pulsoni}, Claudia and {Renzini}, Alvio and {Scaloni}, Letizia and {Shimizu}, Taro T. and {Springel}, Volker and {Sternberg}, Amiel and {Sturm}, Eckhard and {Tozzi}, Giulia and {Wuyts}, Stijn and {{\"U}bler}, Hannah},
        title = "{NOEMA$^{\rm 3D}$: A first kpc resolution study of a $z\sim1.5$ main sequence barred galaxy channeling gas into a growing bulge}",
      journal = {arXiv e-prints},
         year = 2025,
        month = may,
          eid = {arXiv:2505.07925},
        pages = {arXiv:2505.07925},
          doi = {10.48550/arXiv.2505.07925},
archivePrefix = {arXiv},
       eprint = {2505.07925},
 primaryClass = {astro-ph.GA},
       adsurl = {https://ui.adsabs.harvard.edu/abs/2025arXiv250507925P}
}

@ARTICLE{Huang_2023,
       author = {{Huang}, Shuo and {Kawabe}, Ryohei and {Kohno}, Kotaro and {Saito}, Toshiki and {Mizukoshi}, Shoichiro and {Iono}, Daisuke and {Michiyama}, Tomonari and {Tamura}, Yoichi and {Hayward}, Christopher C. and {Umehata}, Hideki},
        title = "{J0107a: A Barred Spiral Dusty Star-forming Galaxy at z = 2.467}",
      journal = {\apjl},
         year = 2023,
        month = dec,
       volume = {958},
       number = {2},
          eid = {L26},
        pages = {L26},
          doi = {10.3847/2041-8213/acff63},
archivePrefix = {arXiv},
       eprint = {2310.01782},
 primaryClass = {astro-ph.GA},
       adsurl = {https://ui.adsabs.harvard.edu/abs/2023ApJ...958L..26H}
}

@ARTICLE{Salcedo_2025,
       author = {{Espejo Salcedo}, Juan M. and {Pastras}, Stavros and {V{\'a}cha}, Josef and {Pulsoni}, Claudia and {Genzel}, Reinhard and {F{\"o}rster Schreiber}, N.~M. and {Jolly}, Jean-Baptiste and {Barfety}, Capucine and {Chen}, Jianhang and {Tozzi}, Giulia and {Liu}, Daizhong and {Lee}, Lilian and {Wuyts}, Stijn and {Tacconi}, Linda and {Davies}, Ric and {{\"U}bler}, Hannah and {Lutz}, Dieter and {Wisnioski}, Emily and {Shangguan}, Jinyi and {Lee}, Minju and {Price}, Sedona H. and {Eisenhauer}, Frank and {Renzini}, Alvio and {Nestor Shachar}, Amit and {Herrera-Camus}, Rodrigo},
        title = "{Galaxy Morphologies at Cosmic Noon with JWST : A Foundation for Exploring Gas Transport with Bars and Spiral Arms}",
      journal = {arXiv e-prints},
         year = 2025,
        month = mar,
          eid = {arXiv:2503.21738},
        pages = {arXiv:2503.21738},
          doi = {10.48550/arXiv.2503.21738},
archivePrefix = {arXiv},
       eprint = {2503.21738},
 primaryClass = {astro-ph.GA},
       adsurl = {https://ui.adsabs.harvard.edu/abs/2025arXiv250321738E}
}

@misc{Géron_2025,
      title={Galaxy Zoo CEERS: Bar fractions up to z~4.0}, 
      author={Tobias Géron and R. J. Smethurst and Hugh Dickinson and L. F. Fortson and Izzy L. Garland and Sandor Kruk and Chris Lintott and Jason Shingirai Makechemu and Kameswara Bharadwaj Mantha and Karen L. Masters and David O'Ryan and Hayley Roberts and B. D. Simmons and Mike Walmsley and Antonello Calabrò and Rimpei Chiba and Luca Costantin and Maria R. Drout and Francesca Fragkoudi and Yuchen Guo and B. W. Holwerda and Shardha Jogee and Anton M. Koekemoer and Ray A. Lucas and Fabio Pacucci},
      year={2025},
      eprint={2505.01421},
      archivePrefix={arXiv},
      primaryClass={astro-ph.GA},
      url={https://arxiv.org/abs/2505.01421}, 
}

@ARTICLE{Fragkoudi_2025,
       author = {{Fragkoudi}, Francesca and {Grand}, Robert J.~J. and {Pakmor}, R{\"u}diger and {G{\'o}mez}, Facundo and {Marinacci}, Federico and {Springel}, Volker},
        title = "{Bar formation and evolution in the cosmological context: inputs from the Auriga simulations}",
      journal = {\mnras},
         year = 2025,
        month = apr,
       volume = {538},
       number = {3},
        pages = {1587-1608},
          doi = {10.1093/mnras/staf389},
archivePrefix = {arXiv},
       eprint = {2406.09453},
 primaryClass = {astro-ph.GA},
       adsurl = {https://ui.adsabs.harvard.edu/abs/2025MNRAS.538.1587F}
}

@article{storn_1997,
author = {Storn, Rainer and Price, Kenneth},
year = {1997},
month = {01},
pages = {341-359},
title = {Differential Evolution - A Simple and Efficient Heuristic for Global Optimization over Continuous Spaces},
volume = {11},
journal = {Journal of Global Optimization},
doi = {10.1023/A:1008202821328}
}

@ARTICLE{Tsukui_2024,
       author = {{Tsukui}, Takafumi and {Wisnioski}, Emily and {Bland-Hawthorn}, Joss and {Mai}, Yifan and {Iguchi}, Satoru and {Baba}, Junichi and {Freeman}, Ken},
        title = "{Detecting a disc bending wave in a barred-spiral galaxy at redshift 4.4}",
      journal = {\mnras},
         year = 2024,
        month = jan,
       volume = {527},
       number = {3},
        pages = {8941-8949},
          doi = {10.1093/mnras/stad3588},
archivePrefix = {arXiv},
       eprint = {2308.14798},
 primaryClass = {astro-ph.GA},
       adsurl = {https://ui.adsabs.harvard.edu/abs/2024MNRAS.527.8941T}
}

@ARTICLE{Amvrosiadis_2025,
       author = {{Amvrosiadis}, A. and {Lange}, S. and {Nightingale}, J.~W. and {He}, Q. and {Frenk}, C.~S. and {Oman}, K.~A. and {Smail}, I. and {Swinbank}, A.~M. and {Fragkoudi}, F. and {Gadotti}, D.~A. and {Cole}, S. and {Borsato}, E. and {Robertson}, A. and {Massey}, R. and {Cao}, X. and {Li}, R.},
        title = "{The onset of bar formation in a massive galaxy at z \raisebox{-0.5ex}\textasciitilde 3.8}",
      journal = {\mnras},
         year = 2025,
        month = feb,
       volume = {537},
       number = {2},
        pages = {1163-1181},
          doi = {10.1093/mnras/staf048},
archivePrefix = {arXiv},
       eprint = {2404.01918},
 primaryClass = {astro-ph.GA},
       adsurl = {https://ui.adsabs.harvard.edu/abs/2025MNRAS.537.1163A}
}

@ARTICLE{LeConte_2024,
       author = {{Le Conte}, Zoe A. and {Gadotti}, Dimitri A. and {Ferreira}, Leonardo and {Conselice}, Christopher J. and {de S{\'a}-Freitas}, Camila and {Kim}, Taehyun and {Neumann}, Justus and {Fragkoudi}, Francesca and {Athanassoula}, E. and {Adams}, Nathan J.},
        title = "{A JWST investigation into the bar fraction at redshifts 1 {\ensuremath{\leq}} z {\ensuremath{\leq}} 3}",
      journal = {\mnras},
         year = 2024,
        month = may,
       volume = {530},
       number = {2},
        pages = {1984-2000},
          doi = {10.1093/mnras/stae921},
archivePrefix = {arXiv},
       eprint = {2309.10038},
 primaryClass = {astro-ph.GA},
       adsurl = {https://ui.adsabs.harvard.edu/abs/2024MNRAS.530.1984L}
}

@ARTICLE{Guo_2025,
       author = {{Guo}, Yuchen and {Jogee}, Shardha and {Wise}, Eden and {Pritchett}, Keith and {McGrath}, Elizabeth J. and {Finkelstein}, Steven L. and {Iyer}, Kartheik G. and {Arrabal Haro}, Pablo and {Bagley}, Micaela B. and {Dickinson}, Mark and {Kartaltepe}, Jeyhan S. and {Koekemoer}, Anton M. and {Papovich}, Casey and {Pirzkal}, Nor and {Yung}, L.~Y. Aaron and {Backhaus}, Bren E. and {Bell}, Eric F. and {Bhatawdekar}, Rachana and {Cheng}, Yingjie and {Costantin}, Luca and {de la Vega}, Alexander and {Giavalisco}, Mauro and {Hathi}, Nimish P. and {Holwerda}, Benne W. and {Kurczynski}, Peter and {Lucas}, Ray A. and {Mobasher}, Bahram and {P{\'e}rez-Gonz{\'a}lez}, Pablo G. and {Pacucci}, Fabio},
        title = "{The Abundance and Properties of Barred Galaxies out to z {\ensuremath{\sim}} 4 Using JWST CEERS Data}",
      journal = {\apj},
         year = 2025,
        month = jun,
       volume = {985},
       number = {2},
          eid = {181},
        pages = {181},
          doi = {10.3847/1538-4357/adc8a7},
archivePrefix = {arXiv},
       eprint = {2409.06100},
 primaryClass = {astro-ph.GA},
       adsurl = {https://ui.adsabs.harvard.edu/abs/2025ApJ...985..181G}
}

@inproceedings{Randolph_2005,
  title={Free-Marginal Multirater Kappa (multirater K[free]): An Alternative to Fleiss' Fixed-Marginal Multirater Kappa.},
  author={Justus J. Randolph},
  year={2005},
  url={https://api.semanticscholar.org/CorpusID:59676845}
}

@ARTICLE{Ormerod_2024,
       author = {{Ormerod}, K. and {Conselice}, C.~J. and {Adams}, N.~J. and {Harvey}, T. and {Austin}, D. and {Trussler}, J. and {Ferreira}, L. and {Caruana}, J. and {Lucatelli}, G. and {Li}, Q. and {Roper}, W.~J.},
        title = "{EPOCHS VI: the size and shape evolution of galaxies since z   8 with JWST Observations}",
      journal = {\mnras},
         year = 2024,
        month = jan,
       volume = {527},
       number = {3},
        pages = {6110-6125},
          doi = {10.1093/mnras/stad3597},
archivePrefix = {arXiv},
       eprint = {2309.04377},
 primaryClass = {astro-ph.GA},
       adsurl = {https://ui.adsabs.harvard.edu/abs/2024MNRAS.527.6110O}
}

@ARTICLE{Erwin_2015,
       author = {{Erwin}, Peter},
        title = "{IMFIT: A Fast, Flexible New Program for Astronomical Image Fitting}",
      journal = {\apj},
         year = 2015,
        month = feb,
       volume = {799},
       number = {2},
          eid = {226},
        pages = {226},
          doi = {10.1088/0004-637X/799/2/226},
archivePrefix = {arXiv},
       eprint = {1408.1097},
 primaryClass = {astro-ph.IM},
       adsurl = {https://ui.adsabs.harvard.edu/abs/2015ApJ...799..226E}
}

@ARTICLE{Liang_2024,
       author = {{Liang}, Xinyue and {Yu}, Si-Yue and {Fang}, Taotao and {Ho}, Luis C.},
        title = "{The robustness in identifying and quantifying high-redshift bars using JWST observations}",
      journal = {\aap},
         year = 2024,
        month = aug,
       volume = {688},
          eid = {A158},
        pages = {A158},
          doi = {10.1051/0004-6361/202348539},
archivePrefix = {arXiv},
       eprint = {2311.04019},
 primaryClass = {astro-ph.GA},
       adsurl = {https://ui.adsabs.harvard.edu/abs/2024A&A...688A.158L}
}

@ARTICLE{Bertin_1996,
       author = {{Bertin}, E. and {Arnouts}, S.},
        title = "{SExtractor: Software for source extraction.}",
      journal = {\aaps},
         year = 1996,
        month = jun,
       volume = {117},
        pages = {393-404},
          doi = {10.1051/aas:1996164},
       adsurl = {https://ui.adsabs.harvard.edu/abs/1996A&AS..117..393B}
}

@ARTICLE{Eskridge_2000,
       author = {{Eskridge}, Paul B. and {Frogel}, Jay A. and {Pogge}, Richard W. and {Quillen}, Alice C. and {Davies}, Roger L. and {DePoy}, D.~L. and {Houdashelt}, Mark L. and {Kuchinski}, Leslie E. and {Ram{\'\i}rez}, Solange V. and {Sellgren}, K. and {Terndrup}, Donald M. and {Tiede}, Glenn P.},
        title = "{The Frequency of Barred Spiral Galaxies in the Near-Infrared}",
      journal = {\aj},
         year = 2000,
        month = feb,
       volume = {119},
       number = {2},
        pages = {536-544},
          doi = {10.1086/301203},
archivePrefix = {arXiv},
       eprint = {astro-ph/9910479},
 primaryClass = {astro-ph},
       adsurl = {https://ui.adsabs.harvard.edu/abs/2000AJ....119..536E}
}

@ARTICLE{Gadotti_2019,
       author = {{Gadotti}, Dimitri A. and {S{\'a}nchez-Bl{\'a}zquez}, Patricia and {Falc{\'o}n-Barroso}, Jes{\'u}s and {Husemann}, Bernd and {Seidel}, Marja K. and {P{\'e}rez}, Isabel and {de Lorenzo-C{\'a}ceres}, Adriana and {Martinez-Valpuesta}, Inma and {Fragkoudi}, Francesca and {Leung}, Gigi and {van de Ven}, Glenn and {Leaman}, Ryan and {Coelho}, Paula and {Martig}, Marie and {Kim}, Taehyun and {Neumann}, Justus and {Querejeta}, Miguel},
        title = "{Time Inference with MUSE in Extragalactic Rings (TIMER): properties of the survey and high-level data products}",
      journal = {\mnras},
         year = 2019,
        month = jan,
       volume = {482},
       number = {1},
        pages = {506-529},
          doi = {10.1093/mnras/sty2666},
archivePrefix = {arXiv},
       eprint = {1810.01425},
 primaryClass = {astro-ph.GA},
       adsurl = {https://ui.adsabs.harvard.edu/abs/2019MNRAS.482..506G}
}

@ARTICLE{Marinova_2007,
       author = {{Marinova}, Irina and {Jogee}, Shardha},
        title = "{Characterizing Bars at z \raisebox{-0.5ex}\textasciitilde 0 in the Optical and NIR: Implications for the Evolution of Barred Disks with Redshift}",
      journal = {\apj},
         year = 2007,
        month = apr,
       volume = {659},
       number = {2},
        pages = {1176-1197},
          doi = {10.1086/512355},
archivePrefix = {arXiv},
       eprint = {astro-ph/0608039},
 primaryClass = {astro-ph},
       adsurl = {https://ui.adsabs.harvard.edu/abs/2007ApJ...659.1176M}
}

@ARTICLE{Menendez_Delmestre_2007,
       author = {{Men{\'e}ndez-Delmestre}, Kar{\'\i}n and {Sheth}, Kartik and {Schinnerer}, Eva and {Jarrett}, Thomas H. and {Scoville}, Nick Z.},
        title = "{A Near-Infrared Study of 2MASS Bars in Local Galaxies: An Anchor for High-Redshift Studies}",
      journal = {\apj},
         year = 2007,
        month = mar,
       volume = {657},
       number = {2},
        pages = {790-804},
          doi = {10.1086/511025},
archivePrefix = {arXiv},
       eprint = {astro-ph/0611540},
 primaryClass = {astro-ph},
       adsurl = {https://ui.adsabs.harvard.edu/abs/2007ApJ...657..790M}
}

@ARTICLE{Erwin_2018,
       author = {{Erwin}, Peter},
        title = "{The dependence of bar frequency on galaxy mass, colour, and gas content - and angular resolution - in the local universe}",
      journal = {\mnras},
         year = 2018,
        month = mar,
       volume = {474},
       number = {4},
        pages = {5372-5392},
          doi = {10.1093/mnras/stx3117},
archivePrefix = {arXiv},
       eprint = {1711.04867},
 primaryClass = {astro-ph.GA},
       adsurl = {https://ui.adsabs.harvard.edu/abs/2018MNRAS.474.5372E}
}

@BOOK{deVaucouleurs_1991,
       author = {{de Vaucouleurs}, Gerard and {de Vaucouleurs}, Antoinette and {Corwin}, Herold G., Jr. and {Buta}, Ronald J. and {Paturel}, Georges and {Fouque}, Pascal},
        title = "{Third Reference Catalogue of Bright Galaxies}",
         year = 1991,
       adsurl = {https://ui.adsabs.harvard.edu/abs/1991rc3..book.....D}
}

@ARTICLE{Erwin_2005,
       author = {{Erwin}, Peter},
        title = "{How large are the bars in barred galaxies?}",
      journal = {\mnras},
         year = 2005,
        month = nov,
       volume = {364},
       number = {1},
        pages = {283-302},
          doi = {10.1111/j.1365-2966.2005.09560.x},
archivePrefix = {arXiv},
       eprint = {astro-ph/0508590},
 primaryClass = {astro-ph},
       adsurl = {https://ui.adsabs.harvard.edu/abs/2005MNRAS.364..283E}
}

@ARTICLE{Gadotti_2011,
       author = {{Gadotti}, Dimitri A.},
        title = "{Secular evolution and structural properties of stellar bars in galaxies}",
      journal = {\mnras},
         year = 2011,
        month = aug,
       volume = {415},
       number = {4},
        pages = {3308-3318},
          doi = {10.1111/j.1365-2966.2011.18945.x},
archivePrefix = {arXiv},
       eprint = {1003.1719},
 primaryClass = {astro-ph.CO},
       adsurl = {https://ui.adsabs.harvard.edu/abs/2011MNRAS.415.3308G}
}

@ARTICLE{Alonso_2018,
       author = {{Alonso}, Sol and {Coldwell}, Georgina and {Duplancic}, Fernanda and {Mesa}, Valeria and {Lambas}, Diego G.},
        title = "{The impact of bars and interactions on optically selected AGNs in spiral galaxies}",
      journal = {\aap},
         year = 2018,
        month = oct,
       volume = {618},
          eid = {A149},
        pages = {A149},
          doi = {10.1051/0004-6361/201832796},
archivePrefix = {arXiv},
       eprint = {1808.05536},
 primaryClass = {astro-ph.GA},
       adsurl = {https://ui.adsabs.harvard.edu/abs/2018A&A...618A.149A}
}

@ARTICLE{Garland_2023,
       author = {{Garland}, Izzy L. and {Fahey}, Matthew J. and {Simmons}, Brooke D. and {Smethurst}, Rebecca J. and {Lintott}, Chris J. and {Shanahan}, Jesse and {Silcock}, Maddie S. and {Smith}, Joshua and {Keel}, William C. and {Coil}, Alison and {G{\'e}ron}, Tobias and {Kruk}, Sandor and {Masters}, Karen L. and {O'Ryan}, David and {Thorne}, Matthew R. and {Wiersema}, Klaas},
        title = "{The most luminous, merger-free AGNs show only marginal correlation with bar presence}",
      journal = {\mnras},
         year = 2023,
        month = jun,
       volume = {522},
       number = {1},
        pages = {211-225},
          doi = {10.1093/mnras/stad966},
archivePrefix = {arXiv},
       eprint = {2304.01260},
 primaryClass = {astro-ph.GA},
       adsurl = {https://ui.adsabs.harvard.edu/abs/2023MNRAS.522..211G}
}

@ARTICLE{Sheth_2008,
       author = {{Sheth}, Kartik and {Elmegreen}, Debra Meloy and {Elmegreen}, Bruce G. and {Capak}, Peter and {Abraham}, Roberto G. and {Athanassoula}, E. and {Ellis}, Richard S. and {Mobasher}, Bahram and {Salvato}, Mara and {Schinnerer}, Eva and {Scoville}, Nicholas Z. and {Spalsbury}, Lori and {Strubbe}, Linda and {Carollo}, Marcella and {Rich}, Michael and {West}, Andrew A.},
        title = "{Evolution of the Bar Fraction in COSMOS: Quantifying the Assembly of the Hubble Sequence}",
      journal = {\apj},
         year = 2008,
        month = mar,
       volume = {675},
       number = {2},
        pages = {1141-1155},
          doi = {10.1086/524980},
archivePrefix = {arXiv},
       eprint = {0710.4552},
 primaryClass = {astro-ph},
       adsurl = {https://ui.adsabs.harvard.edu/abs/2008ApJ...675.1141S}
}

@ARTICLE{Fragkoudi_2018,
       author = {{Fragkoudi}, F. and {Di Matteo}, P. and {Haywood}, M. and {Schultheis}, M. and {Khoperskov}, S. and {G{\'o}mez}, A. and {Combes}, F.},
        title = "{The disc origin of the Milky Way bulge. Dissecting the chemo-morphological relations using N-body simulations and APOGEE}",
      journal = {\aap},
         year = 2018,
        month = sep,
       volume = {616},
          eid = {A180},
        pages = {A180},
          doi = {10.1051/0004-6361/201732509},
archivePrefix = {arXiv},
       eprint = {1802.00453},
 primaryClass = {astro-ph.GA},
       adsurl = {https://ui.adsabs.harvard.edu/abs/2018A&A...616A.180F}
}

@ARTICLE{Bittner_2020,
       author = {{Bittner}, Adrian and {S{\'a}nchez-Bl{\'a}zquez}, Patricia and {Gadotti}, Dimitri A. and {Neumann}, Justus and {Fragkoudi}, Francesca and {Coelho}, Paula and {de Lorenzo-C{\'a}ceres}, Adriana and {Falc{\'o}n-Barroso}, Jes{\'u}s and {Kim}, Taehyun and {Leaman}, Ryan and {Mart{\'\i}n-Navarro}, Ignacio and {M{\'e}ndez-Abreu}, Jairo and {P{\'e}rez}, Isabel and {Querejeta}, Miguel and {Seidel}, Marja K. and {van de Ven}, Glenn},
        title = "{Inside-out formation of nuclear discs and the absence of old central spheroids in barred galaxies of the TIMER survey}",
      journal = {\aap},
         year = 2020,
        month = nov,
       volume = {643},
          eid = {A65},
        pages = {A65},
          doi = {10.1051/0004-6361/202038450},
archivePrefix = {arXiv},
       eprint = {2009.01856},
 primaryClass = {astro-ph.GA},
       adsurl = {https://ui.adsabs.harvard.edu/abs/2020A&A...643A..65B}
}

@ARTICLE{Kim_2021,
       author = {{Kim}, Taehyun and {Athanassoula}, E. and {Sheth}, Kartik and {Bosma}, Albert and {Park}, Myeong-Gu and {Lee}, Yun Hee and {Ann}, Hong Bae},
        title = "{Cosmic Evolution of Barred Galaxies up to z   0.84}",
      journal = {\apj},
         year = 2021,
        month = dec,
       volume = {922},
       number = {2},
          eid = {196},
        pages = {196},
          doi = {10.3847/1538-4357/ac2300},
archivePrefix = {arXiv},
       eprint = {2109.03420},
 primaryClass = {astro-ph.GA},
       adsurl = {https://ui.adsabs.harvard.edu/abs/2021ApJ...922..196K}
}

@ARTICLE{Nair_2010b,
       author = {{Nair}, Preethi B. and {Abraham}, Roberto G.},
        title = "{On the Fraction of Barred Spiral Galaxies}",
      journal = {\apjl},
         year = 2010,
        month = may,
       volume = {714},
       number = {2},
        pages = {L260-L264},
          doi = {10.1088/2041-8205/714/2/L260},
archivePrefix = {arXiv},
       eprint = {1004.0684},
 primaryClass = {astro-ph.CO},
       adsurl = {https://ui.adsabs.harvard.edu/abs/2010ApJ...714L.260N}
}

@ARTICLE{Erwin_2019,
       author = {{Erwin}, Peter},
        title = "{What determines the sizes of bars in spiral galaxies?}",
      journal = {\mnras},
         year = 2019,
        month = nov,
       volume = {489},
       number = {3},
        pages = {3553-3564},
          doi = {10.1093/mnras/stz2363},
archivePrefix = {arXiv},
       eprint = {1908.08423},
 primaryClass = {astro-ph.GA},
       adsurl = {https://ui.adsabs.harvard.edu/abs/2019MNRAS.489.3553E}
}

@ARTICLE{Aguerri_2009,
       author = {{Aguerri}, J.~A.~L. and {M{\'e}ndez-Abreu}, J. and {Corsini}, E.~M.},
        title = "{The population of barred galaxies in the local universe. I. Detection and characterisation of bars}",
      journal = {\aap},
         year = 2009,
        month = feb,
       volume = {495},
       number = {2},
        pages = {491-504},
          doi = {10.1051/0004-6361:200810931},
archivePrefix = {arXiv},
       eprint = {0901.2346},
 primaryClass = {astro-ph.GA},
       adsurl = {https://ui.adsabs.harvard.edu/abs/2009A&A...495..491A}
}

@ARTICLE{Hohl_1971,
       author = {{Hohl}, Frank},
        title = "{Numerical Experiments with a Disk of Stars}",
      journal = {\apj},
         year = 1971,
        month = sep,
       volume = {168},
        pages = {343},
          doi = {10.1086/151091},
       adsurl = {https://ui.adsabs.harvard.edu/abs/1971ApJ...168..343H}
}

@ARTICLE{Kalnajs_1972,
       author = {{Kalnajs}, Agris J.},
        title = "{The Equilibria and Oscillations of a Family of Uniformly Rotating Stellar Disks}",
      journal = {\apj},
         year = 1972,
        month = jul,
       volume = {175},
        pages = {63},
          doi = {10.1086/151538},
       adsurl = {https://ui.adsabs.harvard.edu/abs/1972ApJ...175...63K}
}

@ARTICLE{Ostriker_1973,
       author = {{Ostriker}, J.~P. and {Peebles}, P.~J.~E.},
        title = "{A Numerical Study of the Stability of Flattened Galaxies: or, can Cold Galaxies Survive?}",
      journal = {\apj},
         year = 1973,
        month = dec,
       volume = {186},
        pages = {467-480},
          doi = {10.1086/152513},
       adsurl = {https://ui.adsabs.harvard.edu/abs/1973ApJ...186..467O}
}

@ARTICLE{Sellwood_1993,
       author = {{Sellwood}, J.~A. and {Wilkinson}, A.},
        title = "{Dynamics of barred galaxies}",
      journal = {Reports on Progress in Physics},
         year = 1993,
        month = feb,
       volume = {56},
       number = {2},
        pages = {173-256},
          doi = {10.1088/0034-4885/56/2/001},
archivePrefix = {arXiv},
       eprint = {astro-ph/0608665},
 primaryClass = {astro-ph},
       adsurl = {https://ui.adsabs.harvard.edu/abs/1993RPPh...56..173S}
}

@ARTICLE{Lynden_1972,
       author = {{Lynden-Bell}, D. and {Kalnajs}, A.~J.},
        title = "{On the generating mechanism of spiral structure}",
      journal = {\mnras},
         year = 1972,
        month = jan,
       volume = {157},
        pages = {1},
          doi = {10.1093/mnras/157.1.1},
       adsurl = {https://ui.adsabs.harvard.edu/abs/1972MNRAS.157....1L}
}

@ARTICLE{Regan_2006,
       author = {{Regan}, Michael W. and {Thornley}, Michele D. and {Vogel}, Stuart N. and {Sheth}, Kartik and {Draine}, Bruce T. and {Hollenbach}, David J. and {Meyer}, Martin and {Dale}, Daniel A. and {Engelbracht}, Charles W. and {Kennicutt}, Robert C. and {Armus}, Lee and {Buckalew}, Brent and {Calzetti}, Daniela and {Gordon}, Karl D. and {Helou}, George and {Leitherer}, Claus and {Malhotra}, Sangeeta and {Murphy}, Eric and {Rieke}, George H. and {Rieke}, Marcia J. and {Smith}, J.~D.},
        title = "{The Radial Distribution of the Interstellar Medium in Disk Galaxies: Evidence for Secular Evolution}",
      journal = {\apj},
         year = 2006,
        month = dec,
       volume = {652},
       number = {2},
        pages = {1112-1121},
          doi = {10.1086/505382},
       adsurl = {https://ui.adsabs.harvard.edu/abs/2006ApJ...652.1112R}
}

@ARTICLE{Matteo_2013,
       author = {{Di Matteo}, P. and {Haywood}, M. and {Combes}, F. and {Semelin}, B. and {Snaith}, O.~N.},
        title = "{Signatures of radial migration in barred galaxies: Azimuthal variations in the metallicity distribution of old stars}",
      journal = {\aap},
         year = 2013,
        month = may,
       volume = {553},
          eid = {A102},
        pages = {A102},
          doi = {10.1051/0004-6361/201220539},
archivePrefix = {arXiv},
       eprint = {1301.2545},
 primaryClass = {astro-ph.GA},
       adsurl = {https://ui.adsabs.harvard.edu/abs/2013A&A...553A.102D}
}

@ARTICLE{Coelho_2011,
       author = {{Coelho}, P. and {Gadotti}, D.~A.},
        title = "{Bars Rejuvenating Bulges? Evidence from Stellar Population Analysis}",
      journal = {\apjl},
         year = 2011,
        month = dec,
       volume = {743},
       number = {1},
          eid = {L13},
        pages = {L13},
          doi = {10.1088/2041-8205/743/1/L13},
archivePrefix = {arXiv},
       eprint = {1111.1736},
 primaryClass = {astro-ph.CO},
       adsurl = {https://ui.adsabs.harvard.edu/abs/2011ApJ...743L..13C}
}

@ARTICLE{Alonso_2013,
       author = {{Alonso}, M.~S. and {Coldwell}, G. and {Lambas}, D.~G.},
        title = "{Effect of bars in AGN host galaxies and black hole activity}",
      journal = {\aap},
         year = 2013,
        month = jan,
       volume = {549},
          eid = {A141},
        pages = {A141},
          doi = {10.1051/0004-6361/201220117},
archivePrefix = {arXiv},
       eprint = {1211.5156},
 primaryClass = {astro-ph.CO},
       adsurl = {https://ui.adsabs.harvard.edu/abs/2013A&A...549A.141A}
}

@ARTICLE{Cisternas_2015,
       author = {{Cisternas}, Mauricio and {Sheth}, Kartik and {Salvato}, Mara and {Knapen}, Johan H. and {Civano}, Francesca and {Santini}, Paola},
        title = "{The Role of Bars in AGN Fueling in Disk Galaxies Over the Last Seven Billion Years}",
      journal = {\apj},
         year = 2015,
        month = apr,
       volume = {802},
       number = {2},
          eid = {137},
        pages = {137},
          doi = {10.1088/0004-637X/802/2/137},
archivePrefix = {arXiv},
       eprint = {1409.2871},
 primaryClass = {astro-ph.GA},
       adsurl = {https://ui.adsabs.harvard.edu/abs/2015ApJ...802..137C}
}

@ARTICLE{Abraham_1999,
       author = {{Abraham}, R.~G. and {Merrifield}, M.~R. and {Ellis}, R.~S. and {Tanvir}, N.~R. and {Brinchmann}, J.},
        title = "{The evolution of barred spiral galaxies in the Hubble Deep Fields North and South}",
      journal = {\mnras},
         year = 1999,
        month = sep,
       volume = {308},
       number = {2},
        pages = {569-576},
          doi = {10.1046/j.1365-8711.1999.02766.x},
archivePrefix = {arXiv},
       eprint = {astro-ph/9811476},
 primaryClass = {astro-ph},
       adsurl = {https://ui.adsabs.harvard.edu/abs/1999MNRAS.308..569A}
}

@ARTICLE{Melvin_2014,
       author = {{Melvin}, Thomas and {Masters}, Karen and {Lintott}, Chris and {Nichol}, Robert C. and {Simmons}, Brooke and {Bamford}, Steven P. and {Casteels}, Kevin R.~V. and {Cheung}, Edmond and {Edmondson}, Edward M. and {Fortson}, Lucy and {Schawinski}, Kevin and {Skibba}, Ramin A. and {Smith}, Arfon M. and {Willett}, Kyle W.},
        title = "{Galaxy Zoo: an independent look at the evolution of the bar fraction over the last eight billion years from HST-COSMOS}",
      journal = {\mnras},
         year = 2014,
        month = mar,
       volume = {438},
       number = {4},
        pages = {2882-2897},
          doi = {10.1093/mnras/stt2397},
archivePrefix = {arXiv},
       eprint = {1401.3334},
 primaryClass = {astro-ph.GA},
       adsurl = {https://ui.adsabs.harvard.edu/abs/2014MNRAS.438.2882M}
}

@ARTICLE{Athanassoula_2005,
       author = {{Athanassoula}, E.},
        title = "{Dynamical Evolution of Barred Galaxies}",
      journal = {Celestial Mechanics and Dynamical Astronomy},
         year = 2005,
        month = jan,
       volume = {91},
       number = {1-2},
        pages = {9-31},
          doi = {10.1007/s10569-004-4947-7},
archivePrefix = {arXiv},
       eprint = {astro-ph/0501196},
 primaryClass = {astro-ph},
       adsurl = {https://ui.adsabs.harvard.edu/abs/2005CeMDA..91....9A}
}

@ARTICLE{Knapen_1995,
       author = {{Knapen}, J.~H. and {Beckman}, J.~E. and {Heller}, C.~H. and {Shlosman}, I. and {de Jong}, R.~S.},
        title = "{The Central Region in M100: Observations and Modeling}",
      journal = {\apj},
         year = 1995,
        month = dec,
       volume = {454},
        pages = {623},
          doi = {10.1086/176516},
archivePrefix = {arXiv},
       eprint = {astro-ph/9506098},
 primaryClass = {astro-ph},
       adsurl = {https://ui.adsabs.harvard.edu/abs/1995ApJ...454..623K}
}

@ARTICLE{Cheung_2013,
       author = {{Cheung}, Edmond and {Athanassoula}, E. and {Masters}, Karen L. and {Nichol}, Robert C. and {Bosma}, A. and {Bell}, Eric F. and {Faber}, S.~M. and {Koo}, David C. and {Lintott}, Chris and {Melvin}, Thomas and {Schawinski}, Kevin and {Skibba}, Ramin A. and {Willett}, Kyle W.},
        title = "{Galaxy Zoo: Observing Secular Evolution through Bars}",
      journal = {\apj},
         year = 2013,
        month = dec,
       volume = {779},
       number = {2},
          eid = {162},
        pages = {162},
          doi = {10.1088/0004-637X/779/2/162},
archivePrefix = {arXiv},
       eprint = {1310.2941},
 primaryClass = {astro-ph.CO},
       adsurl = {https://ui.adsabs.harvard.edu/abs/2013ApJ...779..162C}
}

@ARTICLE{Elmegreen_2004,
       author = {{Elmegreen}, Bruce G. and {Elmegreen}, Debra Meloy and {Hirst}, Amelia C.},
        title = "{A Constant Bar Fraction out to Redshift z \raisebox{-0.5ex}\textasciitilde 1 in the Advanced Camera for Surveys Field of the Tadpole Galaxy}",
      journal = {\apj},
         year = 2004,
        month = sep,
       volume = {612},
       number = {1},
        pages = {191-201},
          doi = {10.1086/422407},
archivePrefix = {arXiv},
       eprint = {astro-ph/0407577},
 primaryClass = {astro-ph},
       adsurl = {https://ui.adsabs.harvard.edu/abs/2004ApJ...612..191E}
}

@ARTICLE{Wozniak_1995,
       author = {{Wozniak}, H. and {Friedli}, D. and {Martinet}, L. and {Martin}, P. and {Bratschi}, P.},
        title = "{Disc galaxies with multiple triaxial structures. I. BVRI and H{\ensuremath{\alpha}} surface photometry.}",
      journal = {\aaps},
         year = 1995,
        month = may,
       volume = {111},
        pages = {115},
       adsurl = {https://ui.adsabs.harvard.edu/abs/1995A&AS..111..115W}
}

@ARTICLE{Jogee_2004,
       author = {{Jogee}, Shardha and {Barazza}, Fabio D. and {Rix}, Hans-Walter and {Shlosman}, Isaac and {Barden}, Marco and {Wolf}, Christian and {Davies}, James and {Heyer}, Inge and {Beckwith}, Steven V.~W. and {Bell}, Eric F. and {Borch}, Andrea and {Caldwell}, John A.~R. and {Conselice}, Christopher J. and {Dahlen}, Tomas and {H{\"a}ussler}, Boris and {Heymans}, Catherine and {Jahnke}, Knud and {Knapen}, Johan H. and {Laine}, Seppo and {Lubell}, Gabriel M. and {Mobasher}, Bahram and {McIntosh}, Daniel H. and {Meisenheimer}, Klaus and {Peng}, Chien Y. and {Ravindranath}, Swara and {Sanchez}, Sebastian F. and {Somerville}, Rachel S. and {Wisotzki}, Lutz},
        title = "{Bar Evolution over the Last 8 Billion Years: A Constant Fraction of Strong Bars in the GEMS Survey}",
      journal = {\apjl},
         year = 2004,
        month = nov,
       volume = {615},
       number = {2},
        pages = {L105-L108},
          doi = {10.1086/426138},
archivePrefix = {arXiv},
       eprint = {astro-ph/0408382},
 primaryClass = {astro-ph},
       adsurl = {https://ui.adsabs.harvard.edu/abs/2004ApJ...615L.105J}
}

@ARTICLE{Simmons_2014,
       author = {{Simmons}, B.~D. and {Melvin}, Thomas and {Lintott}, Chris and {Masters}, Karen L. and {Willett}, Kyle W. and {Keel}, William C. and {Smethurst}, R.~J. and {Cheung}, Edmond and {Nichol}, Robert C. and {Schawinski}, Kevin and {Rutkowski}, Michael and {Kartaltepe}, Jeyhan S. and {Bell}, Eric F. and {Casteels}, Kevin R.~V. and {Conselice}, Christopher J. and {Almaini}, Omar and {Ferguson}, Henry C. and {Fortson}, Lucy and {Hartley}, William and {Kocevski}, Dale and {Koekemoer}, Anton M. and {McIntosh}, Daniel H. and {Mortlock}, Alice and {Newman}, Jeffrey A. and {Ownsworth}, Jamie and {Bamford}, Steven and {Dahlen}, Tomas and {Faber}, Sandra M. and {Finkelstein}, Steven L. and {Fontana}, Adriano and {Galametz}, Audrey and {Grogin}, N.~A. and {Gr{\"u}tzbauch}, Ruth and {Guo}, Yicheng and {H{\"a}u{\ss}ler}, Boris and {Jek}, Kian J. and {Kaviraj}, Sugata and {Lucas}, Ray A. and {Peth}, Michael and {Salvato}, Mara and {Wiklind}, Tommy and {Wuyts}, Stijn},
        title = "{Galaxy Zoo: CANDELS barred discs and bar fractions}",
      journal = {\mnras},
         year = 2014,
        month = dec,
       volume = {445},
       number = {4},
        pages = {3466-3474},
          doi = {10.1093/mnras/stu1817},
archivePrefix = {arXiv},
       eprint = {1409.1214},
 primaryClass = {astro-ph.GA},
       adsurl = {https://ui.adsabs.harvard.edu/abs/2014MNRAS.445.3466S}
}

@ARTICLE{Buta_2015,
       author = {{Buta}, Ronald J. and {Sheth}, Kartik and {Athanassoula}, E. and {Bosma}, A. and {Knapen}, Johan H. and {Laurikainen}, Eija and {Salo}, Heikki and {Elmegreen}, Debra and {Ho}, Luis C. and {Zaritsky}, Dennis and {Courtois}, Helene and {Hinz}, Joannah L. and {Mu{\~n}oz-Mateos}, Juan-Carlos and {Kim}, Taehyun and {Regan}, Michael W. and {Gadotti}, Dimitri A. and {Gil de Paz}, Armando and {Laine}, Jarkko and {Men{\'e}ndez-Delmestre}, Kar{\'\i}n and {Comer{\'o}n}, S{\'e}bastien and {Erroz Ferrer}, Santiago and {Seibert}, Mark and {Mizusawa}, Trisha and {Holwerda}, Benne and {Madore}, Barry F.},
        title = "{A Classical Morphological Analysis of Galaxies in the Spitzer Survey of Stellar Structure in Galaxies (S4G)}",
      journal = {\apjs},
         year = 2015,
        month = apr,
       volume = {217},
       number = {2},
          eid = {32},
        pages = {32},
          doi = {10.1088/0067-0049/217/2/32},
archivePrefix = {arXiv},
       eprint = {1501.00454},
 primaryClass = {astro-ph.GA},
       adsurl = {https://ui.adsabs.harvard.edu/abs/2015ApJS..217...32B}
}

@ARTICLE{Guo_2023,
       author = {{Guo}, Yuchen and {Jogee}, Shardha and {Finkelstein}, Steven L. and {Chen}, Zilei and {Wise}, Eden and {Bagley}, Micaela B. and {Barro}, Guillermo and {Wuyts}, Stijn and {Kocevski}, Dale D. and {Kartaltepe}, Jeyhan S. and {McGrath}, Elizabeth J. and {Ferguson}, Henry C. and {Mobasher}, Bahram and {Giavalisco}, Mauro and {Lucas}, Ray A. and {Zavala}, Jorge A. and {Lotz}, Jennifer M. and {Grogin}, Norman A. and {Huertas-Company}, Marc and {Vega-Ferrero}, Jes{\'u}s and {Hathi}, Nimish P. and {Haro}, Pablo Arrabal and {Dickinson}, Mark and {Koekemoer}, Anton M. and {Papovich}, Casey and {Pirzkal}, Nor and {Yung}, L.~Y. Aaron and {Backhaus}, Bren E. and {Bell}, Eric F. and {Calabr{\`o}}, Antonello and {Cleri}, Nikko J. and {Coogan}, Rosemary T. and {Cooper}, M.~C. and {Costantin}, Luca and {Croton}, Darren and {Davis}, Kelcey and {Dekel}, Avishai and {Franco}, Maximilien and {Gardner}, Jonathan P. and {Holwerda}, Benne W. and {Hutchison}, Taylor A. and {Pandya}, Viraj and {P{\'e}rez-Gonz{\'a}lez}, Pablo G. and {Ravindranath}, Swara and {Rose}, Caitlin and {Trump}, Jonathan R. and {de la Vega}, Alexander and {Wang}, Weichen},
        title = "{First Look at z > 1 Bars in the Rest-frame Near-infrared with JWST Early CEERS Imaging}",
      journal = {\apjl},
         year = 2023,
        month = mar,
       volume = {945},
       number = {1},
          eid = {L10},
        pages = {L10},
          doi = {10.3847/2041-8213/acacfb},
archivePrefix = {arXiv},
       eprint = {2210.08658},
 primaryClass = {astro-ph.GA},
       adsurl = {https://ui.adsabs.harvard.edu/abs/2023ApJ...945L..10G}
}

@ARTICLE{Masters_2011,
       author = {{Masters}, Karen L. and {Nichol}, Robert C. and {Hoyle}, Ben and {Lintott}, Chris and {Bamford}, Steven P. and {Edmondson}, Edward M. and {Fortson}, Lucy and {Keel}, William C. and {Schawinski}, Kevin and {Smith}, Arfon M. and {Thomas}, Daniel},
        title = "{Galaxy Zoo: bars in disc galaxies}",
      journal = {\mnras},
         year = 2011,
        month = mar,
       volume = {411},
       number = {3},
        pages = {2026-2034},
          doi = {10.1111/j.1365-2966.2010.17834.x},
archivePrefix = {arXiv},
       eprint = {1003.0449},
 primaryClass = {astro-ph.CO},
       adsurl = {https://ui.adsabs.harvard.edu/abs/2011MNRAS.411.2026M}
}

@BOOK{Vaucouleurs_1991,
       author = {{de Vaucouleurs}, Gerard and {de Vaucouleurs}, Antoinette and {Corwin}, Herold G., Jr. and {Buta}, Ronald J. and {Paturel}, Georges and {Fouque}, Pascal},
        title = "{Third Reference Catalogue of Bright Galaxies}",
    publisher = "{Springer-Verlag}",
         year = 1991,
       adsurl = {https://ui.adsabs.harvard.edu/abs/1991rc3..book.....D}
}

@ARTICLE{Cameron_2011,
       author = {{Cameron}, Ewan},
        title = "{On the Estimation of Confidence Intervals for Binomial Population Proportions in Astronomy: The Simplicity and Superiority of the Bayesian Approach}",
      journal = {\pasa},
         year = 2011,
        month = jun,
       volume = {28},
       number = {2},
        pages = {128-139},
          doi = {10.1071/AS10046},
archivePrefix = {arXiv},
       eprint = {1012.0566},
 primaryClass = {astro-ph.IM},
       adsurl = {https://ui.adsabs.harvard.edu/abs/2011PASA...28..128C}
}

@ARTICLE{Duncan_2019,
       author = {{Duncan}, Kenneth and {Conselice}, Christopher J. and {Mundy}, Carl and {Bell}, Eric and {Donley}, Jennifer and {Galametz}, Audrey and {Guo}, Yicheng and {Grogin}, Norman A. and {Hathi}, Nimish and {Kartaltepe}, Jeyhan and {Kocevski}, Dale and {Koekemoer}, Anton M. and {P{\'e}rez-Gonz{\'a}lez}, Pablo G. and {Mantha}, Kameswara B. and {Snyder}, Gregory F. and {Stefanon}, Mauro},
        title = "{Observational Constraints on the Merger History of Galaxies since z {\ensuremath{\approx}} 6: Probabilistic Galaxy Pair Counts in the CANDELS Fields}",
      journal = {\apj},
         year = 2019,
        month = may,
       volume = {876},
       number = {2},
          eid = {110},
        pages = {110},
          doi = {10.3847/1538-4357/ab148a},
archivePrefix = {arXiv},
       eprint = {1903.12188},
 primaryClass = {astro-ph.GA},
       adsurl = {https://ui.adsabs.harvard.edu/abs/2019ApJ...876..110D}
}

@ARTICLE{Grogin_2011,
       author = {{Grogin}, Norman A. and {Kocevski}, Dale D. and {Faber}, S.~M. and {Ferguson}, Henry C. and {Koekemoer}, Anton M. and {Riess}, Adam G. and {Acquaviva}, Viviana and {Alexander}, David M. and {Almaini}, Omar and {Ashby}, Matthew L.~N. and {Barden}, Marco and {Bell}, Eric F. and {Bournaud}, Fr{\'e}d{\'e}ric and {Brown}, Thomas M. and {Caputi}, Karina I. and {Casertano}, Stefano and {Cassata}, Paolo and {Castellano}, Marco and {Challis}, Peter and {Chary}, Ranga-Ram and {Cheung}, Edmond and {Cirasuolo}, Michele and {Conselice}, Christopher J. and {Roshan Cooray}, Asantha and {Croton}, Darren J. and {Daddi}, Emanuele and {Dahlen}, Tomas and {Dav{\'e}}, Romeel and {de Mello}, Du{\'\i}lia F. and {Dekel}, Avishai and {Dickinson}, Mark and {Dolch}, Timothy and {Donley}, Jennifer L. and {Dunlop}, James S. and {Dutton}, Aaron A. and {Elbaz}, David and {Fazio}, Giovanni G. and {Filippenko}, Alexei V. and {Finkelstein}, Steven L. and {Fontana}, Adriano and {Gardner}, Jonathan P. and {Garnavich}, Peter M. and {Gawiser}, Eric and {Giavalisco}, Mauro and {Grazian}, Andrea and {Guo}, Yicheng and {Hathi}, Nimish P. and {H{\"a}ussler}, Boris and {Hopkins}, Philip F. and {Huang}, Jia-Sheng and {Huang}, Kuang-Han and {Jha}, Saurabh W. and {Kartaltepe}, Jeyhan S. and {Kirshner}, Robert P. and {Koo}, David C. and {Lai}, Kamson and {Lee}, Kyoung-Soo and {Li}, Weidong and {Lotz}, Jennifer M. and {Lucas}, Ray A. and {Madau}, Piero and {McCarthy}, Patrick J. and {McGrath}, Elizabeth J. and {McIntosh}, Daniel H. and {McLure}, Ross J. and {Mobasher}, Bahram and {Moustakas}, Leonidas A. and {Mozena}, Mark and {Nandra}, Kirpal and {Newman}, Jeffrey A. and {Niemi}, Sami-Matias and {Noeske}, Kai G. and {Papovich}, Casey J. and {Pentericci}, Laura and {Pope}, Alexandra and {Primack}, Joel R. and {Rajan}, Abhijith and {Ravindranath}, Swara and {Reddy}, Naveen A. and {Renzini}, Alvio and {Rix}, Hans-Walter and {Robaina}, Aday R. and {Rodney}, Steven A. and {Rosario}, David J. and {Rosati}, Piero and {Salimbeni}, Sara and {Scarlata}, Claudia and {Siana}, Brian and {Simard}, Luc and {Smidt}, Joseph and {Somerville}, Rachel S. and {Spinrad}, Hyron and {Straughn}, Amber N. and {Strolger}, Louis-Gregory and {Telford}, Olivia and {Teplitz}, Harry I. and {Trump}, Jonathan R. and {van der Wel}, Arjen and {Villforth}, Carolin and {Wechsler}, Risa H. and {Weiner}, Benjamin J. and {Wiklind}, Tommy and {Wild}, Vivienne and {Wilson}, Grant and {Wuyts}, Stijn and {Yan}, Hao-Jing and {Yun}, Min S.},
        title = "{CANDELS: The Cosmic Assembly Near-infrared Deep Extragalactic Legacy Survey}",
      journal = {\apjs},
         year = 2011,
        month = dec,
       volume = {197},
       number = {2},
          eid = {35},
        pages = {35},
          doi = {10.1088/0067-0049/197/2/35},
archivePrefix = {arXiv},
       eprint = {1105.3753},
 primaryClass = {astro-ph.CO},
       adsurl = {https://ui.adsabs.harvard.edu/abs/2011ApJS..197...35G}
}

@ARTICLE{Koekemoer_2011,
       author = {{Koekemoer}, Anton M. and {Faber}, S.~M. and {Ferguson}, Henry C. and {Grogin}, Norman A. and {Kocevski}, Dale D. and {Koo}, David C. and {Lai}, Kamson and {Lotz}, Jennifer M. and {Lucas}, Ray A. and {McGrath}, Elizabeth J. and {Ogaz}, Sara and {Rajan}, Abhijith and {Riess}, Adam G. and {Rodney}, Steve A. and {Strolger}, Louis and {Casertano}, Stefano and {Castellano}, Marco and {Dahlen}, Tomas and {Dickinson}, Mark and {Dolch}, Timothy and {Fontana}, Adriano and {Giavalisco}, Mauro and {Grazian}, Andrea and {Guo}, Yicheng and {Hathi}, Nimish P. and {Huang}, Kuang-Han and {van der Wel}, Arjen and {Yan}, Hao-Jing and {Acquaviva}, Viviana and {Alexander}, David M. and {Almaini}, Omar and {Ashby}, Matthew L.~N. and {Barden}, Marco and {Bell}, Eric F. and {Bournaud}, Fr{\'e}d{\'e}ric and {Brown}, Thomas M. and {Caputi}, Karina I. and {Cassata}, Paolo and {Challis}, Peter J. and {Chary}, Ranga-Ram and {Cheung}, Edmond and {Cirasuolo}, Michele and {Conselice}, Christopher J. and {Roshan Cooray}, Asantha and {Croton}, Darren J. and {Daddi}, Emanuele and {Dav{\'e}}, Romeel and {de Mello}, Duilia F. and {de Ravel}, Loic and {Dekel}, Avishai and {Donley}, Jennifer L. and {Dunlop}, James S. and {Dutton}, Aaron A. and {Elbaz}, David and {Fazio}, Giovanni G. and {Filippenko}, Alexei V. and {Finkelstein}, Steven L. and {Frazer}, Chris and {Gardner}, Jonathan P. and {Garnavich}, Peter M. and {Gawiser}, Eric and {Gruetzbauch}, Ruth and {Hartley}, Will G. and {H{\"a}ussler}, Boris and {Herrington}, Jessica and {Hopkins}, Philip F. and {Huang}, Jia-Sheng and {Jha}, Saurabh W. and {Johnson}, Andrew and {Kartaltepe}, Jeyhan S. and {Khostovan}, Ali A. and {Kirshner}, Robert P. and {Lani}, Caterina and {Lee}, Kyoung-Soo and {Li}, Weidong and {Madau}, Piero and {McCarthy}, Patrick J. and {McIntosh}, Daniel H. and {McLure}, Ross J. and {McPartland}, Conor and {Mobasher}, Bahram and {Moreira}, Heidi and {Mortlock}, Alice and {Moustakas}, Leonidas A. and {Mozena}, Mark and {Nandra}, Kirpal and {Newman}, Jeffrey A. and {Nielsen}, Jennifer L. and {Niemi}, Sami and {Noeske}, Kai G. and {Papovich}, Casey J. and {Pentericci}, Laura and {Pope}, Alexandra and {Primack}, Joel R. and {Ravindranath}, Swara and {Reddy}, Naveen A. and {Renzini}, Alvio and {Rix}, Hans-Walter and {Robaina}, Aday R. and {Rosario}, David J. and {Rosati}, Piero and {Salimbeni}, Sara and {Scarlata}, Claudia and {Siana}, Brian and {Simard}, Luc and {Smidt}, Joseph and {Snyder}, Diana and {Somerville}, Rachel S. and {Spinrad}, Hyron and {Straughn}, Amber N. and {Telford}, Olivia and {Teplitz}, Harry I. and {Trump}, Jonathan R. and {Vargas}, Carlos and {Villforth}, Carolin and {Wagner}, Cory R. and {Wandro}, Pat and {Wechsler}, Risa H. and {Weiner}, Benjamin J. and {Wiklind}, Tommy and {Wild}, Vivienne and {Wilson}, Grant and {Wuyts}, Stijn and {Yun}, Min S.},
        title = "{CANDELS: The Cosmic Assembly Near-infrared Deep Extragalactic Legacy Survey{\textemdash}The Hubble Space Telescope Observations, Imaging Data Products, and Mosaics}",
      journal = {\apjs},
         year = 2011,
        month = dec,
       volume = {197},
       number = {2},
          eid = {36},
        pages = {36},
          doi = {10.1088/0067-0049/197/2/36},
archivePrefix = {arXiv},
       eprint = {1105.3754},
 primaryClass = {astro-ph.CO},
       adsurl = {https://ui.adsabs.harvard.edu/abs/2011ApJS..197...36K}
}

@ARTICLE{Planck_2020,
       author = {{Planck Collaboration} and {Aghanim}, N. and {Akrami}, Y. and {Ashdown}, M. and {Aumont}, J. and {Baccigalupi}, C. and {Ballardini}, M. and {Banday}, A.~J. and {Barreiro}, R.~B. and {Bartolo}, N. and {Basak}, S. and {Battye}, R. and {Benabed}, K. and {Bernard}, J. -P. and {Bersanelli}, M. and {Bielewicz}, P. and {Bock}, J.~J. and {Bond}, J.~R. and {Borrill}, J. and {Bouchet}, F.~R. and {Boulanger}, F. and {Bucher}, M. and {Burigana}, C. and {Butler}, R.~C. and {Calabrese}, E. and {Cardoso}, J. -F. and {Carron}, J. and {Challinor}, A. and {Chiang}, H.~C. and {Chluba}, J. and {Colombo}, L.~P.~L. and {Combet}, C. and {Contreras}, D. and {Crill}, B.~P. and {Cuttaia}, F. and {de Bernardis}, P. and {de Zotti}, G. and {Delabrouille}, J. and {Delouis}, J. -M. and {Di Valentino}, E. and {Diego}, J.~M. and {Dor{\'e}}, O. and {Douspis}, M. and {Ducout}, A. and {Dupac}, X. and {Dusini}, S. and {Efstathiou}, G. and {Elsner}, F. and {En{\ss}lin}, T.~A. and {Eriksen}, H.~K. and {Fantaye}, Y. and {Farhang}, M. and {Fergusson}, J. and {Fernandez-Cobos}, R. and {Finelli}, F. and {Forastieri}, F. and {Frailis}, M. and {Fraisse}, A.~A. and {Franceschi}, E. and {Frolov}, A. and {Galeotta}, S. and {Galli}, S. and {Ganga}, K. and {G{\'e}nova-Santos}, R.~T. and {Gerbino}, M. and {Ghosh}, T. and {Gonz{\'a}lez-Nuevo}, J. and {G{\'o}rski}, K.~M. and {Gratton}, S. and {Gruppuso}, A. and {Gudmundsson}, J.~E. and {Hamann}, J. and {Handley}, W. and {Hansen}, F.~K. and {Herranz}, D. and {Hildebrandt}, S.~R. and {Hivon}, E. and {Huang}, Z. and {Jaffe}, A.~H. and {Jones}, W.~C. and {Karakci}, A. and {Keih{\"a}nen}, E. and {Keskitalo}, R. and {Kiiveri}, K. and {Kim}, J. and {Kisner}, T.~S. and {Knox}, L. and {Krachmalnicoff}, N. and {Kunz}, M. and {Kurki-Suonio}, H. and {Lagache}, G. and {Lamarre}, J. -M. and {Lasenby}, A. and {Lattanzi}, M. and {Lawrence}, C.~R. and {Le Jeune}, M. and {Lemos}, P. and {Lesgourgues}, J. and {Levrier}, F. and {Lewis}, A. and {Liguori}, M. and {Lilje}, P.~B. and {Lilley}, M. and {Lindholm}, V. and {L{\'o}pez-Caniego}, M. and {Lubin}, P.~M. and {Ma}, Y. -Z. and {Mac{\'\i}as-P{\'e}rez}, J.~F. and {Maggio}, G. and {Maino}, D. and {Mandolesi}, N. and {Mangilli}, A. and {Marcos-Caballero}, A. and {Maris}, M. and {Martin}, P.~G. and {Martinelli}, M. and {Mart{\'\i}nez-Gonz{\'a}lez}, E. and {Matarrese}, S. and {Mauri}, N. and {McEwen}, J.~D. and {Meinhold}, P.~R. and {Melchiorri}, A. and {Mennella}, A. and {Migliaccio}, M. and {Millea}, M. and {Mitra}, S. and {Miville-Desch{\^e}nes}, M. -A. and {Molinari}, D. and {Montier}, L. and {Morgante}, G. and {Moss}, A. and {Natoli}, P. and {N{\o}rgaard-Nielsen}, H.~U. and {Pagano}, L. and {Paoletti}, D. and {Partridge}, B. and {Patanchon}, G. and {Peiris}, H.~V. and {Perrotta}, F. and {Pettorino}, V. and {Piacentini}, F. and {Polastri}, L. and {Polenta}, G. and {Puget}, J. -L. and {Rachen}, J.~P. and {Reinecke}, M. and {Remazeilles}, M. and {Renzi}, A. and {Rocha}, G. and {Rosset}, C. and {Roudier}, G. and {Rubi{\~n}o-Mart{\'\i}n}, J.~A. and {Ruiz-Granados}, B. and {Salvati}, L. and {Sandri}, M. and {Savelainen}, M. and {Scott}, D. and {Shellard}, E.~P.~S. and {Sirignano}, C. and {Sirri}, G. and {Spencer}, L.~D. and {Sunyaev}, R. and {Suur-Uski}, A. -S. and {Tauber}, J.~A. and {Tavagnacco}, D. and {Tenti}, M. and {Toffolatti}, L. and {Tomasi}, M. and {Trombetti}, T. and {Valenziano}, L. and {Valiviita}, J. and {Van Tent}, B. and {Vibert}, L. and {Vielva}, P. and {Villa}, F. and {Vittorio}, N. and {Wandelt}, B.~D. and {Wehus}, I.~K. and {White}, M. and {White}, S.~D.~M. and {Zacchei}, A. and {Zonca}, A.},
        title = "{Planck 2018 results. VI. Cosmological parameters}",
      journal = {\aap},
         year = 2020,
        month = sep,
       volume = {641},
          eid = {A6},
        pages = {A6},
          doi = {10.1051/0004-6361/201833910},
archivePrefix = {arXiv},
       eprint = {1807.06209},
 primaryClass = {astro-ph.CO},
       adsurl = {https://ui.adsabs.harvard.edu/abs/2020A&A...641A...6P}
}

@ARTICLE{Sheth_2010,
       author = {{Sheth}, Kartik and {Regan}, Michael and {Hinz}, Joannah L. and {Gil de Paz}, Armando and {Men{\'e}ndez-Delmestre}, Kar{\'\i}n and {Mu{\~n}oz-Mateos}, Juan-Carlos and {Seibert}, Mark and {Kim}, Taehyun and {Laurikainen}, Eija and {Salo}, Heikki and {Gadotti}, Dimitri A. and {Laine}, Jarkko and {Mizusawa}, Trisha and {Armus}, Lee and {Athanassoula}, E. and {Bosma}, Albert and {Buta}, Ronald J. and {Capak}, Peter and {Jarrett}, Thomas H. and {Elmegreen}, Debra M. and {Elmegreen}, Bruce G. and {Knapen}, Johan H. and {Koda}, Jin and {Helou}, George and {Ho}, Luis C. and {Madore}, Barry F. and {Masters}, Karen L. and {Mobasher}, Bahram and {Ogle}, Patrick and {Peng}, Chien Y. and {Schinnerer}, Eva and {Surace}, Jason A. and {Zaritsky}, Dennis and {Comer{\'o}n}, S{\'e}bastien and {de Swardt}, Bonita and {Meidt}, Sharon E. and {Kasliwal}, Mansi and {Aravena}, Manuel},
        title = "{The Spitzer Survey of Stellar Structure in Galaxies (S4G)}",
      journal = {\pasp},
         year = 2010,
        month = dec,
       volume = {122},
       number = {898},
        pages = {1397},
          doi = {10.1086/657638},
archivePrefix = {arXiv},
       eprint = {1010.1592},
 primaryClass = {astro-ph.CO},
       adsurl = {https://ui.adsabs.harvard.edu/abs/2010PASP..122.1397S}
}

@ARTICLE{Silva-Lima_2022,
       author = {{Silva-Lima}, Luiz A. and {Martins}, Lucimara P. and {Coelho}, Paula R.~T. and {Gadotti}, Dimitri A.},
        title = "{Revisiting the role of bars in AGN fuelling with propensity score sample matching}",
      journal = {\aap},
         year = 2022,
        month = may,
       volume = {661},
          eid = {A105},
        pages = {A105},
          doi = {10.1051/0004-6361/202142432},
       adsurl = {https://ui.adsabs.harvard.edu/abs/2022A&A...661A.105S}
}

@ARTICLE{Algorry_2017,
       author = {{Algorry}, David G. and {Navarro}, Julio F. and {Abadi}, Mario G. and {Sales}, Laura V. and {Bower}, Richard G. and {Crain}, Robert A. and {Dalla Vecchia}, Claudio and {Frenk}, Carlos S. and {Schaller}, Matthieu and {Schaye}, Joop and {Theuns}, Tom},
        title = "{Barred galaxies in the EAGLE cosmological hydrodynamical simulation}",
      journal = {\mnras},
         year = 2017,
        month = jul,
       volume = {469},
       number = {1},
        pages = {1054-1064},
          doi = {10.1093/mnras/stx1008},
archivePrefix = {arXiv},
       eprint = {1609.05909},
 primaryClass = {astro-ph.GA},
       adsurl = {https://ui.adsabs.harvard.edu/abs/2017MNRAS.469.1054A}
}

@book{Gelman_2003,
    author = {{Gelman}, A. and {Carlin}, J.B. and {Stern}, H.S. and {Rubin}, D.B.},
    title = "{Bayesian Data Analysis (2nd ed.)}",
    publisher = {Chapman and Hall/CRC},
    year = 2003
}

@article{Brown_2001,
author = {Lawrence D. Brown and T. Tony Cai and Anirban DasGupta},
title = {{Interval Estimation for a Binomial Proportion}},
volume = {16},
journal = {Statistical Science},
number = {2},
publisher = {Institute of Mathematical Statistics},
pages = {101 -- 133},
year = {2001},
doi = {10.1214/ss/1009213286},
URL = {https://doi.org/10.1214/ss/1009213286}
}

@ARTICLE{Ferreira_2022,
       author = {{Ferreira}, Leonardo and {Conselice}, Christopher J. and {Sazonova}, Elizaveta and {Ferrari}, Fabricio and {Caruana}, Joseph and {Tohill}, Cl{\'a}r-Br{\'\i}d and {Lucatelli}, Geferson and {Adams}, Nathan and {Irodotou}, Dimitrios and {Marshall}, Madeline A. and {Roper}, Will J. and {Lovell}, Christopher C. and {Verma}, Aprajita and {Austin}, Duncan and {Trussler}, James and {Wilkins}, Stephen M.},
        title = "{The JWST Hubble Sequence: The Rest-frame Optical Evolution of Galaxy Structure at 1.5 < z < 6.5}",
      journal = {\apj},
         year = 2023,
        month = oct,
       volume = {955},
       number = {2},
          eid = {94},
        pages = {94},
          doi = {10.3847/1538-4357/acec76},
archivePrefix = {arXiv},
       eprint = {2210.01110},
 primaryClass = {astro-ph.GA},
       adsurl = {https://ui.adsabs.harvard.edu/abs/2023ApJ...955...94F}
}

@ARTICLE{Buta_1998,
       author = {{Buta}, R. and {Alpert}, Adina J. and {Cobb}, Melinda Lewis and {Crocker}, D.~A. and {Purcell}, Guy B.},
        title = "{An Optical, Near-Infrared, and Kinematic Study of Four Early-Type Resonance Ring Galaxies}",
      journal = {\aj},
         year = 1998,
        month = sep,
       volume = {116},
       number = {3},
        pages = {1142-1162},
          doi = {10.1086/300494},
       adsurl = {https://ui.adsabs.harvard.edu/abs/1998AJ....116.1142B}
}

@ARTICLE{Astropy_2013,
       author = {{Astropy Collaboration} and {Robitaille}, Thomas P. and {Tollerud}, Erik J. and {Greenfield}, Perry and {Droettboom}, Michael and {Bray}, Erik and {Aldcroft}, Tom and {Davis}, Matt and {Ginsburg}, Adam and {Price-Whelan}, Adrian M. and {Kerzendorf}, Wolfgang E. and {Conley}, Alexander and {Crighton}, Neil and {Barbary}, Kyle and {Muna}, Demitri and {Ferguson}, Henry and {Grollier}, Fr{\'e}d{\'e}ric and {Parikh}, Madhura M. and {Nair}, Prasanth H. and {Unther}, Hans M. and {Deil}, Christoph and {Woillez}, Julien and {Conseil}, Simon and {Kramer}, Roban and {Turner}, James E.~H. and {Singer}, Leo and {Fox}, Ryan and {Weaver}, Benjamin A. and {Zabalza}, Victor and {Edwards}, Zachary I. and {Azalee Bostroem}, K. and {Burke}, D.~J. and {Casey}, Andrew R. and {Crawford}, Steven M. and {Dencheva}, Nadia and {Ely}, Justin and {Jenness}, Tim and {Labrie}, Kathleen and {Lim}, Pey Lian and {Pierfederici}, Francesco and {Pontzen}, Andrew and {Ptak}, Andy and {Refsdal}, Brian and {Servillat}, Mathieu and {Streicher}, Ole},
        title = "{Astropy: A community Python package for astronomy}",
      journal = {\aap},
         year = 2013,
        month = oct,
       volume = {558},
          eid = {A33},
        pages = {A33},
          doi = {10.1051/0004-6361/201322068},
archivePrefix = {arXiv},
       eprint = {1307.6212},
 primaryClass = {astro-ph.IM},
       adsurl = {https://ui.adsabs.harvard.edu/abs/2013A&A...558A..33A}
}

@ARTICLE{Finkelstein_2023,
       author = {{Finkelstein}, Steven L. and {Bagley}, Micaela B. and {Ferguson}, Henry C. and {Wilkins}, Stephen M. and {Kartaltepe}, Jeyhan S. and {Papovich}, Casey and {Yung}, L.~Y. Aaron and {Haro}, Pablo Arrabal and {Behroozi}, Peter and {Dickinson}, Mark and {Kocevski}, Dale D. and {Koekemoer}, Anton M. and {Larson}, Rebecca L. and {Le Bail}, Aur{\'e}lien and {Morales}, Alexa M. and {P{\'e}rez-Gonz{\'a}lez}, Pablo G. and {Burgarella}, Denis and {Dav{\'e}}, Romeel and {Hirschmann}, Michaela and {Somerville}, Rachel S. and {Wuyts}, Stijn and {Bromm}, Volker and {Casey}, Caitlin M. and {Fontana}, Adriano and {Fujimoto}, Seiji and {Gardner}, Jonathan P. and {Giavalisco}, Mauro and {Grazian}, Andrea and {Grogin}, Norman A. and {Hathi}, Nimish P. and {Hutchison}, Taylor A. and {Jha}, Saurabh W. and {Jogee}, Shardha and {Kewley}, Lisa J. and {Kirkpatrick}, Allison and {Long}, Arianna S. and {Lotz}, Jennifer M. and {Pentericci}, Laura and {Pierel}, Justin D.~R. and {Pirzkal}, Nor and {Ravindranath}, Swara and {Ryan}, Russell E. and {Trump}, Jonathan R. and {Yang}, Guang and {Bhatawdekar}, Rachana and {Bisigello}, Laura and {Buat}, V{\'e}ronique and {Calabr{\`o}}, Antonello and {Castellano}, Marco and {Cleri}, Nikko J. and {Cooper}, M.~C. and {Croton}, Darren and {Daddi}, Emanuele and {Dekel}, Avishai and {Elbaz}, David and {Franco}, Maximilien and {Gawiser}, Eric and {Holwerda}, Benne W. and {Huertas-Company}, Marc and {Jaskot}, Anne E. and {Leung}, Gene C.~K. and {Lucas}, Ray A. and {Mobasher}, Bahram and {Pandya}, Viraj and {Tacchella}, Sandro and {Weiner}, Benjamin J. and {Zavala}, Jorge A.},
        title = "{CEERS Key Paper. I. An Early Look into the First 500 Myr of Galaxy Formation with JWST}",
      journal = {\apjl},
         year = 2023,
        month = mar,
       volume = {946},
       number = {1},
          eid = {L13},
        pages = {L13},
          doi = {10.3847/2041-8213/acade4},
archivePrefix = {arXiv},
       eprint = {2211.05792},
 primaryClass = {astro-ph.GA},
       adsurl = {https://ui.adsabs.harvard.edu/abs/2023ApJ...946L..13F}
}

@ARTICLE{Adams_2023,
       author = {{Adams}, N.~J. and {Conselice}, C.~J. and {Ferreira}, L. and {Austin}, D. and {Trussler}, J.~A.~A. and {Juod{\v{z}}balis}, I. and {Wilkins}, S.~M. and {Caruana}, J. and {Dayal}, P. and {Verma}, A. and {Vijayan}, A.~P.},
        title = "{Discovery and properties of ultra-high redshift galaxies (9 < z < 12) in the JWST ERO SMACS 0723 Field}",
      journal = {\mnras},
         year = 2023,
        month = jan,
       volume = {518},
       number = {3},
        pages = {4755-4766},
          doi = {10.1093/mnras/stac3347},
archivePrefix = {arXiv},
       eprint = {2207.11217},
 primaryClass = {astro-ph.GA},
       adsurl = {https://ui.adsabs.harvard.edu/abs/2023MNRAS.518.4755A}
}

@ARTICLE{Gadotti_2020,
       author = {{Gadotti}, Dimitri A. and {Bittner}, Adrian and {Falc{\'o}n-Barroso}, Jes{\'u}s and {M{\'e}ndez-Abreu}, Jairo and {Kim}, Taehyun and {Fragkoudi}, Francesca and {de Lorenzo-C{\'a}ceres}, Adriana and {Leaman}, Ryan and {Neumann}, Justus and {Querejeta}, Miguel and {S{\'a}nchez-Bl{\'a}zquez}, Patricia and {Martig}, Marie and {Mart{\'\i}n-Navarro}, Ignacio and {P{\'e}rez}, Isabel and {Seidel}, Marja K. and {van de Ven}, Glenn},
        title = "{Kinematic signatures of nuclear discs and bar-driven secular evolution in nearby galaxies of the MUSE TIMER project}",
      journal = {\aap},
         year = 2020,
        month = nov,
       volume = {643},
          eid = {A14},
        pages = {A14},
          doi = {10.1051/0004-6361/202038448},
archivePrefix = {arXiv},
       eprint = {2009.01852},
 primaryClass = {astro-ph.GA},
       adsurl = {https://ui.adsabs.harvard.edu/abs/2020A&A...643A..14G}
}

@ARTICLE{deSaFreitas_2023,
       author = {{de S{\'a}-Freitas}, Camila and {Fragkoudi}, Francesca and {Gadotti}, Dimitri A. and {Falc{\'o}n-Barroso}, Jes{\'u}s and {Bittner}, Adrian and {S{\'a}nchez-Bl{\'a}zquez}, Patricia and {van de Ven}, Glenn and {Bieri}, Rebekka and {Coccato}, Lodovico and {Coelho}, Paula and {Fahrion}, Katja and {Gon{\c{c}}alves}, Geraldo and {Kim}, Taehyun and {de Lorenzo-C{\'a}ceres}, Adriana and {Martig}, Marie and {Mart{\'\i}n-Navarro}, Ignacio and {Mendez-Abreu}, Jairo and {Neumann}, Justus and {Querejeta}, Miguel},
        title = "{A new method for age-dating the formation of bars in disc galaxies. The TIMER view on NGC1433's old bar and the inside-out growth of its nuclear disc}",
      journal = {\aap},
         year = 2023,
        month = mar,
       volume = {671},
          eid = {A8},
        pages = {A8},
          doi = {10.1051/0004-6361/202244667},
archivePrefix = {arXiv},
       eprint = {2211.07670},
 primaryClass = {astro-ph.GA},
       adsurl = {https://ui.adsabs.harvard.edu/abs/2023A&A...671A...8D}
}

@ARTICLE{Athanassoula_1990,
       author = {{Athanassoula}, E. and {Morin}, S. and {Wozniak}, H. and {Puy}, D. and {Pierce}, M.~J. and {Lombard}, J. and {Bosma}, A.},
        title = "{The shape of bars in early-type barred galaxies.}",
      journal = {\mnras},
         year = 1990,
        month = jul,
       volume = {245},
        pages = {130},
       adsurl = {https://ui.adsabs.harvard.edu/abs/1990MNRAS.245..130A}
}

@ARTICLE{Athanassoula_2002,
       author = {{Athanassoula}, E. and {Misiriotis}, A.},
        title = "{Morphology, photometry and kinematics of N -body bars - I. Three models with different halo central concentrations}",
      journal = {\mnras},
         year = 2002,
        month = feb,
       volume = {330},
       number = {1},
        pages = {35-52},
          doi = {10.1046/j.1365-8711.2002.05028.x},
archivePrefix = {arXiv},
       eprint = {astro-ph/0111449},
 primaryClass = {astro-ph},
       adsurl = {https://ui.adsabs.harvard.edu/abs/2002MNRAS.330...35A}
}

@ARTICLE{Barazza_2009,
       author = {{Barazza}, F.~D. and {Jablonka}, P. and {Desai}, V. and {Jogee}, S. and {Arag{\'o}n-Salamanca}, A. and {De Lucia}, G. and {Saglia}, R.~P. and {Halliday}, C. and {Poggianti}, B.~M. and {Dalcanton}, J.~J. and {Rudnick}, G. and {Milvang-Jensen}, B. and {Noll}, S. and {Simard}, L. and {Clowe}, D.~I. and {Pell{\'o}}, R. and {White}, S.~D.~M. and {Zaritsky}, D.},
        title = "{Frequency and properties of bars in cluster and field galaxies at intermediate redshifts}",
      journal = {\aap},
         year = 2009,
        month = apr,
       volume = {497},
       number = {3},
        pages = {713-728},
          doi = {10.1051/0004-6361/200810352},
archivePrefix = {arXiv},
       eprint = {0902.4080},
 primaryClass = {astro-ph.CO},
       adsurl = {https://ui.adsabs.harvard.edu/abs/2009A&A...497..713B}
}

@ARTICLE{Sanders_1980,
       author = {{Sanders}, R.~H. and {Tubbs}, A.~D.},
        title = "{Gas as a tracer of barred spiral dynamics}",
      journal = {\apj},
         year = 1980,
        month = feb,
       volume = {235},
        pages = {803-820},
          doi = {10.1086/157683},
       adsurl = {https://ui.adsabs.harvard.edu/abs/1980ApJ...235..803S}
}

@ARTICLE{Allard_2006,
       author = {{Allard}, E.~L. and {Knapen}, J.~H. and {Peletier}, R.~F. and {Sarzi}, M.},
        title = "{The star formation history and evolution of the circumnuclear region of M100}",
      journal = {\mnras},
         year = 2006,
        month = sep,
       volume = {371},
       number = {3},
        pages = {1087-1105},
          doi = {10.1111/j.1365-2966.2006.10751.x},
archivePrefix = {arXiv},
       eprint = {astro-ph/0606490},
 primaryClass = {astro-ph},
       adsurl = {https://ui.adsabs.harvard.edu/abs/2006MNRAS.371.1087A}
}

@ARTICLE{Lorenzo_2012,
       author = {{de Lorenzo-C{\'a}ceres}, A. and {Vazdekis}, A. and {Aguerri}, J.~A.~L. and {Corsini}, E.~M. and {Debattista}, Victor P.},
        title = "{Constraining the formation of inner bars: photometry, kinematics and stellar populations in NGC 357}",
      journal = {\mnras},
         year = 2012,
        month = feb,
       volume = {420},
       number = {2},
        pages = {1092-1106},
          doi = {10.1111/j.1365-2966.2011.20100.x},
archivePrefix = {arXiv},
       eprint = {1111.1718},
 primaryClass = {astro-ph.CO},
       adsurl = {https://ui.adsabs.harvard.edu/abs/2012MNRAS.420.1092D}
}

@ARTICLE{Gadotti_2007,
       author = {{Gadotti}, D.~A. and {Athanassoula}, E. and {Carrasco}, L. and {Bosma}, A. and {de Souza}, R.~E. and {Recillas}, E.},
        title = "{Near-infrared surface photometry of a sample of barred galaxies}",
      journal = {\mnras},
         year = 2007,
        month = nov,
       volume = {381},
       number = {3},
        pages = {943-961},
          doi = {10.1111/j.1365-2966.2007.12295.x},
archivePrefix = {arXiv},
       eprint = {0707.4599},
 primaryClass = {astro-ph},
       adsurl = {https://ui.adsabs.harvard.edu/abs/2007MNRAS.381..943G}
}

@Inbook{Athanassoula_2003,
author="Athanassoula, Lia",
title="Angular Momentum Redistribution and the Evolution and Morphology of Bars",
bookTitle="Galaxies and Chaos",
year="2003",
publisher="Springer Berlin Heidelberg",
address="Berlin, Heidelberg",
pages="313--326",
isbn="978-3-540-45040-5",
doi="10.1007/978-3-540-45040-5\_26",
url="https://doi.org/10.1007/978-3-540-45040-5\_26"
}

@misc{Bradley_2022,
author       = {Larry Bradley and
                Brigitta Sipőcz and
                Thomas Robitaille and
                Erik Tollerud and
                Zé Vinícius and
                Christoph Deil and
                Kyle Barbary and
                Tom J Wilson and
                Ivo Busko and
                Axel Donath and
                Hans Moritz Günther and
                Mihai Cara and
                P. L. Lim and
                Sebastian Meßlinger and
                Simon Conseil and
                Azalee Bostroem and
                Michael Droettboom and
                E. M. Bray and
                Lars Andersen Bratholm and
                Geert Barentsen and
                Matt Craig and
                Shivangee Rathi and
                Sergio Pascual and
                Gabriel Perren and
                Iskren Y. Georgiev and
                Miguel de Val-Borro and
                Wolfgang Kerzendorf and
                Yoonsoo P. Bach and
                Bruno Quint and
                Harrison Souchereau},
title        = {astropy/photutils: 1.5.0},
month        = jul,
year         = 2022,
publisher    = {Zenodo},
version      = {1.5.0},
doi          = {10.5281/zenodo.6825092},
url          = {https://doi.org/10.5281/zenodo.6825092}
}

@ARTICLE{Adams_2024,
       author = {{Adams}, Nathan J. and {Conselice}, Christopher J. and {Austin}, Duncan and {Harvey}, Thomas and {Ferreira}, Leonardo and {Trussler}, James and {Juod{\v{z}}balis}, Ignas and {Li}, Qiong and {Windhorst}, Rogier and {Cohen}, Seth H. and {Jansen}, Rolf A. and {Summers}, Jake and {Tompkins}, Scott and {Driver}, Simon P. and {Robotham}, Aaron and {D'Silva}, Jordan C.~J. and {Yan}, Haojing and {Coe}, Dan and {Frye}, Brenda and {Grogin}, Norman A. and {Koekemoer}, Anton M. and {Marshall}, Madeline A. and {Pirzkal}, Nor and {Ryan}, Russell E. and {Maksym}, W. Peter and {Rutkowski}, Michael J. and {Willmer}, Christopher N.~A. and {Hammel}, Heidi B. and {Nonino}, Mario and {Bhatawdekar}, Rachana and {Wilkins}, Stephen M. and {Bradley}, Larry D. and {Broadhurst}, Tom and {Cheng}, Cheng and {Dole}, Herv{\'e} and {Hathi}, Nimish P. and {Zitrin}, Adi},
        title = "{EPOCHS. II. The Ultraviolet Luminosity Function from 7.5 < z < 13.5 Using 180 arcmin$^{2}$ of Deep, Blank Fields from the PEARLS Survey and Public JWST Data}",
      journal = {\apj},
         year = 2024,
        month = apr,
       volume = {965},
       number = {2},
          eid = {169},
        pages = {169},
          doi = {10.3847/1538-4357/ad2a7b},
archivePrefix = {arXiv},
       eprint = {2304.13721},
 primaryClass = {astro-ph.GA},
       adsurl = {https://ui.adsabs.harvard.edu/abs/2024ApJ...965..169A}
}

@ARTICLE{Harvey_2024,
       author = {{Harvey}, Thomas and {Conselice}, Christopher J. and {Adams}, Nathan J. and {Austin}, Duncan and {Juod{\v{z}}balis}, Ignas and {Trussler}, James and {Li}, Qiong and {Ormerod}, Katherine and {Ferreira}, Leonardo and {Lovell}, Christopher C. and {Duan}, Qiao and {Westcott}, Lewi and {Harris}, Honor and {Bhatawdekar}, Rachana and {Coe}, Dan and {Cohen}, Seth H. and {Caruana}, Joseph and {Cheng}, Cheng and {Driver}, Simon P. and {Frye}, Brenda and {Furtak}, Lukas J. and {Grogin}, Norman A. and {Hathi}, Nimish P. and {Holwerda}, Benne W. and {Jansen}, Rolf A. and {Koekemoer}, Anton M. and {Marshall}, Madeline A. and {Nonino}, Mario and {Vijayan}, Aswin P. and {Wilkins}, Stephen M. and {Windhorst}, Rogier and {Willmer}, Christopher N.~A. and {Yan}, Haojing and {Zitrin}, Adi},
        title = "{EPOCHS. IV. SED Modeling Assumptions and Their Impact on the Stellar Mass Function at 6.5 {\ensuremath{\leq}} z {\ensuremath{\leq}} 13.5 Using PEARLS and Public JWST Observations}",
      journal = {\apj},
         year = 2025,
        month = jan,
       volume = {978},
       number = {1},
          eid = {89},
        pages = {89},
          doi = {10.3847/1538-4357/ad8c29},
archivePrefix = {arXiv},
       eprint = {2403.03908},
 primaryClass = {astro-ph.GA},
       adsurl = {https://ui.adsabs.harvard.edu/abs/2025ApJ...978...89H}
}

%%%%%%%%%%%%%%%%%%%%%%%%%%%%%%%%%%%%%%%%%%%%%%%%%%

%%%%%%%%%%%%%%%%% APPENDICES %%%%%%%%%%%%%%%%%%%%%

\appendix
\onecolumn
\newpage
\section{Projected bar length}
\label{App: A}

\begin{figure*}
	\includegraphics[width=\textwidth]{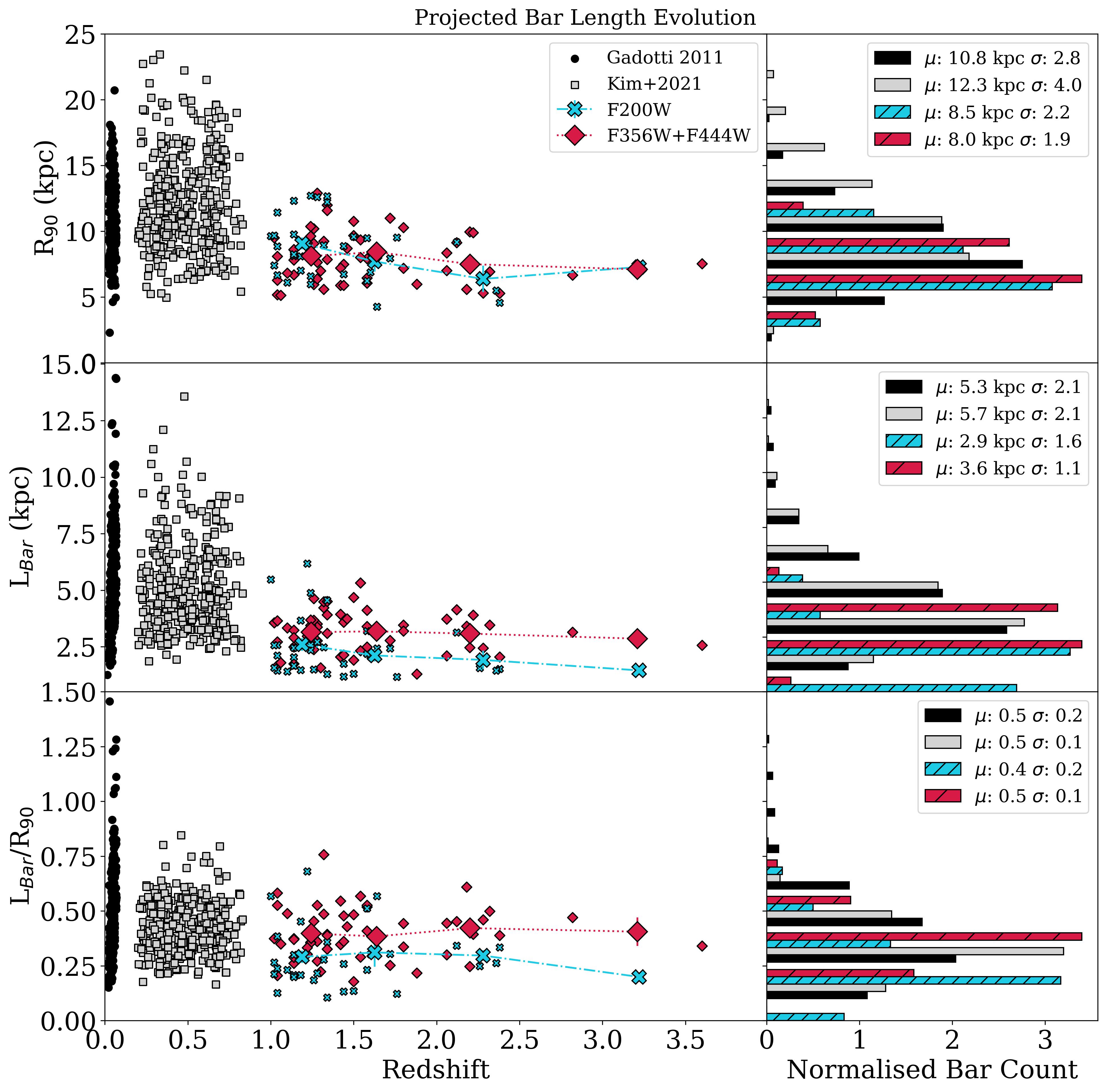}
    \caption{The evolution of the bar length over $0 \leq z \leq 4$. The left panels show lengths measured in this work from JWST NIRCam filters F200W (blue) and F356W+F444W (red) for bars found in galaxies between the redshift range $1 \leq z \leq 4$, with their normalised distribution shown on the right panels. The first row shows the distribution of $R_{90}$, whilst the second row is the projected $L_{bar}$ and the third row is the normalised $L_{bar}$, $L_{bar}/R_{90}$. The high redshift sample is compared against the projected values from a sample of SDSS $i-$band barred galaxies at $z \approx 0$ \citepalias[][black]{Gadotti_2011} and a sample of barred galaxies at $0.2 < z \leq 0.835$ using F814W images from the COSMOS survey \citepalias[][grey]{Kim_2021}. The mean, $\mu$, value for each parameter in each sample is given in the right panel, with the standard deviation, $\sigma$.}
    \label{fig:projected}
\end{figure*}

We emphasise that this study of high-$z$ barred galaxies has focused on the derived deprojected bar lengths. The deprojection technique, described in \S~\ref{Sec: bar length}, is dependent on the inclination of the galaxy, which is derived from the isophotal contours of the disc. This measurement of bar length is critical for making comparative statements to lower redshift studies, which have near-face-on galaxy samples, i.e., \citetalias{Gadotti_2011}. For completeness, we share the evolution of the project bar length from $1 \leq z \leq 4$ in Figure \ref{fig:projected}, in the NIRCam filters F200W and F356W+F444W. No significant differences are found.

%%%%%%%%%%%%%%%%%%%%%%%%%%%%%%%%%%%%%%%%%%%%%%%%%%

% Don't change these lines
\bsp	% typesetting comment
\label{lastpage}
\end{document}